\documentclass[
aps,
amsmath,
amssymb,
aps,
superscriptaddress,
twocolumn,
]{revtex4-2}
\usepackage[T1]{fontenc} 
\usepackage{graphicx}
\usepackage{dcolumn}
\usepackage{bm}
\usepackage{hyperref}
\usepackage{appendix}
\usepackage{xcolor}
\usepackage[normalem]{ulem} %makes underlining possible: \uline

\newcommand{\hide}[1]{}

\newcommand{\ee}{{\rm e}}
\newcommand{\ii}{{\rm i}}
\newcommand{\dd}{{\rm d}}

\DeclareMathOperator*{\argmax}{arg\,max}

\begin{document} 

\title{Many-Body Correlations in Phase-Space Microscopes}

\author{Y. Yang}
\affiliation{TCM Group, Cavendish Laboratory, University of Cambridge, J.J. Thomson Avenue, Cambridge, CB3 0US, United Kingdom\looseness=-1}

\author{C. Weitenberg}
%\email {christof.weitenberg@tu-dortmund.de}
\affiliation{Department of Physics, TU Dortmund University, 44227 Dortmund, Germany}

\author{N. R. Cooper}
\affiliation{TCM Group, Cavendish Laboratory, University of Cambridge, J.J. Thomson Avenue, Cambridge, CB3 0US, United Kingdom\looseness=-1}

%\author{the authors}
%\email {christof.weitenberg@tu-dortmund.de}
%\affiliation{Department of Physics, TU Dortmund University, 44227 Dortmund, Germany}

\begin{abstract}

The recently proposed ``phase-space microscope'' (PSM) extends quantum gas microscopes by allowing joint, single-shot measurements of position and momentum. We show that the $n$-body correlation functions constructed from such phase-space snapshots determine, in principle, all $n$-body observables of a many-body quantum state. This establishes phase-space imaging as a complete probe of strongly correlated matter, beyond what snapshots of position and momentum can provide---a limitation that goes back to a famous question posed by Pauli. Even without the full reconstruction of the density matrices, the PSM provides valuable insight into the quantum states. We demonstrate this power through several applications: witnessing entanglement hidden from both position and momentum, including between spatially separated collective degrees of freedom; resolving pairing structure across the BEC-BCS crossover, including the phase of the pair wavefunction and the reconstruction of its two-body Wigner function; probing inhomogeneous systems and preformed pairs without long-range order, relevant to the pseudogap phase; extracting the interaction energy in a magnetic field; and measurements of the local Chern marker. Furthermore, we discuss phase imprinting in the matter-wave Fourier plane as an alternative method for probing time-reversal-symmetry-broken phases such as chiral $p$-wave superconductors. Together, these results establish the PSM as a versatile new tool for exploring fundamental features of quantum many-body phases in strongly correlated quantum matter.

\end{abstract}

\date{\today}

\maketitle
%\tableofcontents

\subsection{Introduction}
Quantum gas microscopes (QGMs) have become a central experimental tool in the use of ultracold atomic gases as quantum simulators~\cite{Bakr2010,Sherson2010}.
With the ability to take snapshots in position space with single atom resolution, QGMs provide direct measurements of density distributions and their correlation functions, giving valuable insight into the structure of many-body states~\cite{Endres2011,Gross2021}.
Furthermore, by using a pair of ``matter-wave lenses''---pulsed quarter-period evolution in harmonic traps---a magnification of the density distribution allows very high effective spatial resolution ~\cite{Asteria2021,Brandstetter2025}.
This Fourier imaging also gives access to the matter-wave Fourier space, which can be exploited for further manipulation~\cite{Murthy2019,Bruggenjurgen2026}.
Recently, modified versions of these microscopes were developed to go beyond density measurements, such as to probe local phases in optical lattices \cite{Murthy2019,Bruggenjurgen2026,Weitenberg2026}. 

A new proposal for a ``phase-space'' quantum gas microscope takes advantage of these technologies and paves the way for mapping out the phase-space structure of quantum states~\cite{cooper_phase-space_2026}.
This phase-space microscope (PSM) encodes information about the momentum distribution in some auxiliary dimension or spin degree of freedom while largely preserving the position space structure (subject to the Heisenberg uncertainty relation). As a result, the PSM can take snapshots of the system that encode both position and momentum information for each particle. This is a practical implementation of an abstract theoretical setup based on positive operator-valued measures (POVMs)~\cite{Peres1995} to make measurements of incompatible (non-commuting) observables~\cite{Arthurs1965}. 
Separately, position-velocity correlations were previously also studied in cold atomic gases in the classical regime using Raman velocity selection~\cite{Afek2017}.

Position and momentum distributions do not, in general, determine a quantum state---a question that goes back to Pauli~\cite{pauli_1933,corbett_pauli_2006,zheng_pauli_2024}.
Crucially, the PSM provides strictly more information than position- and momentum-space measurements combined, and the phase space information is enough in principle to reconstruct the state.
More precisely, as discussed further below, the PSMs allow measurements of the Husimi Q distribution for each particle~\cite{husimi1940density}.
We therefore call this the Husimi Q mode of the PSM~\cite{cooper_phase-space_2026}.
The Husimi Q distribution is a representation of the phase-space structure of the quantum state, akin to the Wigner distribution~\cite{leonhardt2010essential}.
It is characterized by a resolution parameter $\sigma$, which sets the relative resolution with which position and momentum are simultaneously resolved, subject to the Heisenberg uncertainty principle.
In terms of applications, the PSM has been discussed for studying vortices in superfluids, performing local thermometry, and measuring the local Tan contact characterizing two-body collisions~\cite{cooper_phase-space_2026}. 
\begin{figure}
    \centering
    \includegraphics[width=0.8\linewidth]{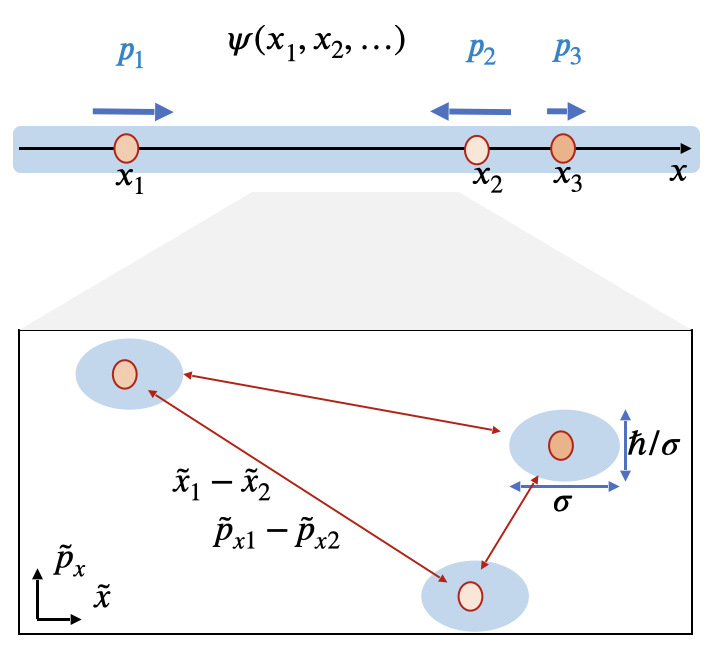}
    \caption{Illustration of the snapshot in phase space for a one-dimensional system. A PSM measures both position and momentum of each particle. The positional resolution, $\sigma$, is controlled by the mapping of the phase space to the auxiliary dimension \cite{cooper_phase-space_2026}. These snapshots allow the extraction of many-body correlations that are studied in this article.}
    \label{fig:phasespacesnapshot}
\end{figure}

In this paper, we take a very general view and ask what information can be extracted about the quantum state from the  multi-point correlations in the phase-space snapshots produced by the PSM (Fig.~\ref{fig:phasespacesnapshot})---samples of the Husimi Q distribution of the many-body state.
We first explain how phase-space correlations contain all information knowable about correlations in general. We then give five example applications where the PSM allows one to go beyond traditional QGM measurements in the investigation of strongly correlated quantum phases of matter, even without the reconstruction of the full quantum states: studying entanglement, studying the crossover from a Bose-Einstein condensate (BEC) to a Bardeen-Cooper-Schrieffer (BCS) superfluid, studying spatially inhomogeneous states, studying the interaction energy in a magnetic field, and measuring the local Chern marker.

More particularly, we show that the PSM can reveal hidden entanglement between two particles, established even when their density distributions, measured separately in position and momentum space, suggest an unentangled state, and that standard entanglement witnesses can be readily adapted to phase-space data, including between spatially separated groups of particles.
Within the BEC-BCS crossover, we show that PSMs enable direct visualizations or measurements of the relative angular momentum of tightly bound pairs and of more complex pairing structures, e.g., of higher angular momentum.
They can also recover the phase of the pair wavefunction, allowing the reconstruction of its two-body Wigner function, which is an object that becomes naturally defined with the introduction of the PSM.
We further show that the PSM can locally resolve order in spatially inhomogeneous systems, for example identifying superfluid puddles in a Bose glass and detecting preformed pairs in the pseudogap phase, where conventional momentum-space signatures can be washed out.
We then show that two-body phase-space correlations in the presence of a (synthetic) magnetic field allow Tan's contact, and hence the interaction energy, to be extracted from a single measurement of the population in a low Landau level.
Finally, we show that the local Chern marker can be reliably obtained through two-body phase-space correlations in a number-conserving Gaussian state.

Separately, we show that the matter-wave Fourier plane on which the PSM is built can also be used to imprint momentum-dependent phases on two-body wavefunctions.
We demonstrate how this can be used to detect the chirality of $p$-wave superconductivity and, combined with the auxiliary dimension of the PSM, to measure BCS wavefunctions with phase information.

Together, these results establish the PSM as a versatile new tool for the tomography and certification of strongly correlated quantum matter.

\subsection{General features of phase-space correlations}
\label{sec:general}
A physical system is completely specified by its quantum state (the wavefunction or the density matrix), so knowledge of the latter grants one the ability to predict all measurement outcomes on the former.
Unfortunately, unless the system is simple or there are symmetries constraining the wavefunctions, measuring the quantum state in experiment is often quite difficult.

It is the goal of quantum state tomography (QST) to accurately reconstruct the quantum state from experimentally available data~\cite{czerwinskiSelectedConceptsQuantum2022,cramerEfficientQuantumState2010,lundeenDirectMeasurementQuantum2011,hofmannQuantumstateTomographySpinl2004,Kanem2005,Brown2023}.
Indeed, the question recalled in the introduction was one of the first problems in QST: Pauli~\cite{pauli_1933} asked if a wavefunction is uniquely determined by the probability distributions in both position and momentum.
It is known that this is generally not the case~\cite{corbett_pauli_2006,zheng_pauli_2024}, and thus it is not always possible to reconstruct the quantum state in a cold-atom experiment from separate QGM data in position space and momentum space.
This is also true in the many-body case, where one is interested not in reconstructing the entire many-body wavefunction but in the $n$-body density matrices \footnote{In simple cases, such as the ground state or states with symmetry constraints, position and momentum space data can be sufficient.}.

For any finite, non-zero $\sigma$, the PSM ($0< \sigma < \infty$) provides enough information to reconstruct a one-particle wavefunction completely.
As stated above, the PSM measures the Husimi Q distributions of all particles.
Focusing for now on a single particle in a one-dimensional system, with coordinate $\hat{x}$ and momentum $\hat{p}$, the Husimi Q distribution is defined as the overlap squared between the quantum state of the particle, $|\Psi\rangle$, and a coherent state in phase space, $Q(\tilde{x}, \tilde{p}) \equiv |\langle \tilde{x}, \tilde{p}|\Psi \rangle|^2$.
For this 1D example, the phase-space coherent state centred at $(\tilde{x}, \tilde{p})$ has wavefunction $\langle x | \tilde{x}, \tilde{p} \rangle = \ee^{-(x-\tilde x)^2/2\sigma^2 + \ii \tilde p x/\hbar}/(\pi\sigma^2)^{1/4}$ with some resolution $\sigma$~\cite{husimi1940density}.
Here and beyond, we use the tilde symbol to denote the phase-space coordinates.
Through this distribution, one can compute the density matrix or, equivalently, the Wigner function of a single particle.
The Wigner function is obtained from the Husimi Q function by~\cite{Zachos2005}
\begin{multline}
    W(\tilde X, \tilde P) \propto \int Q(\tilde x, \tilde p) \ee^{2\ii\tilde \zeta(\tilde p-\tilde P)/\hbar}\ee^{\tilde \zeta^2/\sigma^2} \times {} \\
    \ee^{-\ii \tilde k(\tilde x-\tilde X)} \ee^{\tilde k^2\sigma^2/4} \, \dd \tilde p \, \dd \tilde x \, \dd \tilde k \, \dd\tilde \zeta.
    \label{eq:wigner}
\end{multline}
[See Appendix~\ref{app:nrdm_qst} for further details.]
In practice, due to the multiplication by the inverse Gaussian factors $\ee^{+\tilde k^2\sigma^2/4}$, such evaluations on experimental data can be difficult, and accurate reconstruction requires very precise measurements over all coherent states, suitably chosen resolution $\sigma$, and a Wigner function that has a small bandwidth in Fourier space (\textit{i.e.}, without high frequency oscillations in position or momentum).
The possibility of reconstructing the state is in principle independent of the choice of $\sigma$, though the practicality of it is strongly affected by such choices.
When $\sigma$ is chosen large or small, noise along the position axis or the momentum axis is amplified more, respectively.
Indeed, with $\sigma \to 0$ or $\sigma \to \infty$, PSM measurements reduce to position or momentum space QGM measurements, and reconstruction becomes impossible.
Hence, it is important to choose $\sigma$ according to the bandwidth of the Husimi Q function along the position and momentum axes.
[See Sec.~\ref{sec:pair} where this procedure is carried out for a BCS state.]
Despite this difficulty, in principle, PSMs can be used to measure all properties of a particle.
We also emphasize again that the PSM is useful even without these inversions, as we showcase in the applications (Secs.~\ref{sec:entangle}-\ref{sec:imprint}).

The conclusion generalizes readily to the many-body case, where the PSM can determine $n$-body density matrices, defined as
\begin{multline}
    \rho^{(n)}(x_1, \dots, x_n; x_1', \dots, x_n') \equiv \frac{N!}{(N-n)!}\times {} \\
    \int \Psi(x_1, \dots, x_n, \dots) \Psi^*(x_1', \dots, x_n', \dots) \, \prod_{i=n+1}^N \mathrm dx_i
\end{multline}
where $\Psi(x_1,\dots, x_N)$ is the many-body wavefunction for $N$ particles.
In particular, the one-body and two-body density matrices (OBDM and TBDM) can be obtained.
All one-body observables can be computed from the OBDM, two-body observables from the TBDM, and so on.

A PSM provides samples from the many-particle Husimi Q distribution. That is, it measures diagonal entries of $n$-body density matrices in phase-space coherent states, \textit{e.g.}, $\langle \tilde x, \tilde p| \hat\rho^{(1)} | \tilde x, \tilde p \rangle$ via density in phase space and $\langle \tilde x, \tilde p| \langle \tilde x', \tilde p'|  \hat\rho^{(2)} | \tilde x, \tilde p \rangle | \tilde x', \tilde p' \rangle $ via two-body correlation functions in phase space.
The full snapshot in phase space measures $|\langle \tilde {x}_1, \tilde {p}_1| \dots \langle \tilde {x}_N, \tilde {p}_N| \Psi \rangle|^2$.
These are proportional to $n$-body Husimi Q functions, defined as $Q(\{\tilde{\bm x}_i, \tilde{\bm p}_i\}) \equiv \langle \{\tilde{\bm x}_i, \tilde{\bm p}_i \}| \hat\rho^{(n)} | \{\tilde{\bm x}_i, \tilde{\bm p}_i \} \rangle$, where $i=1\dots n$ and $| \{\tilde{\bm x}_i, \tilde{\bm p}_i \} \rangle\equiv | \tilde{\bm x}_1, \tilde{\bm p}_1 \rangle \dots | \tilde{\bm x}_n, \tilde{\bm p}_n \rangle$.
This is similar to the generalizations proposed in Ref.~\cite{sugita_moments_2003}.

Analogously to the one-particle case, the OBDM is not determined by the density or momentum distributions, but can be obtained through the expectation values in all coherent states.
The TBDM can similarly be obtained through expectation values in tensor products of coherent states, $|\tilde x, \tilde p\rangle |\tilde x',\tilde p'\rangle$.
This can be proven with general theorems in QST~\cite{czerwinskiSelectedConceptsQuantum2022}, or one can view it as a reconstruction in a space with double the dimensionality (see Appendix~\ref{app:nrdm_qst}).
These conclusions apply to general $n$-body density matrices, too: measurements of $n$-body correlation functions in phase space in principle determine the $n$-body density matrices completely.

Here, we also mention that, specifically for a Gaussian state where Wick's theorem applies, one can obtain the off-diagonal elements of the OBDM by measuring the two-body Husimi Q function, without requiring the explicit inversion.
In a particle-number conserving state, one obtains
\begin{multline}
\langle \tilde{\bm x}_1, \tilde{\bm p}_1| \langle \tilde{\bm x}_2, \tilde{\bm p}_2| \hat{\rho}^{(2)} | \tilde{\bm x}_1, \tilde{\bm p}_1 \rangle | \tilde{\bm x}_2, \tilde{\bm p}_2 \rangle = \\
\langle \tilde{\bm x}_1, \tilde{\bm p}_1| \hat{\rho}^{(1)} | \tilde{\bm x}_1, \tilde{\bm p}_1 \rangle \langle \tilde{\bm x}_2, \tilde{\bm p}_2| \hat{\rho}^{(1)} | \tilde{\bm x}_2, \tilde{\bm p}_2 \rangle \pm {}\\
|\langle \tilde{\bm x}_1, \tilde{\bm p}_1| \hat{\rho}^{(1)} | \tilde{\bm x}_2, \tilde{\bm p}_2 \rangle|^2,
\end{multline}
with plus (minus) sign for bosons (fermions).
This thus allows the determination of the magnitudes of the (off-diagonal) matrix elements $|\langle \tilde{\bm x}_1, \tilde{\bm p}_1| \hat{\rho}^{(1)} | \tilde{\bm x}_2, \tilde{\bm p}_2 \rangle|$, and we show now that the phase can be retrieved more stably than from $Q(\tilde{\bm x}, \tilde{\bm p})$ alone.
We outline one possible way that utilizes the analyticity of these matrix elements below.
To do so, we label the coherent states with complex coordinates $\alpha, \alpha^*$ instead of $\tilde{\bm x}, \tilde{\bm p}$.
As discussed in Appendix~\ref{app:nrdm_qst}, $\varrho(\alpha, {\alpha'}^*) \equiv \ee^{(|\alpha|^2+|\alpha'|^2)/2} \langle \alpha | \hat{\rho}^{(1)} | \alpha' \rangle$ is sesquianalytic---\textit{i.e.}, holomorphic in its first argument and anti-holomorphic in its second.
Thus its logarithm $\ln\varrho = \ln|\varrho| + \ii\varphi$, where $\varphi=\arg\varrho$ is the phase of $\varrho$, is also locally sesquianalytic, so one can apply the Cauchy--Riemann equations to relate derivatives of $\varphi$ to those of the measured $\ln|\varrho| = (|\alpha|^2+|\alpha'|^2)/2 + \ln |\langle \alpha | \hat{\rho}^{(1)} | \alpha' \rangle|$.
There is no longer a multiplication with an inverse Gaussian factor, and noise only enters through the relative errors in $|\langle \alpha | \hat{\rho}^{(1)} | \alpha' \rangle|$.
Thus, the OBDM can be more stably recovered in Gaussian states.

Two remarks on the scope of what follows are in order. First, beyond reconstructing $n$-body density matrices from their Husimi Q functions, the PSM can more generally be used to probe components or general functions of momentum together with position, as discussed in Ref.~\cite{cooper_phase-space_2026} (see also Appendix~\ref{app:prac}).
By components we mean measuring position together with a subset of the  momentum components [e.g.\ $(x,\tilde y,\tilde p_y)$], which is easier to realize than measurements of the full $2d$-dimensional phase space; by functions we mean coupling to a chosen function  of momentum (e.g.,\ $|\tilde{\bm p}|^2$), which gives a reduced view of phase space  tailored to the question at hand. We make use of this flexibility in several of the applications below.
Second, while we focus throughout on continuum systems, the same principles apply to lattice systems, either by switching off the lattice suddenly or by using band mapping to isolate correlation effects from single-particle lattice effects (see Appendix~\ref{app:prac}).

\subsection{Entanglement}
\label{sec:entangle}
One direct application of the PSM is to ascertain the entanglement between two particles where conventional QGM measurements fail to do so, offering a new route to certify quantum correlations in strongly interacting many-body systems.
We construct here an example that illustrates this important point, which becomes relevant with the new technology.

Consider first the density matrix $\rho$ of a pure state defined on the Hilbert space $\mathcal{H} = \mathcal{H}_A \otimes \mathcal{H}_B$.
The density matrix is separable iff one can write $\rho = \rho_A \otimes \rho_B$, where $\rho_{A,B}$ are defined on $\mathcal{H}_{A,B}$, respectively.
The state is entangled if the density matrix is not separable in such fashion.

If the density matrix is separable, then the density distributions in position space and momentum space are also separable, since $\langle x| \langle x'| \rho |x\rangle |x'\rangle = \langle x| \rho_A |x\rangle \langle x'| \rho_B |x'\rangle$.
Thus, if the distributions are not separable in position or momentum space, the state is entangled.
However, the converse is not true: a state can have density distributions that are separable in both position and momentum space, yet still be entangled.

With the PSM, the correlation function in phase space must similarly be separable if the density matrix is separable, but the converse is also true, as can be seen from an explicit reconstruction in the fashion of Eq.~\eqref{eq:wigner}.
Therefore, the PSM can establish hidden entanglement even when position and momentum space measurements suggest an unentangled state. 

Here, we give an explicit example in one dimension.
Consider a real wavefunction with parity symmetry, such that it is even or odd $\psi(-x) = \pm \psi(x)$.
We then define two wavefunctions $L(x) = \sqrt{2}\psi(x)\Theta(-x)$ and $R(x) = \sqrt{2}\psi(x)\Theta(x)$ where $\Theta(x)$ is the Heaviside theta function, such that $L$ and $R$ are only non-zero on the left / right half line.
Furthermore, we define another spatial wavefunction $\bar{L}(x)$ to have the shape of the Fourier transform of $L(x)$, or
\begin{equation}
    \bar{L}(x) = \frac{1}{\sqrt{2\pi a^2}}\int L(x') \ee^{-\ii xx'/a^2}\, \mathrm dx'
\end{equation}
for some arbitrary constant $a$, and we define $\bar{R}$ similarly. While $\bar{L}(x) \neq \bar{R}(x)$, due to the parity symmetry, $|\bar{L}(x)| = |\bar{R}(x)|$.

Now, we consider the two-particle state $|\Psi\rangle = (|L \rangle |\bar{L}\rangle + |R \rangle |\bar{R}\rangle)/\sqrt{2}$.
This is not a product state and is entangled. Its density distribution in position space is given by
\begin{multline}
    |\langle x|\langle x'|\Psi\rangle |^2 =  (|L(x)|^2 |\bar{L}(x')|^2 + |R(x)|^2 |\bar{R}(x')|^2)/2 \\ + \operatorname{Re} L(x)\bar{L}(x') R^*(x) \bar{R}^*(x').
\end{multline}
But $L(x)R^*(x)=0$ since the two functions have disjoint support, and using $|\bar{L}(x')|^2 = |\bar{R}(x')|^2$ we find
\begin{equation}
    |\langle x|\langle x'|\Psi\rangle |^2 = (|L(x)|^2+|R(x)|^2) |\bar{L}(x')|^2/2,
\end{equation}
which is separable. The same applies when we Fourier transform into momentum space, as we merely exchange $L$ and $\bar{L}$.

The PSM, by contrast, does not see a separable correlation function.
A coherent state $|\tilde x,\tilde p\rangle$ of finite width $\sigma$ has nonzero overlap with both $|L\rangle$ and $|R\rangle$ (and with both $|\bar L\rangle$ and $|\bar R\rangle$) simultaneously, unlike the sharp position or momentum eigenstates used above.
The interference term $\langle \tilde x,\tilde p|L\rangle\langle \tilde x,\tilde p|R\rangle^*$ therefore no longer vanishes, so the measured phase-space correlation function $Q(\tilde x,\tilde p;\tilde x',\tilde p')$ does not factorize, directly revealing the entanglement that the separate position- and momentum-space measurements would miss.

For a mixed state, separability is defined as the ability to write $\rho = \sum_i p_i\rho_{A,i} \otimes \rho_{B,i}$ for some set of classical probabilities $\{p_i\}$.
Clearly, a separation of the position- and momentum-space correlation functions in this fashion still does not entail separability of the state.
On the other hand, separability in terms of Husimi Q functions such that $Q(\tilde x, \tilde p, \tilde x', \tilde p') = \sum_i p_i Q_{A,i}(\tilde x, \tilde p) Q_{B,i}(\tilde x', \tilde p')$ is exactly equivalent to the separability of density matrices.
The challenge with this is that it can be difficult to verify if the phase-space correlation function is separable in this way.
However, one can appeal to ``entanglement witnesses'' to test the separability of states, which apply equally well to pure and mixed states.

Because of the one-to-one correspondence between Husimi Q functions and density matrices, existing entanglement witnesses can be straightforwardly adapted for use in PSM measurements.
Many entanglement witnesses rely upon the partial transpose of a (bipartite) density matrix.
The Husimi Q function of the transpose of the density matrix is given by flipping the momentum, or $Q_{\rho^\intercal}(\tilde x, \tilde p) = Q_{\rho}(\tilde x, -\tilde p)$, as first noted in the context of Wigner functions~\cite{simon_peres-horodecki_2000}.
As such, partial transposes in the Husimi Q language amounts to flipping the ``momentum'' of part of the system.
The measurements of various moments in the partial transposed density matrix then very straightforwardly translate to measurements of correlations in the momentum-flipped phase space with a PSM, allowing easy implementation of many entanglement witnesses experimentally.

For instance, the EPR-style inequalities $\Delta(\hat{x}_a + \hat x_b) \,\Delta(\hat p_a-\hat p_b) \geq \hbar$ developed in Refs.~\cite{duan_inseparability_2000,mancini_entangling_2002}, whose violation certifies entanglement in any state and which are necessary and sufficient for Gaussian states, can be translated into the language of Husimi Q functions.
To do so, we define
\begin{equation}
    \langle f(\{\tilde x_i, \tilde p_i\}) \rangle \equiv \int Q(\{\tilde x_i, \tilde p_i\}) f(\{\tilde x_i, \tilde p_i\}) \, \prod_{i=1}^N\frac{\dd \tilde x_i \, \dd \tilde p_i}{2\pi\hbar}\,
\end{equation}
in the $N$-particle case, and denote $\tilde u\equiv \tilde x_a+\tilde x_b$ and $\tilde v\equiv \tilde p_a-\tilde p_b$ (note that these are not operators).
In a separable state, the standard deviations of the measured $\tilde u$ and $\tilde v$ satisfy
\begin{equation}
\Delta \tilde u \,\Delta \tilde v \geq 2\hbar\,,
\label{eq:witness}
\end{equation}
so a violation of Eq.~\eqref{eq:witness} certifies entanglement.
The factor of 2 has a simple origin.
Moments of $Q$ are anti-normally ordered, so reordering the ladder operators formed from $\hat{x}$ and $\hat{p}$ adds the noise $\sigma^2$ and $\hbar^2/\sigma^2$ to the variances of $\tilde u$ and $\tilde v$.
This is the fundamental noise of a joint measurement of position and momentum~\cite{Arthurs1965}.
Minimising over the unknown quantum variances turns the separability bound $\Delta\hat u\,\Delta\hat v\ge\hbar$ into Eq.~\eqref{eq:witness}.
These procedures generalize readily to higher-order entanglement witnesses~\cite{agarwal_inseparability_2005,ShchukinVogel2005,Miranowicz2009}.
Since these higher-order criteria do not assume Gaussianity, they offer a plausible route to detecting the kind of entanglement constructed in the 1D example above, which could be invisible to the second moments alone.

Finally, the PSM is not restricted to entanglement between individual particles. 
Because it records the phase-space position of every atom in a single shot, one can construct collective operators of the particles found in any chosen region of space, for example the total position and 
momentum of all atoms in a region $A$, $\hat X_A=\sum_{i\in A}\hat x_i$ and $\hat P_A=\sum_{i\in A}\hat p_i$  (or, in a multicomponent gas, collective spin operators), and likewise for a spatially separated region $B$. 
Witnesses akin to Eq.~\eqref{eq:witness}, or  their higher-order generalizations, can then be developed for $\hat X_A\pm\hat X_B$ and $\hat P_A\mp\hat P_B$.
This allows one to probe entanglement between spatially separated groups of particles, of the kind studied in split Bose--Einstein condensates and atomic ensembles~\cite{Julsgaard2001,Peise2015,Fadel2018,Kunkel2018,Lange2018,Tura2014,Frerot2022}, but now with the regions, and hence the collective operators, chosen freely. 
Since expectation values of arbitrary operators within spatially resolved regions are available from the same data set, no separate measurement is needed for each choice of regions or witness.
Unlike the two-copy interference protocol of Ref.~\cite{Islam2015}, these witnesses require only a single copy of the system and low-order moments, whose statistical cost does not grow exponentially with subsystem size.

\begin{figure*}
    \centering
    \includegraphics[width=0.9\textwidth]{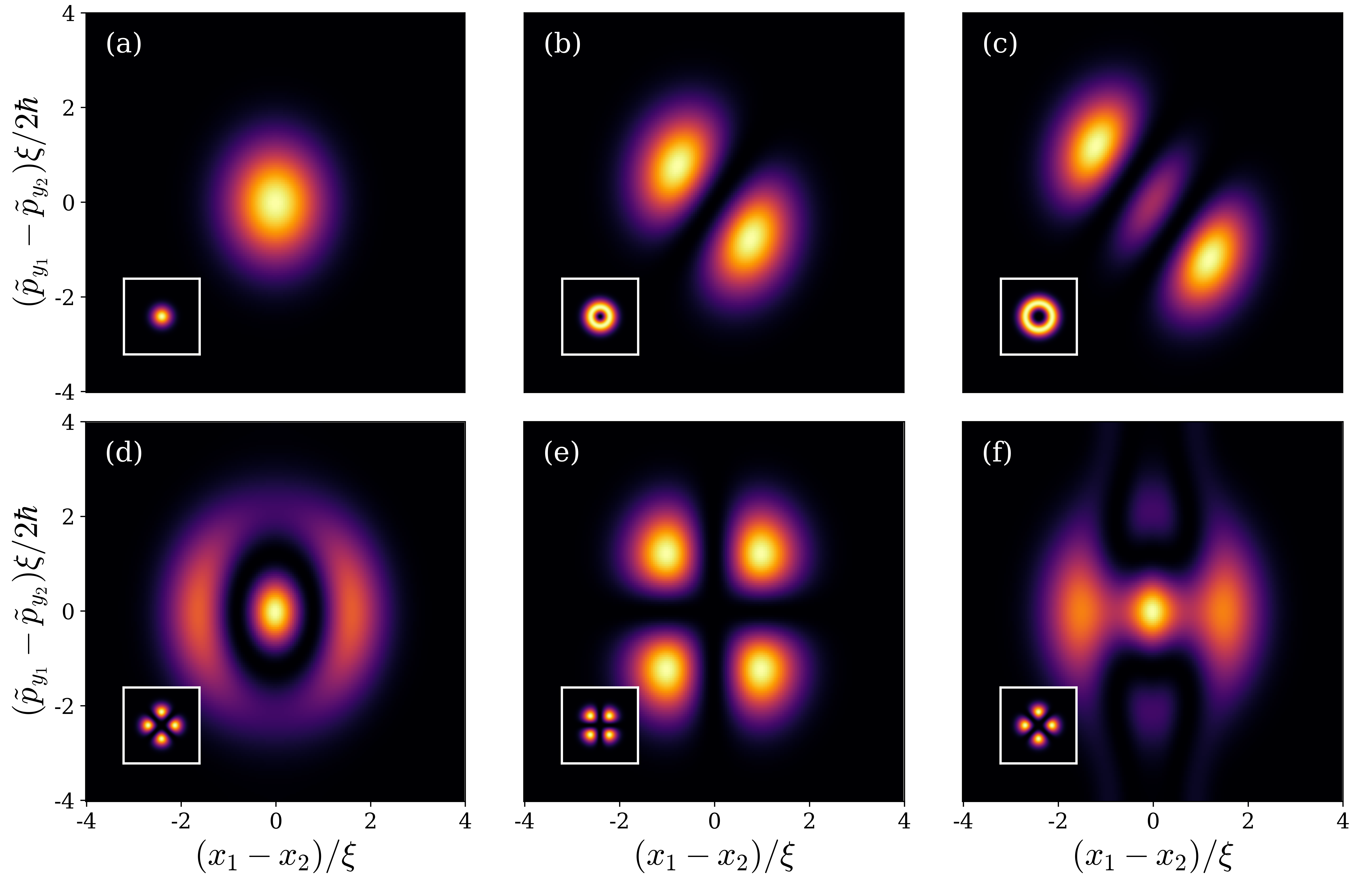}
    \caption{The correlation in phase space $\langle \hat{\rho} (x_1, \tilde y_1=0, \tilde p_{y_1} = p_y) \hat{\rho} (x_2, \tilde y_2=0, \tilde p_{y_2} = -p_y) \rangle_\text{c}$ per unit area measured in a two-body state with relative wavefunction $\psi_\text{pair}({\bm r})$ with momentum space resolution $\sigma=\xi$.
    The values in each panel are scaled to be of similar magnitudes.
    Here, we assume that the auxiliary dimension is coupled only to the $y$-momentum, and thus position space resolution in the $x$-direction is not affected.
    This resolution is chosen so that the PSM can spatially resolve individual particles, while being able to reveal momentum space structures. 
    Note that, because the state is not translationally invariant in momentum space, the correlation function depends on both $\tilde p_{y_1} - \tilde p_{y_2}$ and $\tilde p_{y_1} + \tilde p_{y_2}$.
    As such, we fixed $\tilde p_{y_1} + \tilde p_{y_2}=0$ here.
    On the top row, from left to right, the three panels (a), (b), (c) show $\ell=0, 1, 2$ for $\psi_\text{pair}({\bm r}) \propto (x+\ii y)^\ell \ee^{-r^2/2\xi^2}$, respectively.
    The number of lobes in the correlation function corresponds to the angular momentum, and the orientation of the lobes indicates the chirality of the state or the sign of $\ell$.
    The insets show the density-density correlation functions in the same states, where the value and the sign of $\ell$ cannot be inferred.
    On the bottom row, from left to right, the panels show (d) $\psi_\text{pair}({\bm r}) \propto (x^2-y^2) \ee^{-r^2/2\xi^2}$ for a realistic $d_{x^2-y^2}$ molecule, (e) $\psi_\text{pair}({\bm r}) \propto 2xy \ee^{-r^2/2\xi^2}$ or the $d$-wave molecule rotated by 45 degrees, and (f) $\psi_\text{pair}({\bm r}) \propto |x^2-y^2| \ee^{-r^2/2\xi^2}$.
    Again, the phase-space correlations can distinguish two states that look identical in position-space correlations (panel (d) and panel (f)).}
    \label{fig:bec_povm}
\end{figure*}

\subsection{BEC-BCS crossover}

We now illustrate the use of the PSM to measure strongly-correlated quantum many-body systems.
We focus on the canonical example of the crossover of the superconducting groundstate of spin-1/2 fermions as the interaction strength is varied, from the weakly attractive BCS regime to a BEC of tightly bound pairs~\cite{Ketterle2008,chen_when_2024}.
We consider the states in two-dimensions, so that the original coordinates are position ${\bm r}=(x,y)$ or momentum ${\bm p}=(p_x,p_y)$.
We work in BCS--Leggett mean field theory throughout, which describes a condensate of fermion pairs with zero centre-of-mass momentum.
The state is thus described by a BCS ansatz, defined by the coefficients $u_{\bm p}$ and $v_{\bm p}$ as $|\text{BCS}\rangle = \prod_{\bm{p}} (u_{\bm{p}} + v_{\bm{p}} \hat c_{\bm{p}\uparrow}^\dagger \hat c_{-\bm{p}\downarrow}^\dagger) |\text{vac}\rangle\,$.
The $\hat{c}^\dagger$’s are fermionic creation operators.
In the spirit of QST, if one can measure the \textit{complex} coefficients $u_{\bm p}$ and $v_{\bm p}$, then the state is completely determined.
In particular, one can obtain the structure of the superconducting gap function $\Delta$ through the quantity $u_{\bm p} v_{\bm p}^* = \Delta({\bm p})/2E_{\bm p}$ where $E_{\bm p} = \sqrt{(p^2/2M - \mu)^2 + |\Delta({\bm p})|^2}$ is the quasiparticle energy, $M$ is the particle mass, and $\mu$ is the chemical potential.
For the weakly-coupled BCS regime, we focus on the case of singlet pairing, but the results translate straightforwardly to triplet pairing via correlation measurements of other spin combinations.

As an aside, we note that, while we focus on BCS mean field theory in this section, the PSM can also identify where the mean field approximation fails.
One particularly simple way involves the comparison of kinetic energy-density correlations and density-density correlations, as we explain in Appendix~\ref{app:avg}.

\subsubsection{Angular momentum of BEC pairs}

On the BEC side, measurements of two-body correlations can give detailed insight into the two-body bound states~\cite{Yao2025}.
Consider the wavefunction for a {\it pair} of particles to be $\psi_{\rm pair}({\bm r})$, where ${\bm r} = {\bm r}_1-{\bm r}_2$ is the relative coordinate.
The conventional density-density correlation function $\langle \hat\rho({\bm r}_1)\hat\rho({\bm r}_2)\rangle$ measured by a QGM provides the modulus squared $|\psi_{\rm pair}({\bm r})|^2$ but lacks phase information.
Similarly, correlations in momentum-space imaging yield $|\bar{\psi}_{\rm pair}({\bm p})|^2$, where $\bar{\psi}_\text{pair}({\bm p})$ is the Fourier transform of $\psi_{\rm pair}({\bm r})$.
Quantities like the relative angular momentum within the pair are not easily determined from these magnitude data alone.
If these were all the information available, to measure the relative angular momentum, one can, for instance, look at how $|\psi_{\rm pair}({\bm r})|^2$ vanishes around $\bm{r}=\bm{0}$, since $|\psi_{\rm pair}({\bm r})|^2 \sim r^{2|\ell|}$ for a state with angular momentum $\ell$, but this requires very good sampling of the small densities.

The Husimi Q mode PSM, on the other hand, can directly image the short-range motion in the bound state. In the 2D system considered, we measure for each particle the position $(x,\tilde y)$ along with just one component of the momentum $\tilde p_y$. We thus obtain the correlation function $\langle \hat{\rho}(x_1, \tilde y_1,\tilde p_{y_1}) \hat{\rho} (x_2, \tilde y_2, \tilde p_{y_2}) \rangle$.

For a single molecule, it may appear that a PSM is not required since $\hat{x}$ and $\hat{p}_y$ are not conjugate variables.
However, PSMs are necessary when a realistic gas of BEC molecules is considered because one must do joint measurements of $\hat{y}$ and $\hat{p}_y$ to reliably identify the molecules from the snapshots.
Without $\hat{y}$ information, one cannot tell which two atoms form a pair.
As such, we should choose a resolution $\sigma \sim \xi$ the molecular size, while assuming that $\xi$ is much less than the intermolecular spacing.
The outcome of this measurement is shown in Fig.~\ref{fig:bec_povm} for $\tilde y_1=\tilde y_2=0$ and different $\ell$.

These phase-space correlations show concrete evidence of the relative motions within the pair (for instance, they show how the relative $p_y$ depends on relative $x$) and give easy access to the magnitude of $\ell$ without requiring fine sampling of the near-zero densities.
Moreover, they provide the sign of $\ell$, something unobtainable with position and momentum measurements alone~\cite{Bergschneider2019,Lunt2024,Becher2020}.

\begin{figure*}
    \includegraphics[width=0.9\linewidth]{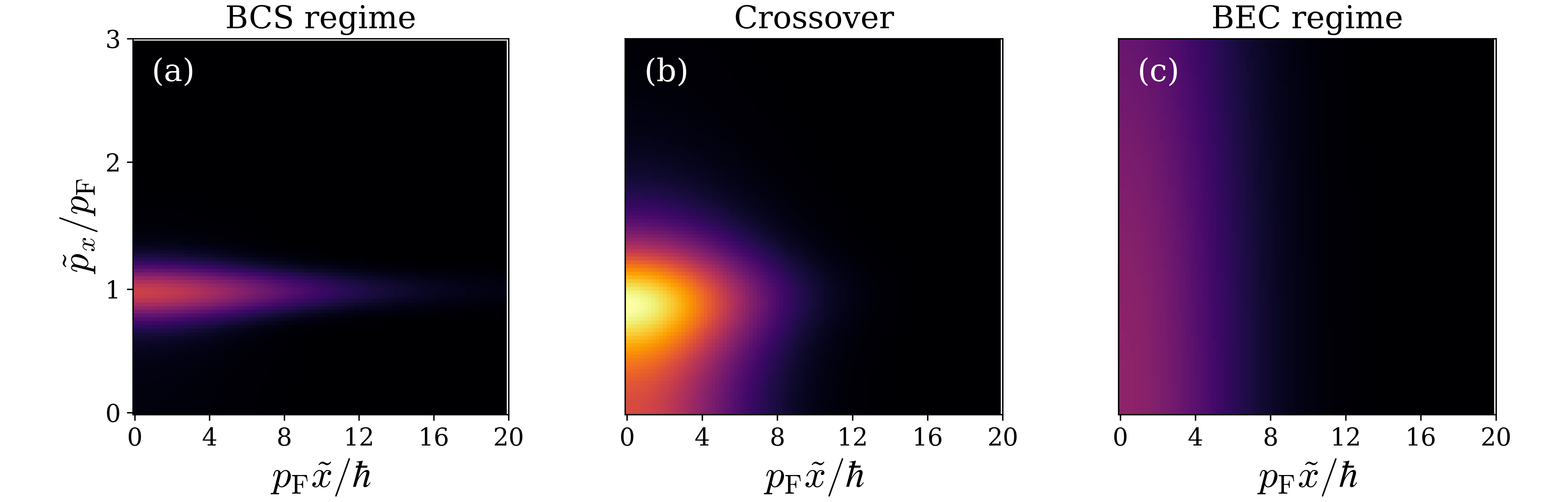}
    \\
    \includegraphics[width=0.9\linewidth]{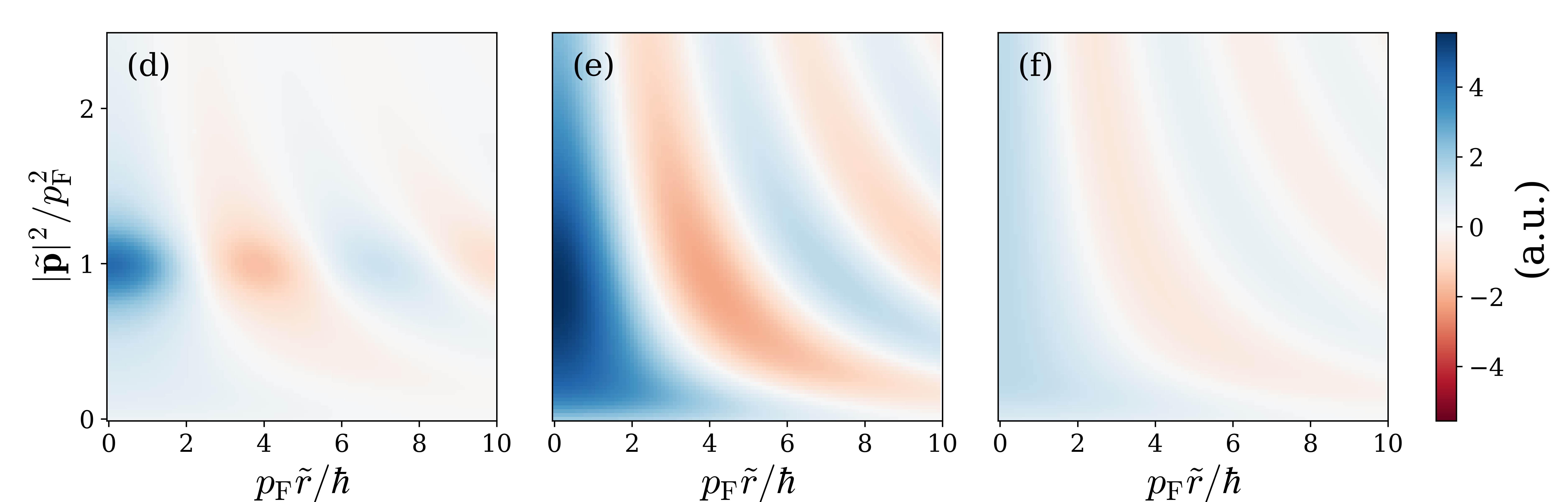}

    \caption{The phase-space correlation functions measured by the PSM across the BEC-BCS crossover of a mean-field $s$-wave superconductor. The left column is the BCS limit ($\Delta/E_\text{F} = 0.10$), the centre is the crossover ($\Delta/E_\text{F} = 0.82$), the right is the BEC limit ($\Delta/E_\text{F} = 15$).
    The top panels (a), (b), and (c) show $\langle \hat\rho_\uparrow (\tilde x, \tilde y=0, \tilde p_x, \tilde p_y=0) \hat\rho_\downarrow (0, 0, -\tilde p_x, 0) \rangle_\text{c}$ as computed via Eq.~\eqref{eq:bcs_povm}.
    The three panels do not have the same normalization; the BCS and BEC limits [panels (a) and (c)] are enhanced to make them visible.
    The resolution is $\sigma=0.7\lambda_\text{F}$ in position space, where $\lambda_\text{F} \equiv 2\pi\hbar/p_\text{F}$ is the Fermi wavelength.
    The bottom panels (d), (e), and (f) show $\langle \hat\rho_\uparrow(\tilde r, |\tilde {\bm p}|^2) \hat\rho_\downarrow(0, |\tilde {\bm p}|^2) \rangle_\text{c}$ as computed via Eq.~\eqref{eq:ring}, without the modulus squared. We plot here the sign of the integral in Eq.~\eqref{eq:ring} for illustration. An experiment would measure the modulus squared, where the fringes would still be visible.
    The resolution is $\sigma=1.0\lambda_\text{F}$ in position space.
    In both measurements, we see that, in the BCS limit, correlation only exists near the Fermi surface $|\tilde{\bm{p}}|\approx p_{\text{F}}$.
    Furthermore, with the ring protocol, the correlations oscillate at wavevector $|\tilde{\bm{p}}|/\hbar$ in spatial separations.
    In the BEC limit, correlations are concentrated near zero separations.
    Note that, in the ring protocol, the oscillations in spatial separations persist even into the BEC regime.
    This is sensitive to the phase coherence of the pairs and could be a useful local test in the pseudogap phase that we discuss in Sec.\ref{sec:inhom}.
    Additionally, such oscillations persist because the momentum space resolution used was high; with lower resolutions (smaller $\sigma$), the oscillations in the BCS regime remain present, but decay faster with $\tilde r$ in the BEC regime.}
    \label{fig:cross}
\end{figure*}

\subsubsection{Imaging pairing in the BCS regime}
\label{sec:pair}

In the BCS limit, the two conventional measurements probe the pair amplitude  $\chi(\bm p)\equiv u_{\bm p}v_{\bm p}^*$ in complementary bases: momentum-space  correlations give $\langle \hat c^\dagger_{\bm{p}\uparrow} \hat c_{\bm{p}\uparrow} \hat c^\dagger_{\bm{p}'\downarrow} \hat c_{\bm{p}'\downarrow} \rangle_{\text{c}} = |\chi(\bm p)|^2 \delta_{\bm{p}+\bm{p}'}$~\cite{Holten2022}, while position-space correlations give  $|g_{\uparrow\downarrow}(\bm r)|^2$, the modulus squared of the Fourier transform of $\chi$~\cite{obeso-jureidini_density_2022}.
Here and henceforth, all correlation functions refer to the connected part only, denoted by $\langle\rangle_{c}$, where $\langle \hat c^\dagger_{\bm{p}\uparrow} \hat c_{\bm{p}\uparrow} \hat c^\dagger_{\bm{p}'\downarrow} \hat c_{\bm{p}'\downarrow} \rangle_{\text{c}} \equiv \langle \hat c^\dagger_{\bm{p}\uparrow} \hat c_{\bm{p}\uparrow} \hat c^\dagger_{\bm{p}'\downarrow} \hat c_{\bm{p}'\downarrow} \rangle - \langle \hat c^\dagger_{\bm{p}\uparrow} \hat c_{\bm{p}\uparrow} \rangle \langle \hat c^\dagger_{\bm{p}'\downarrow} \hat c_{\bm{p}'\downarrow} \rangle$, so the single-body contributions are excluded.
This is precisely Pauli's problem~\cite{pauli_1933}, now for the pair amplitude, with $\chi$ playing the role of a wavefunction.
As in the single-particle case, the two moduli do not in general fix $\chi$.
The relative phases between different momenta, which encode the symmetry of the pairing, are lost.
Since the PSM measures position and momentum jointly, its correlations involve $\chi$ itself and not only magnitudes.  As we show below, $\chi$ (including its phase) can then be reconstructed.

If full phase-space POVMs were performed, we would obtain the correlation function
$\langle \hat\rho_\uparrow (\tilde {\bm r}, \tilde {\bm p}) \hat\rho_\downarrow (\tilde {\bm r}', \tilde {\bm p}') \rangle_\text{c}$,
whose connected part evaluates to
\begin{equation}
\propto \left| \int\mathrm d^2\bm{q}\, \mathrm e^{-\sigma^2(\tilde {\bm p}-{\bm q})^2/2\hbar^2-\sigma^2(\tilde {\bm p}'+{\bm q})^2/2\hbar^2} \mathrm e^{\mathrm i\bm q\cdot (\tilde {\bm r}-\tilde {\bm r}')/\hbar} u_{\bm q}^{\phantom{*}}v_{\bm q}^* \right|^2.
\label{eq:bcs_povm}
\end{equation}
This retains information about the pairing between $+\bm p$ and $-\bm p$.
This is an explicit example of how the joint measurement of conjugate variables allows us to choose the distribution of uncertainties: by tuning $\sigma$, these phase-space measurements interpolate between pure momentum-space ($\sigma\to\infty$, measuring $|\chi({\bm p})|^2$) and pure position-space measurements ($\sigma\to0$, measuring $|g_{\uparrow\downarrow}({\bm r})|^2$).

On the top row of Fig.~\ref{fig:cross}, we show some POVM measurement results in this full 4D phase space. The correlation function is peaked around the Fermi surface in the BCS limit ($|\tilde {\bm p}| \approx p_\text{F}$ the Fermi momentum) and is confined to small separations in the BEC limit.
The control of resolution $\sigma$ in phase-space measurements is also helpful when tracking pairs and pair sizes across the BEC-BCS crossover. 
The resolution here is chosen to highlight the pairing around the Fermi surface in the BCS regime and to show the crossover regime as transitioning between BCS and BEC limits.
In the BEC regime, the pairs are tightly bound molecules, but the correlation function in Fig.~\ref{fig:cross}(c) is resolution-limited, so one does not observe a sharp peak near zero separations.
When one chooses a finer position space resolution, the correlation functions in the BEC regime become more strongly peaked near $\tilde x=0$ and eventually qualitatively approach that shown in Fig.~\ref{fig:bec_povm}(a).
We show in Appendix~\ref{sec:cross} an example of this.
We also discuss there how the crossover regime is signaled by being tightly confined in both position and momentum.

Measuring the full Husimi Q function in four-dimensional phase space may be practically challenging in experiment. However, following a suggestion in Ref.~\cite{cooper_phase-space_2026}, one can also choose to measure $\gamma|\tilde{\bm{p}}|^2$ for some constant $\gamma$ along with position.\footnote{This can be achieved by applying a de-focused Gaussian beam in the Fourier plane, where a potential of the form $Az + Br^2 + Czr^2$ is created (where $A,B,C$ are constants). This contains the required term $V(x,y,z) =C (x^2+y^2)z$, while the gradient $Az$ simply leads to a global shift in $z$ direction and the quadratic term $Br^2$ can be compensated by the length of the matter-wave lens.} Then $\langle \hat\rho_\uparrow(\tilde {\bm r}, |\tilde {\bm p}|^2) \hat\rho_\downarrow(\tilde {\bm r}', |\tilde {\bm p}'|^2) \rangle_\text{c}$ evaluates to
\begin{multline}
    \propto \biggl| \int\mathrm d^2\bm{q}\, \mathrm e^{-\sigma^2 \gamma^2 (|\tilde {\bm p}|^2 - |{\bm q}|^2)^2/2\hbar^2 - \sigma^2 \gamma^2 (|\tilde {\bm p}'|^2-|{\bm q}|^2)^2/2\hbar^2} \times {}\\
    {\mathrm e}^{{\mathrm i}{\bm q}\cdot(\tilde {\bm r}-\tilde {\bm r}')/\hbar} u_{\bm q} v_{\bm q}^* \biggr|^2.
    \label{eq:ring}
\end{multline}
This picks out a ring in momentum space and makes the states contained therein interfere with each other.
In the BCS limit, an example of such a correlator is shown in Fig.~\ref{fig:cross}(d).
This is shown in terms of the radius $\tilde r = |\tilde {\bm r}-\tilde {\bm r}'|$ separating the two particles, and setting $|\tilde {\bm p}|^2=|\tilde {\bm p}'|^2$, since otherwise the correlation is suppressed.
Along the $|\tilde {\bm p}|^2$ axis, one clearly sees that pairing is most concentrated near the Fermi surface.
Along the spatial axis, $\tilde r$, pairs at $\gamma |\tilde {\bm{p}}|^2$ have oscillations at wavevector $|\tilde {\bm{p}}|/\hbar$ and amplitudes that decay exponentially with separation, as is the case in conventional density-density correlation functions~\cite{obeso-jureidini_density_2022}.
A single PSM correlator thus shows both the Fermi-surface pairing, normally visible only in momentum space, and the spatial oscillation and decay of pair correlations, normally visible only in position space.

Similar measurements can be made across the BEC-BCS crossover, and within mean field theory one observes the smooth transition from correlations near the Fermi surface (BCS limit) to correlations near zero separations (BEC limit).
See Fig.~\ref{fig:cross}(d)-(f).

\begin{figure}
    \centering
    \includegraphics[width=0.9\linewidth]{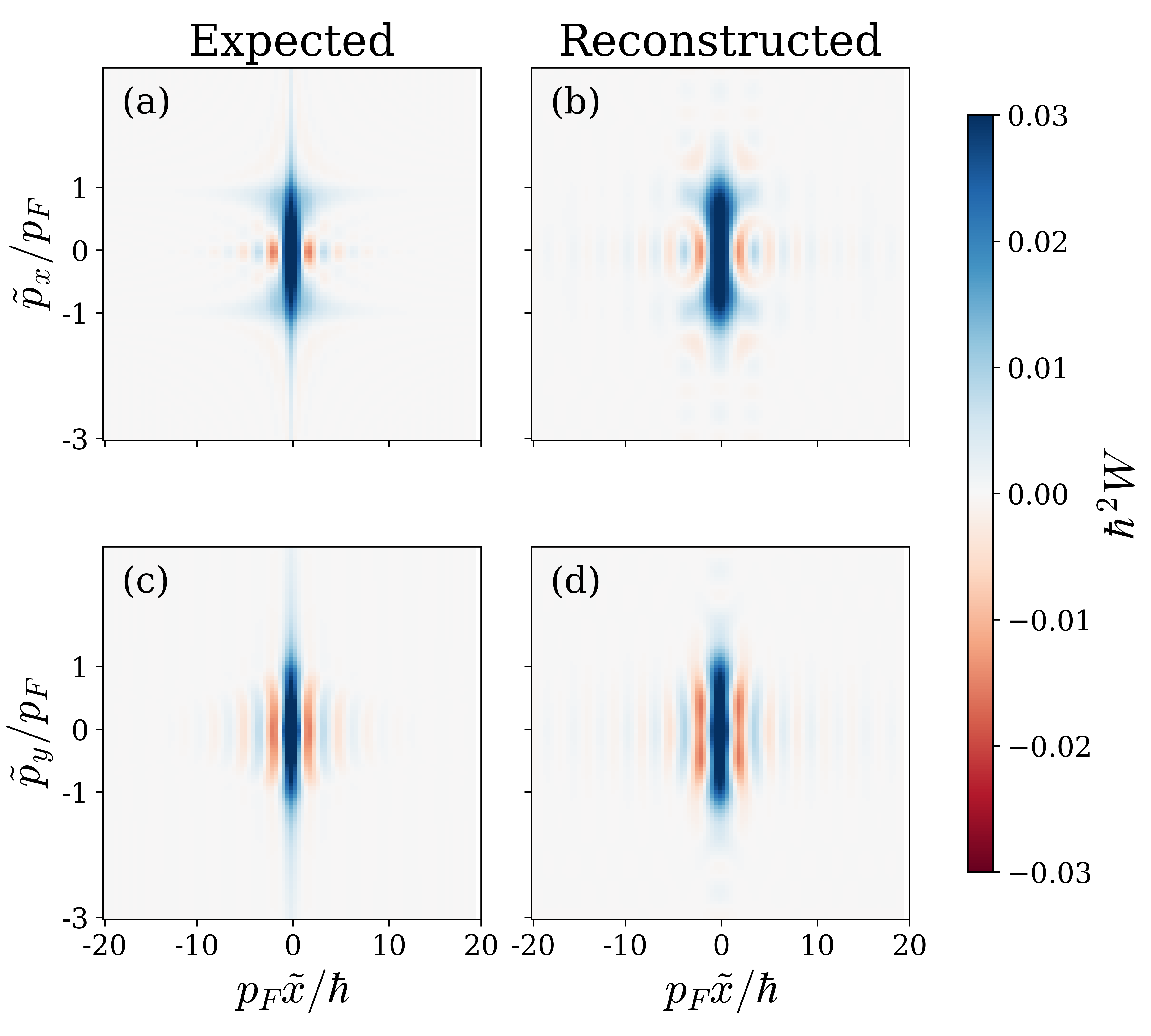}
    \caption{The Wigner function calculated directly from $u_{\bm p} v_{\bm p}^*$ and that reconstructed from the measured two-body correlation function in phase space. Note that the cutoff procedure explained in the main text means that the reconstructed version has a smaller magnitude; as such, they are multiplied here by an arbitrary factor of 1.8 to make them comparable in magnitude to the direct calculations. The top row shows $W(\tilde x, \tilde y=0, \tilde p_x, \tilde p_y=0)$; the bottom row shows $W(\tilde x, \tilde y=0, \tilde p_x=0, \tilde p_y)$. The state is in the BCS regime with $\Delta/E_\text{F} = 0.27$ and clearly yields strongly negative Wigner function values. The resolution used in the phase-space correlation measurement is $\sigma = 0.2\lambda_\text{F}$. The oscillations in the Wigner function are correctly captured in the reconstructed versions.
    }
    \label{fig:wigner}
\end{figure}

The correlators above show where in phase space the pairs are found, but do not directly measure the phase of $\chi(\bm p)$.
To show that this phase is accessible, we reconstruct the Wigner function of the BCS pair amplitude from the measured correlators.
This also raises a question that, to our knowledge, is posed naturally only by the PSM: what is the Wigner function of a many-body state, and is it negative?
In general, introducing relative and centre-of-mass (CoM) coordinates, the two-body Wigner function is defined in terms of the TBDM as
\begin{multline}
    W(\tilde {\bm r}_\text{rel}, \tilde {\bm p}_\text{rel}; \tilde {\bm R}, \tilde {\bm P}) \propto \\
    \int \rho^{(2)}(\tilde {\bm r}_\text{rel}-{\bm h}, \tilde {\bm R}-{\bm H}; \tilde {\bm r}_\text{rel}+{\bm h}, \tilde {\bm R}+{\bm H}) \times {} \\
    \ee^{2\ii {\bm h} \cdot \tilde {\bm p}_\text{rel}/\hbar} \ee^{2\ii {\bm H} \cdot \tilde {\bm P}/\hbar} \, \dd^2 h \, \dd^2 H.
\end{multline}
For the BCS state, we look at the connected, singlet part, which is independent of $\tilde{\bm R}$ and proportional to $\delta(\tilde{\bm P})$, since the pairs condense at zero centre-of-mass momentum. Its relative-coordinate Wigner function becomes (we drop the subscript rel for clarity)
  \begin{equation}
      W(\tilde {\bm r}, \tilde {\bm p}) = \frac{1}{(\pi\hbar)^2} \int
      u_{\tilde {\bm p} + \bm q}^* v_{\tilde {\bm p} + \bm q}\, u_{\tilde {\bm p} - \bm q} v_{\tilde {\bm p} - \bm q}^*
      \, \ee^{-2\ii {\bm q} \cdot \tilde {\bm r}/\hbar} \, \dd^2 q.
  \end{equation}
Comparing with Eq.~\eqref{eq:wigner}, this has exactly the structure of a single-particle Wigner transform, with $u_{\bm p} v_{\bm p}^*$ playing the role of a single-particle wavefunction in momentum space; the single-particle Husimi-to-Wigner reconstruction procedure therefore applies directly to it.
In the BCS limit and the crossover regime, we find this to be oscillatory in $\tilde {\bm{r}}$ and to be negative in parts of phase space, signaling genuinely quantum behaviors.
By explicit construction, we show in Fig.~\ref{fig:wigner} that this structure is recovered from the phase-space correlators (Eq.~\eqref{eq:bcs_povm}) using Eq.~\eqref{eq:wigner}.
In the BEC limit, we find the Wigner function to show much less negativity (for simple $s$-wave pairing functions).

Concretely, according to Eq.~\eqref{eq:wigner}, we Fourier transform the Husimi Q function, multiply by the inverse Gaussian factor $\ee^{+\tilde \zeta^2/\sigma^2 + \tilde k^2\sigma^2/4}$, then inverse Fourier transform to obtain the Wigner function. The key difficulty is that the inverse Gaussian factor will greatly amplify any high frequency noise. As such, we have to impose cutoffs in Fourier space, beyond which we set the Fourier transform to zero.
In practice, the cutoff can be chosen to be the largest value that does not lead to blow ups after multiplication by the inverse Gaussian factor.
Stronger noise means that a smaller cutoff has to be chosen, and more information will be lost.
Similarly, measurements being precise in position space or momentum space (small or large $\sigma^2$) also limits the reconstruction.

A full reconstruction of the Wigner function of a 2D system requires measurements in 4D space.
Nevertheless, the same procedure can also be applied to a BCS state in 1D~\cite{Fuchs2004}, which may be easier to implement experimentally; one expects similar results to the top row shown in Fig.~\ref{fig:wigner}, again yielding negative Wigner functions that can be observed experimentally, allowing a direct application of quantum optics concepts to quantum many-body states.

\subsubsection{Complex pairing structures}
Going beyond $s$-wave, for superconductors with more complex pairing structures, it could be difficult (or indeed impossible) to ascertain their identities on the basis of position- and momentum-space measurements.
Consider a chiral $d$-wave superconductor ($\Delta (\bm{p}) =\Delta_2  (p/p_\text{F})^2 \ee^{2\ii\theta_{\bm p}}$) and a $g$-wave superconductor ($\Delta (\bm{p}) = \Delta_4 (p/p_\text{F})^4 \ee^{4\ii\theta_{\bm p}}$).
Here, $\theta_{\bm p}$ is the azimuthal angle in momentum space.
Note that while the superconducting gap scales as $(p/p_\text{F})^\ell$ for small momenta, such a power law must break down at some larger momentum that we denote by $\hbar\Lambda$, because the gap cannot grow indefinitely.
Since the momentum space correlation functions only depend on $|\Delta(\bm{p})|^2$, one cannot distinguish between the two superconductors using only data around the Fermi surface.
We explore here if position-space correlation functions are more helpful.
They measure
\begin{equation}
    |g_{\uparrow\downarrow}({\bm r})|^2 \propto \Biggl| \int_0^{\hbar\Lambda} \dd p \, \frac{ p (p/p_\text{F})^\ell J_\ell(pr/\hbar)}{\sqrt{(p^2/2M-\mu)^2+\Delta_\ell^2 (p/p_\text{F})^{2\ell}}}\Biggr|^2,
\end{equation}
where $J_\ell$ are Bessel functions of order $\ell$.
At short distances, below the cutoff value $r<1/\Lambda$, the correlation function behaves as $\propto r^4$ in the $d$-wave case and $\propto r^8$ in the $g$-wave case, but this can be very difficult to probe, requiring accurate measurements at short distances.
The correlation function has similar features in both cases for $r\gtrsim 1/\Lambda$.
Explicitly, as $r\to1/\Lambda$ from above, $g_{\uparrow\downarrow}({\bm r})$ contains contributions from high momentum states near the cutoff $ \sim \int^{\hbar\Lambda} \sqrt{p/r} \cos(pr/\hbar + \varphi_{\ell}) \, \dd p$, or $\sim \hbar\sqrt{\Lambda/r^3}\sin(\Lambda r+\varphi_\ell)$ for some phase $\varphi_\ell$.
Hence, $d$-wave and $g$-wave superconductivity can be difficult to tell apart in practice with position and momentum measurements.

In contrast, the PSM can readily distinguish the two cases.
Consider again measuring $\gamma|\tilde{\bm{p}}|^2$ and denote the relative position $\tilde{\bm r}_{\rm rel} = \tilde{\bm r}-\tilde{\bm r}'$ in terms of its polar coordinates, $(\tilde{r}_{\rm rel},\tilde{\theta}_{\rm rel})$. 
In the limit $\sigma\to\infty$, we write $u_{\bm p} v_{\bm p}^* = \sum_\ell c_\ell {\mathrm e}^{{\mathrm i} \ell\theta_{\bm p}}$ as a Fourier series around the ring selected via $\gamma |\tilde {\bm{p}}|^2$. The correlator in Eq.~\eqref{eq:ring} then becomes
\begin{equation}
    \propto \biggl| \sum_{\ell=-\infty}^{+\infty} c_\ell \mathrm e^{{\mathrm i}\ell\tilde \theta_{\rm rel}} {\mathrm i}^\ell J_\ell \bigl(|\tilde {\bm{p}}| \tilde r_{\rm rel} / \hbar \bigr)\biggr|^2.
\end{equation}
A pure phase winding $\mathrm e^{\mathrm i\ell\theta_{\bm p}}$ can be easily identified through the order of the Bessel function.
Choosing the ring to be around the Fermi surface, one can easily distinguish $J_2(p_\text{F} \tilde r_{\rm rel} / \hbar)^2$ and $J_4(p_\text{F} \tilde r_{\rm rel} / \hbar)^2$ with values at distances $p_\text{F} \tilde r_{\rm rel} / \hbar \sim \mathcal{O}(1)$.
The signal is cleaner because the ring selection excludes the states near the cutoff $\Lambda$ that dominate position-space correlations.

\subsection{Inhomogeneous states} \label{sec:inhom}

By virtue of their ability to make joint measurements in real space and momentum space, PSMs allow direct probing of the local order in many inhomogeneous states.
In each case below, the spatial resolution $\sigma$ must be smaller than the length scale over which the order varies, while $\hbar/\sigma$ must still resolve the relevant momentum structure. We begin with one-body order (symmetry-breaking domains and a Bose glass), and then turn to two-body correlations (a pair-density wave and preformed pairs).

The simplest case is a BEC in a lattice that condenses into one of two energy minima in momentum space, spontaneously breaking the symmetry. For example, a BEC on a triangular lattice with frustrated hopping and weak repulsive interactions breaks inversion symmetry, corresponding to condensation in either the K or the K$'$ points of the Brillouin zone~\cite{Struck2011}.  
Imagine that various domains form in the system, with the BEC condensing into different minima at different spatial locations. A time-of-flight image of the momentum distribution will show a spatially averaged response, with all domains contributing, which can wash out the signal of symmetry breaking~\cite{hasan2025,Ozawa2023}.
However, the domain structure can straightforwardly be picked out with the PSM.
It requires only that the spatial resolution $\sigma$ be smaller than the domain size, which readily allows the differences in momentum-space to be distinguished locally, provided the domains are more than a few lattice spacings in size.

\begin{figure}
    \centering
    \includegraphics[width=0.9\linewidth]{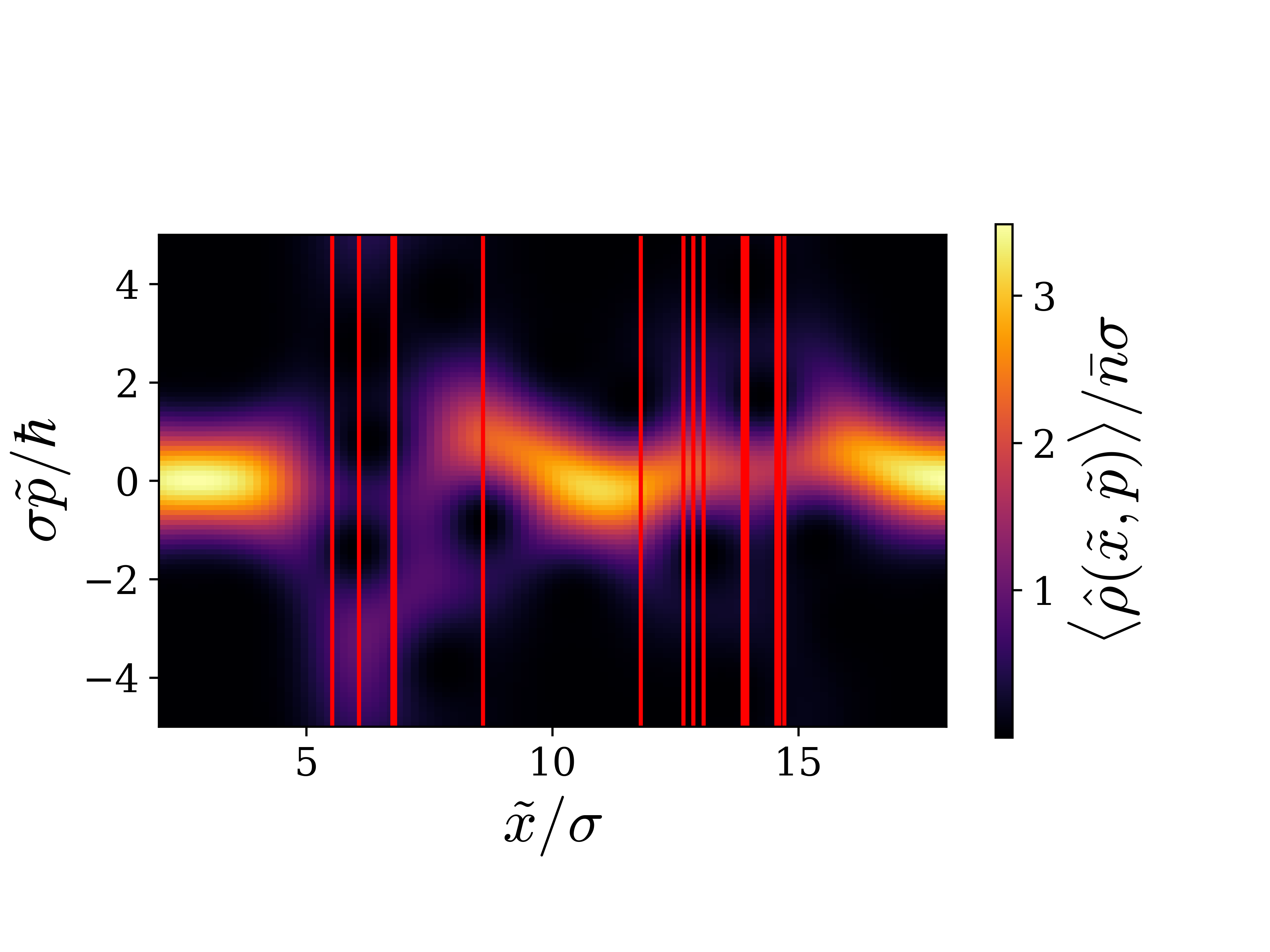}
    \caption{The one-body POVM in the Bose glass toy model as given by Eq.~\eqref{eq:bose_glass}. The red lines indicate the positions of the domain walls $a_n$, chosen such that their spacings follow an exponential distribution here. The phases $\phi_n$ are chosen from a uniform distribution for each domain. It can be seen that, away from domain walls, $\langle \hat\rho({\tilde x,\tilde p}) \rangle$ is concentrated near $\tilde p=0$, as expected for a superfluid, limited only by the resolution $\hbar/\sigma$. Near regions where domain walls are concentrated, $\langle \hat\rho({\tilde x,\tilde p}) \rangle$ becomes much more diffuse and deviates away from $\tilde p=0$. Indeed, the momentum reflects the local phase gradient at any point in space. All of these are achievable within a single experimental realization. The PSM thus powerfully shows local phase order and allows one to go beyond global momentum space imaging.}
    \label{fig:bose_glass}
\end{figure}

Generalizing this, one can envisage using the PSM to image a Bose glass,
which is a phase in the Bose--Hubbard model in the presence of disorder, featuring superfluid puddles in a Mott insulator~\cite{Yu2024,Koehn2026}.
A PSM can identify the positions of those superfluid puddles, as the densities in phase space are concentrated near zero momentum in those puddles.
We consider the following toy model for a Bose glass: divide a one-dimensional state into different domains in intervals $[a_n, a_{n+1}]$ bounded by domain walls at positions $a_n$, and let the order parameter within the $n$th interval be $\sqrt{\bar{n}} \ee^{\ii\phi_n}$ where $\bar{n}$ is the mean particle density.
The density in phase space is
\begin{equation}
\langle \hat\rho({\tilde x,\tilde p}) \rangle = \frac{\bar{n}}{\sqrt{\pi}\sigma}\biggl| \sum_n \int_{a_n}^{a_{n+1}} \ee^{\ii\phi_n} \ee^{-(x'-\tilde x)^2/2\sigma^2-\ii \tilde p x'/\hbar} \,\dd x' \biggr|^2,
\label{eq:bose_glass}
\end{equation}
which contains information about the positions of domain walls and demonstrates phase coherence within puddles (see Fig.~\ref{fig:bose_glass}).

The examples so far involve one-body order. Naturally, the PSM is also capable of probing spatially inhomogeneous forms of correlated many-body phases.
An example is the investigation of superconductors with spatially varying order parameters, such as Fulde-Ferrell-Larkin-Ovchinnikov (FFLO) states~\cite{fulde_superconductivity_1964,larkin_nonuniform_1964}.

\begin{figure}
    \centering
    \includegraphics[width=0.9\linewidth]{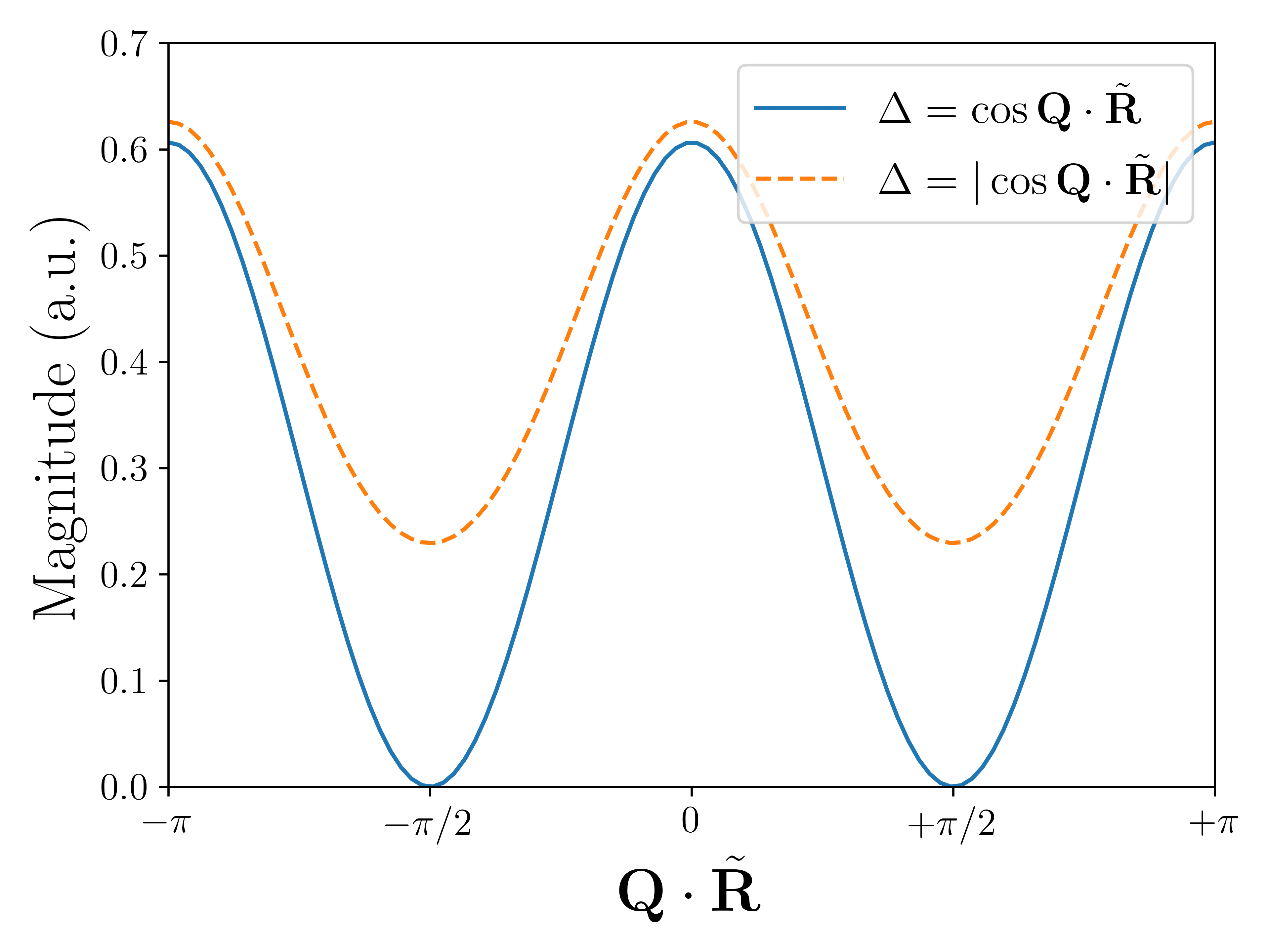}
    \caption{The CoM part of the correlation function $F_\text{CoM}(\tilde {\bm R}, \tilde {\bm P}={\bm 0})$ for the PDW $\Delta(\tilde {\bm R}) = \Delta_0 \cos({\bm Q}\cdot \tilde {\bm R})$ (blue solid line) and another state without phase changes $\Delta(\tilde {\bm R}) = |\Delta_0 \cos({\bm Q}\cdot \tilde {\bm R})|$ (orange dashed line), computed under the same approximation that the CoM part separates from the relative part, as in Eq.~\eqref{eq:com_sep}.
    {Specifically, we wrote down $\langle \hat{c}_{{\bm q}\uparrow}^\dagger \hat{c}_{{\bm q}'\downarrow}^\dagger \rangle = F({\bm q}-{\bm q}') \sum_{n} c_n \delta_{{\bm q} + {\bm q}'- n\hbar {\bm Q}}$ with $c_n$ being the Fourier components of $|\cos({\bm Q}\cdot \tilde {\bm R})|$.}
    The resolution is chosen to be $\sigma = 1/|{\bm Q}|$ in this case.
    The magnitude of correlations goes to zero at the nodes (${\bm Q}\cdot \tilde {\bm R} = \pm \pi/2$) in the PDW with phase changes; it does not go to zero if the order parameter does not change sign.
    This is because the two lobes on two sides of the node cancel each other, while they do not cancel in the PDW without phase changes.}
    \label{fig:pdw_phase}
\end{figure}

We discuss the phase-space correlation function of a general inhomogeneous superconductor in Appendix~\ref{app:inhom_bcs}, and here we focus on two particularly illustrative cases: a pair-density wave (PDW) and a pseudogap phase.
Firstly, we consider Larkin-Ovchinnikov (LO) states or unidirectional PDW states, where the order parameter is given by $\Delta({\bm r}) = \Delta_0 \cos({\bm Q}\cdot {\bm r})$ for some wavevector $\bm Q$~\cite{agterberg_pdw_2020}.
{Due to the periodic structure, the anomalous expectation value in momentum space $\langle \hat{c}_{{\bm q}\uparrow}^\dagger \hat{c}_{{\bm q}'\downarrow}^\dagger \rangle$ in general can have non-zero values when the centre-of-mass momentum ${\bm q} + {\bm q}' = (2n+1)\hbar {\bm Q}$ for integer $n$, or $\langle \hat{c}_{{\bm q}\uparrow}^\dagger \hat{c}_{{\bm q}'\downarrow}^\dagger \rangle = \sum_{n} F_n({\bm q} - {\bm q}') \delta_{{\bm q} + {\bm q}'- (2n+1)\hbar {\bm Q}}$.}
We model these states with two condensates of pairs whose centre-of-mass momenta are $\pm{\bm Q}$. %\old{and whose internal structures take the same functional form as in BCS theory}.
Concretely, we {truncate to the lowest harmonics and} assume $\langle \hat{c}_{{\bm q}\uparrow}^\dagger \hat{c}_{{\bm q}'\downarrow}^\dagger \rangle = u_{({\bm q}-{\bm q}')/2} v_{({\bm q}-{\bm q}')/2}^* (\delta_{{\bm q}+{\bm q}'-\hbar {\bm Q}} + \delta_{{\bm q}+{\bm q}'+\hbar {\bm Q}})$.
{The combination $u_{({\bm q}-{\bm q}')/2} v_{({\bm q}-{\bm q}')/2}^*$ used here is not generally expected to be the same as the uniform BCS result, but more importantly, here the expectation value factorizes into relative and CoM parts.}
{Consequently, }in this model, the phase-space correlators $\langle \hat\rho_\uparrow (\tilde {\bm r}, \tilde {\bm p}) \hat\rho_\downarrow (\tilde {\bm r}', \tilde {\bm p}') \rangle_\text{c}$ also separate into relative and CoM parts, such that
\begin{equation}
    \label{eq:com_sep}
    \langle \hat\rho_\uparrow (\tilde {\bm r}, \tilde {\bm p}) \hat\rho_\downarrow (\tilde {\bm r}', \tilde {\bm p}') \rangle_\text{c} = F_\text{CoM}(\tilde {\bm R}, \tilde {\bm P}) \times f_\text{rel}(\tilde {\bm r}_\text{rel}, \tilde {\bm p}_\text{rel}),
\end{equation}
where we defined $\tilde {\bm R} \equiv (\tilde {\bm r} + \tilde {\bm r}')/2$, $\tilde {\bm r}_\text{rel} \equiv \tilde {\bm r} - \tilde {\bm r}'$, $\tilde {\bm P} \equiv \tilde {\bm p} + \tilde {\bm p}'$, $\tilde {\bm p}_\text{rel} \equiv (\tilde {\bm p} - \tilde {\bm p}')/2$.
{For our particular PDW,} the CoM part is given by
\begin{multline}
    F_\text{CoM}(\tilde {\bm R}, \tilde {\bm P}) = \ee^{-\sigma^2 (\tilde {\bm P}/\hbar + {\bm Q})^2/2} + \ee^{-\sigma^2 (\tilde {\bm P}/\hbar - {\bm Q})^2/2} \\
    + 2\ee^{-\sigma^2 (\tilde {\bm P}/\hbar + {\bm Q})^2/4} \ee^{-\sigma^2 (\tilde {\bm P}/\hbar - {\bm Q})^2/4} \cos(2{\bm Q} \cdot \tilde {\bm R}).
\end{multline}
When choosing a real space resolution larger than the coherence length and smaller than the PDW wavelength, we can observe the pairing structure in the relative part as before and observe the position space modulation through the CoM part.
This is illustrated in Fig.~\ref{fig:pdw_phase}, showing how the PSM can identify the nodes in the gap $\Delta({\bm r})$, thus establishing that the gap changes sign across it.

{We also note that} any accompanying charge density wave orders can also be observed through the single-particle densities~\cite{scalettar_phase_1989,rosenbergUltracoldAtomsSquare2017}.

By choosing the position-space resolution $\sigma \approx a$ to be on the order of the lattice spacing $a$, and hence the momentum-space resolution $\sim 1/\sigma \sim 1/a \sim p_\text{F}/\hbar$, a PSM could show CDW order in position space while also resolving the coarse structure of the  pairing around the Fermi surface, since $\hbar/\sigma \sim p_\text{F}$; finer detail at the Fermi surface would require a larger $\sigma$. As such, it offers a way to image both orders in the same shot, informing the debate of stripe CDW phases vs. $d$-wave superconductivity in Fermi-Hubbard models~\cite{bourgund_formation_2025}.

Secondly, we consider the pseudogap phase seen in the BEC-BCS crossover and high-$T_{\text{c}}$ superconductors, and for which cold-atoms experiments promise crucial insights~\cite{Kendrick2025}.
The pseudogap phase is often explained in terms of pre-formed pairs, where fermions form pairs without condensing~\cite{chen_bcsbec_2005, ohashi_bcsbec_2020}.
Outside of the BEC limit, the preformed pairs cannot easily be identified with QGMs.
The pairs are not condensed, so there is no long-range phase coherence between different pairs. In a global momentum-space measurement this weakens and  broadens the pair-correlation peaks, since the pair amplitudes of regions with different phases add incoherently.
We illustrate this with a simple model {similar to the Bose glass}, where we assume the state is made up of different domains $i=1\dots N$ of size $L$ in real space, each being an independent superconductor with a gap $\Delta_i$.
The global momentum distribution can be computed to be $\langle \hat c^\dagger_{\bm{p}\uparrow} \hat c_{\bm{p}\uparrow} \hat c^\dagger_{\bm{p}'\downarrow} \hat c_{\bm{p}'\downarrow} \rangle_{\text{c}} = | \frac{1}{N} \sum_i \Delta_i/2E_{\bm{p}} |^2 \delta_{\bm{p}+\bm{p}'}$.
Thus, for $\Delta_i$ with independent random phases, 
%\old{the sum diminishes as $\sim 1/N$, making the signal difficult to measure, although information on the coherence length can be inferred from the standard deviation of this quantity over many experimental realizations.} 
the peak height diminishes as $\sim 1/N$;  because each domain has finite size $L$, the delta function above is in fact broadened to a width $\sim\hbar/L$, and information on the coherence length can therefore also be inferred from the width, or from the standard deviation of this quantity over many experimental realizations.
In order to show the existence of these preformed pairs, one can use the PSM and choose the spatial resolution $\sigma$ to be smaller than the coherence length of the pairs (the domain size in the model above).
Within this spatial patch, the correlation functions would again show peaks, as in Fig.~\ref{fig:cross}, since the pairs within a patch are phase coherent. Varying $\sigma$ then gives the length scale over which the pairs are coherent.

\subsection{Interaction energies in a uniform magnetic field}
\label{sec:mag}
In this section, we discuss the application of the PSM to states in a uniform (synthetic) magnetic field.
In particular, we focus on extracting the interaction energy of a Bose gas in the magnetic field. As we shall explain, this is possible because the PSM can be used to measure the overlap of a state with any chosen wavefunction.

First we note that, PSMs still provide full information on the state even in the presence of the magnetic field.
In a cold atom experiment where an effective gauge field is produced through rotation~\cite{Cooper2008}, the PSM can be directly applied in the laboratory frame.
The momentum thus measured is related to that in the rotating frame by the coordinate transformation of rigid body rotation.
If the synthetic gauge field is generated through other means, they can be switched off before the system is subjected to the PSM.
Doing so, the PSM jointly measures the canonical momentum and the position.
Kinetic momentum and other quantities can be obtained through appropriate coordinate transformations in the 4D phase space.

The single-body problem in a uniform magnetic field is solved by Landau levels, and the ground state of a non-interacting Bose gas resides in the lowest Landau level.
On the other hand, interactions will generally drive Landau level mixing and populate the higher Landau levels.
By targeting these excitations onto higher Landau levels, the PSM can be used to measure the interaction energy and shed light on the nature of the interaction.
Note that, some states, such as the much sought-after Laughlin state, do genuinely live in the lowest Landau level and have zero interaction energies with contact interactions.
The PSM can thus also show that a Laughlin state is achieved when zero interaction energy and Landau level mixing are measured.

To do so, we note that generally, one can compute the overlap between the experimentally prepared state with any known state, by integrating the Husimi Q function measured with the PSM against the Glauber--Sudarshan P distribution of the known state, or
\begin{equation}
    \operatorname{tr} \hat\rho_\psi \hat\rho_\phi = \int \frac{\dd \tilde{x} \, \dd \tilde{p}}{2\pi\hbar} \, P_\psi(\tilde{x}, \tilde{p}) Q_\phi(\tilde{x}, \tilde{p}),
    \label{eq:overlap}
\end{equation}
where $\hat\rho_\psi$ and $\hat\rho_\phi$ are the density matrices of the states.
See Appendix~\ref{app:ll_weight}.

These P distributions are easily obtained for states in a uniform magnetic field.
Using these, we now apply the PSM to study the interaction energy of contact-interacting particles in Landau levels, which is encoded in Tan's ``contact''~\cite{tan_large_2008}.
Previous work discussed how the ``averaged-mode'' PSM (see Appendix~\ref{app:avg}) allows one to access the contact by measuring the (one-body) quartic momentum density $\hat\rho_\text{quartic}({\bm r}) \equiv \sum_i \hat{p}_i^2 \delta({\bm r}-\hat{\bm r}_i) \hat{p}_i^2$~\cite{cooper_phase-space_2026}.
As we now show, studying two-body correlations with the PSM allows one to better target the contact in experimental measurements, arising from the fact that it is fundamentally a result of two-body interactions.

To see this, we first solve the two-body problem of two contact-interacting particles in a uniform magnetic field~\cite{hofmann_scale_2024}.
The contact interaction couples the zero relative angular momentum states across Landau levels and leaves all other states unperturbed.
We thus focus on the zero {relative} angular momentum sector and write the state as $|\Psi\rangle = \sum_n c_n |n\rangle$, where $|n\rangle$ is the zero angular momentum state in the $n$th Landau level.
Solving the Schr\"odinger equation, we obtain the spectrum~\cite{hofmann_scale_2024,whitehead_pseudopotential_2016}
\begin{equation}
    \psi_{\rm d}\biggl(\frac12 - \frac{E}{\hbar \omega}\biggr) = \ln \frac{E_\text{B}}{\hbar \omega},
\end{equation}
where $\psi_{\rm d}$ is the digamma function, $E_\text{B}$ is the bound state energy without the magnetic field, and $\omega$ is the cyclotron frequency $\omega\equiv qB/M$.
The eigenstates are given by
\begin{equation}
    |E\rangle \propto \sum_{n}\frac{1}{E-E_n} |n\rangle,
    \label{eq:mag_eig}
\end{equation}
where $E_n$ are the energies of the $n$th Landau level of the relative motion.
The weight of a state in the $n$th Landau level can be extracted from the measured phase-space correlation functions, using Eq.~(\ref{eq:overlap}) with $P_\psi$ formed for higher Landau levels. (See Appendix~\ref{app:ll_weight}.)
In practice, this can be numerically difficult for large $n$ and is equivalent to retrieving the wavefunction from the Husimi Q function.
Nevertheless, especially for $n$ small, the Landau level states have a small extent in phase space, and a reconstruction similar to what we demonstrated in Sec.\ref{sec:pair} is possible.
Thus the small $n$ weights $|c_n|^2$ can be measured.

On the other hand, the large $n$ behavior is $|c_n|^2 \sim 1/n^2$, as predicted by the theory of contact in a strictly 2D scenario.
This applies provided we restrict to 2D momenta $k \ll 1/\ell_z$, where $\ell_z$ is the trap length in the $z$-direction making a quasi-two-dimensional system~\cite{he_contact_2019}.
Relating Tan's contact $C$ to the high momentum tail $N_k \sim C/k^4$, and connecting the Landau level index can be related to momentum by $2n\sim k^2\ell_\text{B}^2$ where $\ell_\text{B}^2 \equiv \hbar/M\omega$ is the magnetic length, the high Landau level tail is $N_n \sim 2\pi k N_k (\dd k/ \dd n) /(2\pi)^2$ or $N_n \sim C\ell_\text{B}^2/8\pi n^2$ per unit area.
The exact eigenstate Eq.~\eqref{eq:mag_eig} relates the weights at high $n$ (used to define the contact) to the weights at low $n$ in a precise way, so knowing the value at a single $n$ is sufficient to infer the entire distribution.
That is, the contact can be obtained from a single measurement at low $n$, without appealing to the large $n$ values.
In particular, the weight of $|E\rangle$ in the first Landau level $|c_1|^2$ gives the contact by
\begin{equation}
    C = \frac{|c_1|^2}{A} 2\pi\ell_\text{B}^2 \biggl(\frac{2M(E-E_1)}{\hbar^2}\biggr)^2,
    \label{eq:mag_con}
\end{equation}
where $A$ is the area.
If the value of the energy $E$ is not known, measurements of weights in two Landau levels are sufficient to determine both $C$ and $E$.

In a many-body case, we first note that the interaction energy $\frac{g}{2}\int \langle \hat c_{\bm r}^\dagger \hat c_{\bm r}^\dagger \hat c_{\bm r} \hat c_{\bm r} \rangle \, \dd^2r = \frac{g}{2} \int \rho^{(2)}({\bm r},{\bm r};{\bm r},{\bm r})\, \dd^2r$, where $g$ is the 2D interaction strength, can be obtained exactly since the Husimi Q function of the TBDM is measured, though the numerically unstable inversion process is still involved.
If the interactions are weak compared to the cyclotron energy ($gn_\text{2D} \ll \hbar\omega$ where $n_\text{2D}$ is the density), then the relation Eq.~\eqref{eq:mag_con} directly applies, and the contact can still be obtained from the first LL population measured with the PSM, setting $E-E_1\approx -\hbar\omega$.
This is consistent with the contact being $C = (Mn_\text{2D} g/\hbar^2)^2$ in the ground state of a weakly interacting Bose gas in 2D~\cite{hofmann_scale_2024} as $c_1 \sim gn_\text{2D}$ in the perturbative regime.
Separately, it is known that the contact is obtained via the total energy and the adiabatic sweep theorem in 2D as follows~\cite{tan_large_2008, hofmann_scale_2024}:
\begin{equation}
    C=\frac{4\pi M}{\hbar^2} \frac{\dd E}{\dd \ln a_\text{2D}},
\end{equation}
where $E$ is the total energy in the system and $a_\text{2D}$ is the {2D} scattering length.
Using also that the interaction energy is $E_\text{int} = \dd E/ \dd \ln(g/2)$ by the Hellmann--Feynman theorem, one finds $E_\text{int}$ to $C$ by
\begin{equation}
    E_\text{int} = \frac{\hbar^2}{4\pi M}\frac{\dd \ln a_\text{2D}}{\dd \ln(g/2)} C.
\end{equation}
Thus, the interaction energy and the contact can both be obtained from the analysis of the two-body problem or by inverting the Husimi Q function of the TBDM.

Note that, while we studied states in a uniform magnetic field, Eq.~\eqref{eq:overlap} is completely general and applies to all states.
The PSM can thus also be used in scenarios where overlaps with trial wavefunctions are of interest.
For example, one can measure effects of imperfect preparations by taking overlaps with unwanted states.

Finally, we comment on the relation between the PSM and conventional QGMs in a strict LLL limit.
It is known that, for states in the lowest Landau level, current-density correlation functions do not give information beyond what is contained in density-density correlation functions~\cite{Spasic-Mlacak2025}.
In fact, the density-density correlation functions contain enough information to reconstruct the TBDM in the LLL.
This can be shown using arguments very similar to how the TBDM can be reconstructed from the Husimi Q function by invoking the analytic properties of the LLL~\cite{macdonald_density_1988}.
{See Appendix~\ref{app:nrdm_qst}.}
Indeed, the $x$ and $y$ components of guiding center motion are conjugate variables like position and momentum without the magnetic field, and they are being imaged simultaneously by the QGM.
A manifestation of this is that, for LLL states, the momentum distribution is a rotated and rescaled copy of the real-space density, so that images in momentum space are equivalent to images in real space~\cite{Read2003}.
Thus, the QGM in the LLL limit is a PSM. 
{It should be noted that, as with general PSM measurements, full reconstruction of the TBDM in the LLL using only density-density correlation information is likely susceptible to noise.
The application of an actual PSM (with auxiliary degrees of freedom) allows more direct targeting of the quantities in phase space and could possibly reduce experimental noise.}
Additionally, as we have seen, when one goes beyond the LLL limit, density-density correlation functions no longer fully characterize the state, whereas the PSM allows one to study Landau level mixing directly.

\subsection{Phase-space Chern marker}
\label{sec:chern}
Typically, the topology of Chern bands realized in ultracold atomic gas systems is measured through global responses~\cite{aidelsburger_measuring_2015,asteria_measuring_2019}, though the Berry curvature can be measured through the tomography of the Bloch bands~\cite{Price_2012,flaschner_experimental_2016}.
An interesting alternative is to use a {\it local} probe of topology, through the local Chern marker~\cite{bianco_local_2011,caio_topological_2019,irsigler_chern_2019}
\begin{equation}
    c({\bm r}) \equiv 4\pi \operatorname{Im} \langle {\bm r} | \hat{P} \hat{x} \hat{P} \hat{y} \hat{P} | {\bm r} \rangle.
\end{equation}
Here, $\hat{P}$ is the projection operator onto occupied bands, which is equal to $\hat{\rho}^{(1)}$.
It has been measured through the Hall drift and tomographical protocols were proposed, but the measurement of the marker is still experimentally challenging~\cite{irsigler_chern_2019,Chalopin2020}.
The PSM provides an alternative way to measure these Chern markers, using their abilities to stably measure the OBDM of a Gaussian state through two-body correlations in phase space (see Sec.~\ref{sec:general}).
As described in Appendix~\ref{app:prac}, we assume the lattice is instantaneously switched off before proceeding with PSM measurements.

Due to the intrinsic spatial resolution $\sigma$ of the PSM, we instead consider the {\it local phase-space Chern marker}
\begin{equation}
    \tilde{c}(\tilde{\bm r}, \tilde{\bm p}) \equiv 4\pi \operatorname{Im} \langle \tilde{\bm r}, \tilde{\bm p} | \hat{P} \hat{x} \hat{P} \hat{y} \hat{P} | \tilde{\bm r}, \tilde{\bm p} \rangle.
\end{equation}
Using the Glauber--Sudarshan P representation of the position operators
\begin{equation}
    \hat{x} = \int \frac{\dd^2 \tilde{r} \, \dd^2 \tilde{p}}{(2\pi\hbar)^2} \, \tilde{x} | \tilde{\bm r}, \tilde{\bm p} \rangle \langle \tilde{\bm r}, \tilde{\bm p} |,
\end{equation}
we can relate the phase-space Chern marker to the one-body density matrix in phase space
\begin{multline}
    \tilde{c}(\tilde{\bm r}, \tilde{\bm p}) = 4\pi \operatorname{Im} \int  \frac{\dd^2 \tilde{r}' \, \dd^2 \tilde{p}'}{(2\pi\hbar)^2} \frac{\dd^2 \tilde{r}'' \, \dd^2 \tilde{p}''}{(2\pi\hbar)^2} \, \tilde{x}' \tilde{y}'' \times {} \\
    \rho^{(1)}(\tilde{\bm r}, \tilde{\bm p}, \tilde{\bm r}', \tilde{\bm p}') \rho^{(1)}(\tilde{\bm r}', \tilde{\bm p}', \tilde{\bm r}'', \tilde{\bm p}'') \rho^{(1)}(\tilde{\bm r}'', \tilde{\bm p}'', \tilde{\bm r}, \tilde{\bm p}).
    \label{eq:marker}
\end{multline}
This is the phase-space equivalent of the position space expression in Ref.~\cite{irsigler_chern_2019}.
As explained in Sec.~\ref{sec:general}, the one-body density matrix  $\rho^{(1)}(\tilde{\bm r}, \tilde{\bm p}, \tilde{\bm r}', \tilde{\bm p}')$ can be stably reconstructed for non-interacting particles by using the PSM to measure the two-body Husimi Q distribution.

This topological marker in phase space has a wide applicability.
Integrating $\tilde{c}(\tilde{\bm r}, \tilde{\bm p})$ with respect to $\tilde{\bm p}$ yields $c({\bm r})$, albeit blurred over the lengthscale $\sigma$.
The blurring is unimportant if the resolution $\sigma$ is smaller than any inhomogeneities of interest.
In the limit $\sigma\to 0$, Eq.~\eqref{eq:marker} recovers exactly the position-space Chern marker, though of course the phase retrieval becomes problematic.

If instead we integrate Eq.~\eqref{eq:marker} over $\tilde{\bm r}$, we resolve the Berry curvature in momentum space.
The local Chern marker $c(\bm r)$ was originally obtained from the $k$-space Chern number by substituting $\hat{\bm r}\to\ii\nabla_{\bm k}$~\cite{bianco_local_2011}.
Read in reverse, for a filled band of a crystal $-2\pi\ii[\hat P\hat x\hat P,\hat P\hat y\hat P]=2\pi\,\Omega(\hat{\bm k})\hat P$, with the Berry curvature normalized so that $C=(2\pi)^{-1}\int_{\rm BZ}\Omega\,\dd^2k$.
Integrating $|\tilde{\bm r},\tilde{\bm p}\rangle\langle\tilde{\bm r},\tilde{\bm p}|$ over $\tilde{\bm r}$ projects onto states in a region of width $\hbar/\sigma$ around momentum $\tilde{\bm p}$. So, for $\sigma$ large compared to the lattice constant this is smaller than any momentum-space features of interest, and  
$A^{-1}\int\dd^2\tilde r\,\tilde c(\tilde{\bm r},\tilde{\bm p})\to2\pi\,\Omega(\bm k)\,|u_{\bm k}(\bm G)|^2$ at $\tilde{\bm p}=\hbar(\bm k+\bm G)$ with ${\bm k}$ in the first Brillouin zone and ${\bm G}$ being a lattice vector, and $|u_{\bm k}(\bm G)|^2$ being the components of the Bloch wavefunction at different lattice vectors.
Summing over $\bm G$ gives $2\pi\Omega(\bm k)$.

Therefore, the phase-space Chern marker Eq.~\eqref{eq:marker} interpolates between the local Chern marker (in position space) and the Berry curvature (in momentum space), giving a local measure of the Chern number in phase space.

\subsection{Phase imprinting}
\label{sec:imprint}
In the above, we focused on the Husimi Q mode of the PSM, whose correlation functions are able to show many features of complex quantum systems where traditional methods are inadequate.
On the other hand, operations on the matter-wave Fourier plane in a quantum gas magnifier can reveal much information about quantum systems, even without using the auxiliary degree of freedom needed for a PSM.
One such way is the phase contrast microscopy technique~\cite{Murthy2019,Bruggenjurgen2026}, where a phase mask imprinted on the matter-wave Fourier plane allows imaging of phase variations, e.g. vortices, in Bose-Einstein condensates.
Here, we examine the consequences of phase imprinting for strongly correlated quantum states. In particular, we show that measurements of density-density correlation functions under such an operation can be used to study pair wavefunctions in superconducting states.

We first summarize the phase-imprinting procedure of Refs.~\cite{Murthy2019,Bruggenjurgen2026}.
This involves turning $\hat{c}_{\bm{p} \eta}^\dagger \to \ee^{\ii f_{\bm{p} \eta}} \hat{c}_{\bm{p} \eta}^\dagger$ with a local phase factor $f_{\bm{p}}$, where $\eta$ is a spin index.
The momentum-dependent phase is imprinted in the matter-wave Fourier plane between the two matter-wave lenses (Appendix~\ref{app:prac}), with no auxiliary degree of freedom: the second lens returns a magnified position-space image, modified by the phase mask. To imprint the phase, one can use two-photon transitions or time evolution under the light shift from an off-resonant laser~\cite{Denschlag2000, Murthy2019}.
In the case where $f_{\bm p} = n\theta_{\bm p}$ for integer $n$, this can also be realized more simply with a transition with one Gaussian beam and one copropagating Laguerre--Gaussian beam~\cite{Andersen2006,Moulder2012}.
These methods realize more easily spin-independent phases, and below we consider $f_{{\bm p}} = f_{{\bm p}\uparrow} = f_{{\bm p}\downarrow}$.

While this is a single-body operation, the phase imprinting procedure also affects relative wavefunctions. In general, for a pair of particles with wavefunction $\Psi({\bm p}_1 \eta_1, {\bm p}_2 \eta_2)$, phase imprinting turns the wavefunction into $\Psi({\bm p}_1 \eta_1, {\bm p}_2 \eta_2) \ee^{\ii (f_{{\bm p}_1\eta_1} + f_{{\bm p}_2\eta_2})}$. Note that the centre-of-mass motion can also be affected.
Importantly, in the BCS theory, this procedure affects the relative phase between different pairs. Specifically, the superconducting gap $\Delta_{\bm p} \propto \langle \hat{c}_{{\bm p} \uparrow}^\dagger \hat{c}_{-{\bm p}\downarrow}^\dagger \rangle$ picks up a phase $\ee^{\ii (f_{{\bm p}\uparrow} + f_{-{\bm p}\downarrow})}$ under this operation.
Thus, properties of the pair wavefunction can also be studied with phase imprinting.

\begin{figure}
    \centering
    \includegraphics[width=0.8\linewidth]{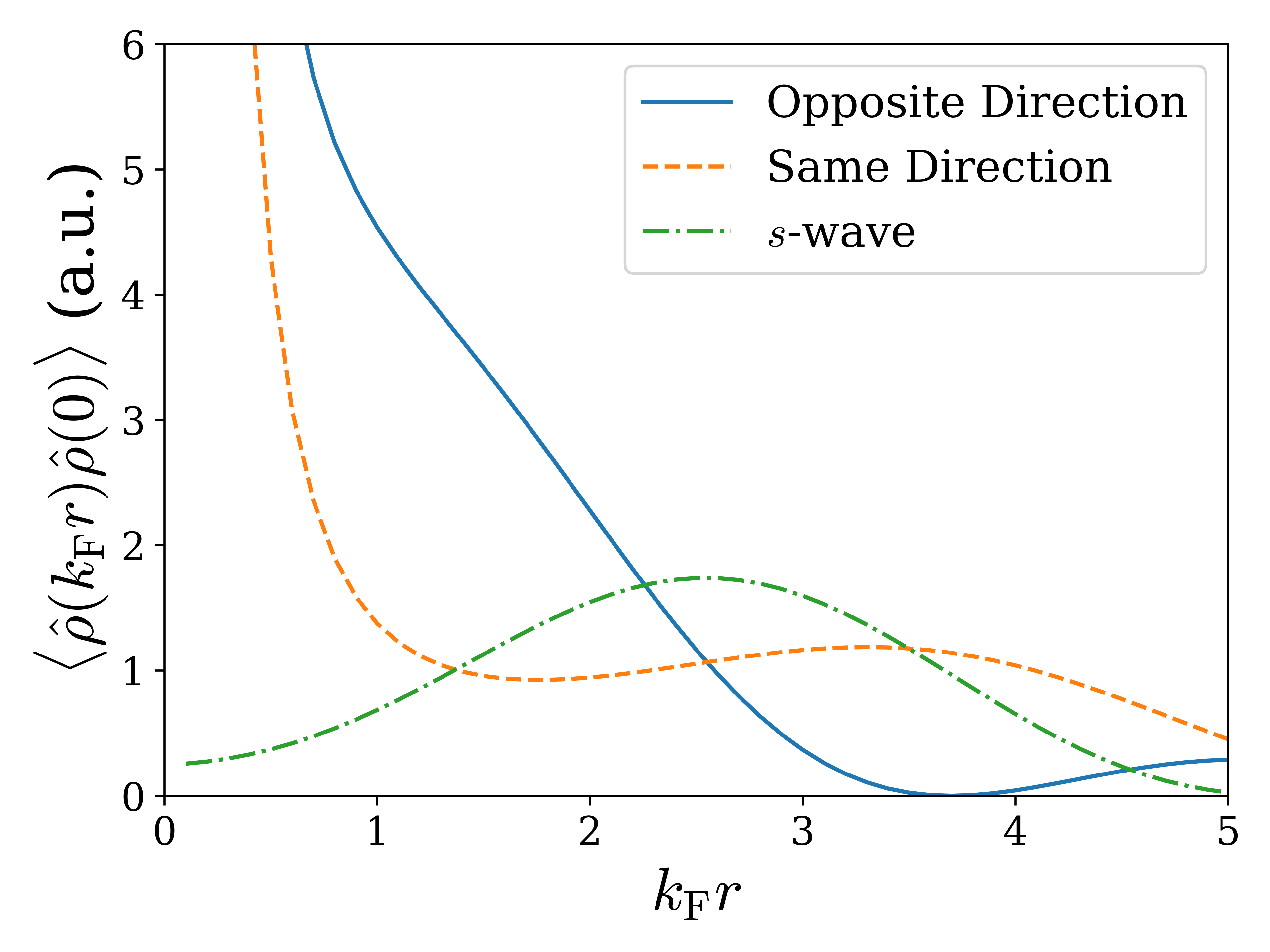}
    \caption{The density-density correlation functions produced in a $p$-wave superconductor after imprinting a phase winding in the Fourier plane. The gap at the Fermi surface is $|\Delta(p_\text{F})|/E_\text{F}=0.1$. When the applied winding is opposite to the original winding, an extra zero appears in the density-density correlations. Any $s$-wave region can also be identified in such a measurement.}
    \label{fig:phase_imprint}
\end{figure}

As a specific example, the phase imprinting procedure allows one to distinguish between clockwise and anticlockwise windings (positive and negative $\ell$), providing a direct, unambiguous signature of time-reversal symmetry breaking that can be elusive to establish conclusively in candidate chiral superconductors. 
We noted earlier that this is possible with the Husimi Q mode.
Phase imprinting is another way, and one that can also be applied to inhomogeneous states.
Take the chiral $\ell=1$ BCS state as an example, where $\Delta(\bm p) \propto p \ee^{\ii\theta_{\bm p}}$.
Applying $f_{\bm p} = \pm \theta_{\bm{p}}$ in the matter-wave Fourier plane, the original phase winding $\ee^{\ii \theta_{\bm p}}$ picks up an additional phase $-\ee^{\pm 2 \ii \theta_{\bm p}}$, turning into $-\ee^{3\ii \theta_{\bm p}}$ and $-\ee^{-\ii \theta_{\bm p}}$, depending on whether the additional phase winding is aligned with or opposite to the initial winding.
The amplitude $|u_{\bm p} v_{\bm p}^*|$ is not altered by such a procedure.
Note that the singularity at the origin is unimportant as $u_{\bm p} v_{\bm p}^*$ is peaked around the Fermi surface; this also means that precise alignment between the applied and initial windings is experimentally uncritical.
In the final image plane, simple density-density correlators $\langle \hat\rho_\uparrow(\bm{r}) \hat\rho_\downarrow(\bm{0}) \rangle_\text{c}$ have (azimuthally-symmetric) position dependencies $\sim |\int \mathrm{d}p \, p^2 J_{1,3}(pr/\hbar) / E_p|^2$, where the order of the Bessel function is either $1$ or $3$ for applied windings either opposite to or in the same direction as the initial winding, respectively.
These two cases can be readily distinguished due to the very different functional forms at $k_{\rm F}r \sim 1$ (see  Fig.~\ref{fig:phase_imprint}).
Thus, the direction of the original winding can be found by exploiting these Fourier-plane capabilities.
It could be used to study the random chiralities in different runs of the experiment or, in a spatially-inhomogeneous state, random chiralities in different regions.

The above discussion pertains to phase imprinting within the quantum gas magnifier.
Combining it with the auxiliary dimension of a full PSM is more powerful.
As we now demonstrate, it can provide a practical way to measure the full phase structure of a wavefunction, without the need to invert a Husimi Q function.
To do so, note that the freedom to choose the function of momentum to measure enables us to select arbitrary regions in momentum space to send to a specific auxiliary coordinate $\tilde{p}_\text{a}$.
In particular, one may choose to let particles within two small disks in momentum space interfere.
This can be achieved, for example, by the use of a double-well potential during the kick procedure in the matter-wave Fourier plane.
In the BCS example, we could measure $\langle \hat\rho_\uparrow(\tilde{\bm{r}},\tilde{p}_\text{a}) \hat\rho_\downarrow(\bm{0},\tilde{p}_\text{a}) \rangle_\text{c} \sim |\int_\Omega\mathrm{d}^2p \, \ee^{\ii\bm{p}\cdot\tilde{\bm{r}}/\hbar} u_{\bm{p}} v_{\bm{p}}^*|^2$ where $\Omega$ is the union of two momentum space disks.
Assuming that $u_{\bm{p}} v_{\bm{p}}^*$ varies on scales larger than the size of the disks, this reduces to $|u_{\bm{p}_1} v_{\bm{p}_1}^* + u_{\bm{p}_2} v_{\bm{p}_2}^*|^2$ if we take $\tilde{\bm{r}}=\bm{0}$.
If we optically apply a relative phase of $\pi/2$ between these two disks, $|u_{\bm{p}_1} v_{\bm{p}_1}^* + \ii u_{\bm{p}_2} v_{\bm{p}_2}^*|^2$ would be measured.
These two pieces of information determine the relative phase.
Repeating the procedure for different disks, the entire phase structure can be mapped out.
This is thus similar to the many-body interferometer considered in Ref.~\cite{Weitenberg2026}, which uses two Raman beams in the matter-wave Fourier plane to allow the state to interfere with a version of itself uniformly displaced in position space.
The protocol introduced here provides more freedom in the choice of the regions to be interfered.
Moreover, since different pairs of regions can be sent to different values of $\tilde{p}_\text{a}$, several relative phases can be measured from the same set of snapshots.

\subsection{Conclusion and Outlook}
In conclusion, we discussed how correlation functions in phase space, as obtained via the Husimi Q mode phase-space microscope, go beyond separate position- and momentum-space measurements.
They provide full information on two-body density matrices, allowing quantum state tomography. We showed how the PSM can establish the presence of entanglement in a system, including entanglement hidden from separate position- and momentum-space measurements, via an explicit construction and adaptations of standard entanglement witnesses between particles or separated collective degrees of freedom.
We showed how the PSM measures relative angular momentum of bound pairs, realizes direct observations of pairing features around the Fermi surface, distinguishes between complex pairing structures, and tracks pairs across the BEC-BCS crossover, and that the two-body Wigner function of a paired state can be reconstructed from phase-space correlators, revealing negativity as a signature of nonclassicality on the BCS side and in the crossover. While we have studied here the features of mean-field states, the PSM measurements would be sensitive to beyond-mean-field features in the states, as we discuss in Appendix~\ref{app:avg}.
 
We also showed how the PSM can resolve local order in inhomogeneous states -- symmetry-breaking domains, superfluid puddles in a Bose glass, and the nodal structure of pair-density-wave superconductors -- and detect preformed pairs in the pseudogap phase. In a magnetic field, we showed that Tan's contact for short-range-interacting particles can be extracted from the population of a single excited Landau level.
For Gaussian states, we showed that the PSM can reliably measure the OBDM, and that this can be used to compute the local Chern marker.
By applying phase masks in the matter-wave Fourier plane, one can measure the chirality of $p$-wave superconducting states. This will be particularly interesting in spatially inhomogeneous states, where the local chirality varies. One may also measure the full phase structure of gap functions by combining phase imprinting and the use of auxiliary dimensions.

Looking beyond the specific examples studied here, the same logic of phase-space tomography applies broadly to strongly correlated states beyond those considered above: superconductors with Rashba spin-orbit coupling~\cite{shiRashbaSpinOrbitCoupling2016}, Tonks--Girardeau gases~\cite{girardeau_relationship_1960,lieb_exact_1963}, (interacting) states with non-trivial topology, and other spatially inhomogeneous phases such as supersolids~\cite{rosenbergUltracoldAtomsSquare2017}. In each case, the same principle applies as throughout this work: because it jointly resolves position and momentum, the PSM captures strictly more information than position- and momentum-space measurements combined, and so can reveal features of the state that position- or momentum-space measurements cannot.

The protocols developed here can themselves be extended. Inducing interactions (e.g., via Rydberg dressing) in the matter-wave Fourier plane would allow genuinely two-body phase imprinting, broadening the class of pair correlations directly accessible without inversion.
Even without changing the protocol, more generally, higher-order phase-space correlation functions in principle determine all $N$-body observables; finding features in these correlation functions robust to the noise amplification encountered throughout this work, remains an open problem. Developing explicit higher-order entanglement witnesses in the Husimi Q language, capable of detecting entanglement invisible to second moments, is a natural first step in this direction.
  
Taken together, these directions point toward the PSM as a general platform for phase-space quantum state tomography of strongly correlated matter.
  
{\bf Acknowledgements}
We thank Aashish Clerk for stimulating discussions, including a question about the Wigner function of the BCS state, and Shiwei Zhang for helpful and insightful comments. We acknowledge funding from the EPSRC (Grant Nos. EP/V062654/1 and EP/Y01510X/1), from a Simons Investigator Award (Grant No. 511029), from the Cambridge Trust, and from the Deutsche Forschungsgemeinschaft (DFG, German Research Foundation) via the Research Units FOR 5919 (Project No. 550495627) and FOR 5688 (Project No. 521530974) (C.W.).
%%This work was funded by Deutsche Forschungsgemeinschaft (DFG, German Research Foundation) via Research Unit FOR 5688, Project No. 521530974.


\newpage
\appendix
\section{Proof of QST of $n$-body density matrices}
\label{app:nrdm_qst}

We first present a direct computation showing that the OBDM can be obtained from the corresponding Husimi Q distribution.
\begin{multline}
    Q(\tilde x, \tilde p) \propto \int \rho(x_1, x_2) \ee^{-\ii \tilde p(x_1-x_2)/\hbar} \times {}\\
    \ee^{-(x_1-\tilde x)^2/2\sigma^2} \ee^{-(x_2-\tilde x)^2/2\sigma^2}  \,\dd x_1 \, \dd x_2,
\end{multline}
so going to the ``centre-of-mass'' coordinates where $(x_1, x_2) \equiv (\tilde X+\tilde \zeta, \tilde X-\tilde \zeta)$, by inverse Fourier transforming with respect to $\tilde p$ we find
\begin{multline}
    \ee^{+\tilde \zeta^2/\sigma^2} \int Q(\tilde x, \tilde p) \ee^{2\ii \tilde \zeta \tilde p/\hbar}\, \dd \tilde p \propto \\
    \int \rho(\tilde X+\tilde \zeta, \tilde X-\tilde \zeta) \ee^{-(\tilde X-\tilde x)^2/\sigma^2} \, \dd \tilde X.
\end{multline}
This remaining integral is a convolution with a Gaussian and can be undone, giving
\begin{multline}
    \rho(\tilde X+\tilde \zeta, \tilde X-\tilde \zeta) \propto \int Q(\tilde x, \tilde p) \ee^{2\ii \tilde \zeta \tilde p/\hbar} \ee^{\tilde \zeta^2/\sigma^2} \times {}\\
    \ee^{-\ii \tilde k(\tilde x-\tilde X) + \tilde k^2\sigma^2/4} \, \dd \tilde p \, \dd \tilde x \, \dd \tilde k.
\end{multline}
These manipulations essentially compute the Wigner quasiprobability distribution from the Husimi Q distribution and Fourier transforms to yield the density matrix.
Indeed, the Wigner quasiprobability distribution is given by (as we noted in the main text in Eq.~\eqref{eq:wigner})
\begin{multline}
    W(\tilde X, \tilde P) \propto \int Q(\tilde x, \tilde p) \ee^{2\ii\tilde \zeta(\tilde p-\tilde P)/\hbar}\ee^{\tilde \zeta^2/\sigma^2} \times {} \\
    \ee^{-\ii \tilde k(\tilde x-\tilde X)} \ee^{\tilde k^2\sigma^2/4} \, \dd \tilde p \, \dd \tilde x \, \dd \tilde k \, \dd\tilde \zeta.
\end{multline}

A non-constructive proof can also be given based on a general theorem in QST~\cite{bandeira_phase_2014, czerwinskiSelectedConceptsQuantum2022}, which states that a vector $|\psi\rangle \in \mathbb{C}^d$ can be reconstructed from intensity data $|\langle \mu_i | \psi \rangle|^2$ for an overcomplete set of states $\{|\mu_i\rangle\}, \,i=1\dots N$, if and only if the linear space
\begin{equation}
    \{\hat{A}\in \mathbb{C}^{d\times d}\,|\, \langle \mu_1|\hat{A}|\mu_1 \rangle = \dots = \langle \mu_N|\hat{A}|\mu_N \rangle =0\}
\end{equation}
does not contain any non-zero Hermitian matrix of rank $\leq 2$.
In our case, the states $\{|\mu_i\rangle\}$ are phase-space coherent states, which are indeed overcomplete.
But if an operator $\hat{A}$ satisfies $\langle \tilde x, \tilde p|\hat{A}| \tilde x, \tilde p\rangle = 0$ for all $\tilde x, \tilde p$, it must be uniformly zero.
Hence, the phase-space coherent states satisfy the theorem and are sufficient to reconstruct any quantum state.

We now give a simple proof that PSM measurements of phase-space correlation functions are sufficient to reconstruct the $n$-body density matrices.
We use the TBDM as an example, but the proof generalizes trivially to the $n$-body case.
First imagine the particles to be distinguishable, then the (non-symmetrized, normalized, many-body) coherent states $|\tilde x, \tilde p\rangle |\tilde x',\tilde p'\rangle$ form an overcomplete basis of the (two-body) Hilbert space.
Expectation values in these coherent states are sufficient to reproduce $\rho^{(2)}$ in exactly the same way that expectation values in $|x, p\rangle$ reproduce $\rho^{(1)}$: one simply needs to take $(x_1, x_2)$ as a combined coordinate with double the dimensions, and any proof carries over.
Indistinguishability poses no extra challenge, since it is only a restriction on the symmetry of the (reduced) density matrices.
The PSM still measures $\langle \tilde x, \tilde p| \langle \tilde x', \tilde p'|  \hat\rho^{(2)} | \tilde x, \tilde p \rangle | \tilde x', \tilde p' \rangle$, despite $| \tilde x, \tilde p \rangle | \tilde x', \tilde p' \rangle$ not being a state in the physical (anti-)symmetric Hilbert space.
Reconstruction can thus be carried out identically to the distinguishable case.

For comparison with the reconstruction within the LLL, here we also mention a complex analytic proof of the reconstruction.
Here, we label the coherent states with $\alpha, \alpha^*$ instead of $\tilde x, \tilde p$.
Consider matrix elements of the density matrix in the unnormalized coherent state basis $\varrho(\alpha, {\alpha'}^*) = \langle \alpha | \ee^{|\alpha|^2/2} \hat{\rho} \ee^{|\alpha'|^2/2} | \alpha' \rangle$.
This is sesquianalytic, meaning that it is holomorphic in its first argument and anti-holomorphic in its second.
Such functions are uniquely determined by their (real analytic) diagonal elements $\varrho(\alpha, \alpha^*) = \ee^{|\alpha|^2} Q(\alpha)$ by analytic continuation.
Writing out the analytic continuation explicitly,
\begin{equation}
    \varrho(\alpha, {\alpha'}^*) = \sum_{mn} c_{mn} \alpha^m ({\alpha'}^*)^n,
\end{equation}
{while $\ee^{|\alpha|^2} Q(\alpha) = \sum_{mn} c_{mn} \alpha^m (\alpha^*)^n$.
To get the coefficients, one evaluates derivatives $\frac{1}{m!n!}\partial_\alpha^m \partial_{\alpha^*}^n \ee^{|\alpha|^2} Q |_{\alpha=\alpha^*=0}$.
Note that this is susceptible to the same noise as explicit reconstructions using Eq.~\eqref{eq:wigner} because evaluating high-order derivatives is noisy.
This derivation also generalizes easily to the $n$-body case.}

{This is the same mathematics as used to show that $n$-point density-density correlation functions are sufficient to reconstruct the $n$-body reduced density matrix in the LLL~\cite{macdonald_density_1988}, where the reduced density matrices are sesquianalytic up to Gaussian factors.
For instance, in the two-body case, writing the density-density correlation function as a (real) analytic function $\langle \hat\rho (\bm r_1) \hat\rho (\bm r_2) \rangle_\text{c} \equiv f(z_1, z_2; z_1^*, z_2^*)$ where $z_i\equiv x_i -\ii y_i$, in the symmetric gauge, the TBDM is given by the analytic continuation of $f$ as}
\begin{multline}
    \rho^{(2)}({\bm r}_1, {\bm r}_2; {\bm r}_1', {\bm r}_2') = \prod_{i=1}^2 \ee^{-(|z_i|^2+|z_i'|^2-2z_i{z_i'}^*)/4\ell_\text{B}^2} \times {}\\
    f(z_1, z_2; {z_1'}^*, {z_2'}^*).
\end{multline}
{Thus, while density-density correlation functions are sufficient to reconstruct $n$-body reduced density matrices in the LLL, they are also prone to noise in exactly the same way as reconstructing Wigner functions from Husimi Q functions.}

\section{Practical Observables and Procedures}
\label{app:prac}
The PSM is a powerful experimental tool for probing many-body correlation features that one does not see through conventional means, even when one does not wish to or cannot perform full QST.

Below, we summarize the principles of the Husimi Q mode PSM~\cite{cooper_phase-space_2026}. Consider a one-dimensional system (along $x$) realized by employing tight potential traps along orthogonal spatial dimensions $y$ and $z$. We apply the unitary operator $\ee^{\ii G(\hat{p}_x)\hat{z}/\hbar}$, where $G$ is an arbitrary real scalar function, to encode the $x$-direction momentum in the momenta along one of the extra dimensions. Finally, measuring position along $x$ and momentum along $z$ together achieves the joint measurement of $x$ and $G(p_x)$. Experimentally, this can be achieved with a three-step procedure: 
\begin{enumerate}
    \item Use a matter-wave lens in the $x$-direction to map $p_x$ to $x$. We call the space after this operation the ``matter-wave Fourier plane''.
    \item Apply a pulsed potential $V(x,z) = G(x)z$ in the matter-wave Fourier plane for a short time.
    \item Use a matter-wave lens in the $x$- and $z$-directions to restore $x$ and to map $p_z$ to $z$.
\end{enumerate}

Generalizing this, one can imagine $d$-dimensional systems with $d_\text{a}$ auxiliary dimensions. Denote the position and momentum in the $d$-dimensions $\bm{r}$ and $\bm{p}$ and those in the auxiliary dimensions $\vec{r}_\text{a}$ and $\vec{p}_\text{a}$. Then we can define a $d_\text{a}$-dimensional function $\vec{G}(\bm p)$ such that the unitary operator to be applied is $\ee^{\ii \vec{G}({\bm p}) \cdot \vec{r}_\text{a}/\hbar}$.

For the one-dimensional system, the function $G(p_x)$ can be chosen to be $p_x$, so that the phase space is imaged in a 2D plane.
%If we further assume initial harmonic confinement along $z$, then the expectation values of the density operators $\langle \hat\rho(r, p_z) \rangle$ gives the Husimi Q function for the OBDM.
For two-dimensional systems, one may still imagine using two extra (synthetic) dimensions so that the full momentum can be captured, \textit{i.e.}, choosing $\vec{G}({\bm p}) = {\bm p}$.
But even without full momentum information, as we argue later, the ability to choose an arbitrary $\vec{G}(\bm{p})$ can be very powerful.
Common choices are $\vec{G}(\bm{p})=p_y$ and $\vec{G}(\bm{p})=\gamma p^2$ for some constant $\gamma$, implementable with appropriately shaped potentials~\cite{cooper_phase-space_2026}.

%\old{In the Husimi Q mode where momentum information is stored along a continuous auxiliary dimension $z$, the PSM measures density operators $\hat{\rho}({\bm r},p_z)\equiv \hat{c}_{{\bm r},p_z}^\dagger \hat{c}_{{\bm r},p_z}$, achieving simultaneous measurements of position and momentum information encoded in said auxiliary dimension.}
In many-body systems, it is convenient to write the measurement outcomes of the PSM in second quantized notation. The PSM measures density operators $\hat{\rho}({\bm r}, \vec{p}_\text{a})\equiv \hat{c}_{{\bm r},\vec{p}_\text{a}}^\dagger \hat{c}_{{\bm r},\vec{p}_\text{a}}$ where $\hat{c}, \hat{c}^\dagger$ are ladder operators in the full $(d+d_\text{a})$-dimensional space.
In terms of ladder operators $\hat{c}_{\bm p}$ for the state without the auxiliary dimensions, and assuming an initial wavefunction $\phi(\vec{p}_\text{a})$ along the auxiliary dimensions, the operators $\hat{c}_{{\bm r},\vec{p}_\text{a}}$ are given by $\hat{c}_{{\bm r},\vec{p}_\text{a}} = \int \hat{c}_{{\bm p}} \ee^{\ii {\bm p}\cdot{\bm r} / \hbar} \phi(\vec{p}_\text{a}-\vec{G}({\bm p})) \,\frac{\dd^d p}{(2\pi\hbar)^d}$.
In this paper, we specialized to the Gaussian case, where $\phi(\vec{p}_\text{a} - \vec{G}({\bm p})) = (\sigma^2/\pi\hbar^2)^{d_\text{a}/4} \ee^{-\sigma^2 (\vec{p}_\text{a} - \vec{G}({\bm p}))^2/2\hbar^2}$ and $\sigma$ is the oscillator length.

Naturally, the PSM also measures correlation functions $\langle \Psi| \hat\rho(\bm{r}, \vec{p}_\text{a}) \hat\rho(\bm{r}', \vec{p}_\text{a}^{\,\prime}) | \Psi\rangle$ in the state of interest $|\Psi\rangle$, which are correlations in phase space.
With spin resolution, the density operators carry spin indices, as do the correlation functions.

While we focus on continuum systems, the principles also apply to lattice systems.
If one suddenly switches off the lattice and directly applies the PSM, then the wavefunction in the lattice is directly probed.
There, the creation operators create Bloch wavefunctions, which are sums over reciprocal lattice vectors, $\hat{c}_{n\bm{q}}^\dagger = \sum_{\bm{G}} f_{n\bm{G}} \hat{c}_{\bm{q}+\bm{G}}^\dagger$ where $n$ is the band index, $\bm{q}$ is the quasi-momentum, $\bm{G}$ are reciprocal lattice vectors and $f_{n\bm{G}}$ are Fourier coefficients.
Different $\bm{G}$ sectors can be addressed separately in the matter-wave Fourier plane.
Kinetic energy measurements in such a quench scenario would correctly give the kinetic energies of particles in the lattice.

One can also opt to remove the Bloch wavefunction information with the technique of band mapping, which maps Bloch states onto plane wave states by adiabatically turning off the optical lattice~\cite{greiner_exploring_2001}.
This effectively maps Bloch states onto plane wave states, so $\hat{c}_{n\bm{q}}^\dagger \to \hat{c}_{\bm{p}}^\dagger$, with the different bands $n$ appearing in different Brillouin zones.
Measurements with a PSM in this case could more cleanly show correlation effects without the single-body lattice effects.

\section{Averaged mode PSM in the BEC-BCS crossover}
\label{app:avg}

Instead of a continuous auxiliary dimension, the averaged mode PSM uses a discrete auxiliary degree of freedom, such as a spin, to measure moments of the momentum distribution \cite{cooper_phase-space_2026}.
This gives access to the operators such as kinetic energy density 
\begin{equation}
\hat\rho_\text{KE}(\bm{r}) = \sum_{{\bm p} {\bm p}'} \frac{{\bm p}\cdot {\bm p}'}{2M} \frac{\ee^{\ii ({\bm p}' - {\bm p})\cdot {\bm r}/\hbar}}{V} \hat{c}_{\bm p}^\dagger \hat{c}_{{\bm p}'}
\end{equation}
where $V$ is the volume of the system. %, $\hat\rho_\text{quartic}(\bm{r})$ with four powers of $\bm{p}$, \textit{etc.}
By measuring the correlation between different auxiliary spin components and position space coordinates, one can measure the correlators $\langle \hat\rho_\text{KE}(\bm{r}) \hat\rho(\bm{0}) \rangle$ and $\langle \hat\rho_\text{KE}(\bm{r}) \hat\rho_\text{KE}(\bm{0}) \rangle$. %, $\langle \hat\rho_\text{quartic}(\bm{r}) \hat\rho(\bm{0}) \rangle$, \textit{etc}.
Below, we describe the application of the averaged mode PSM to the systems we have discussed.

\begin{figure}
    \centering 
\includegraphics[width=0.9\linewidth]{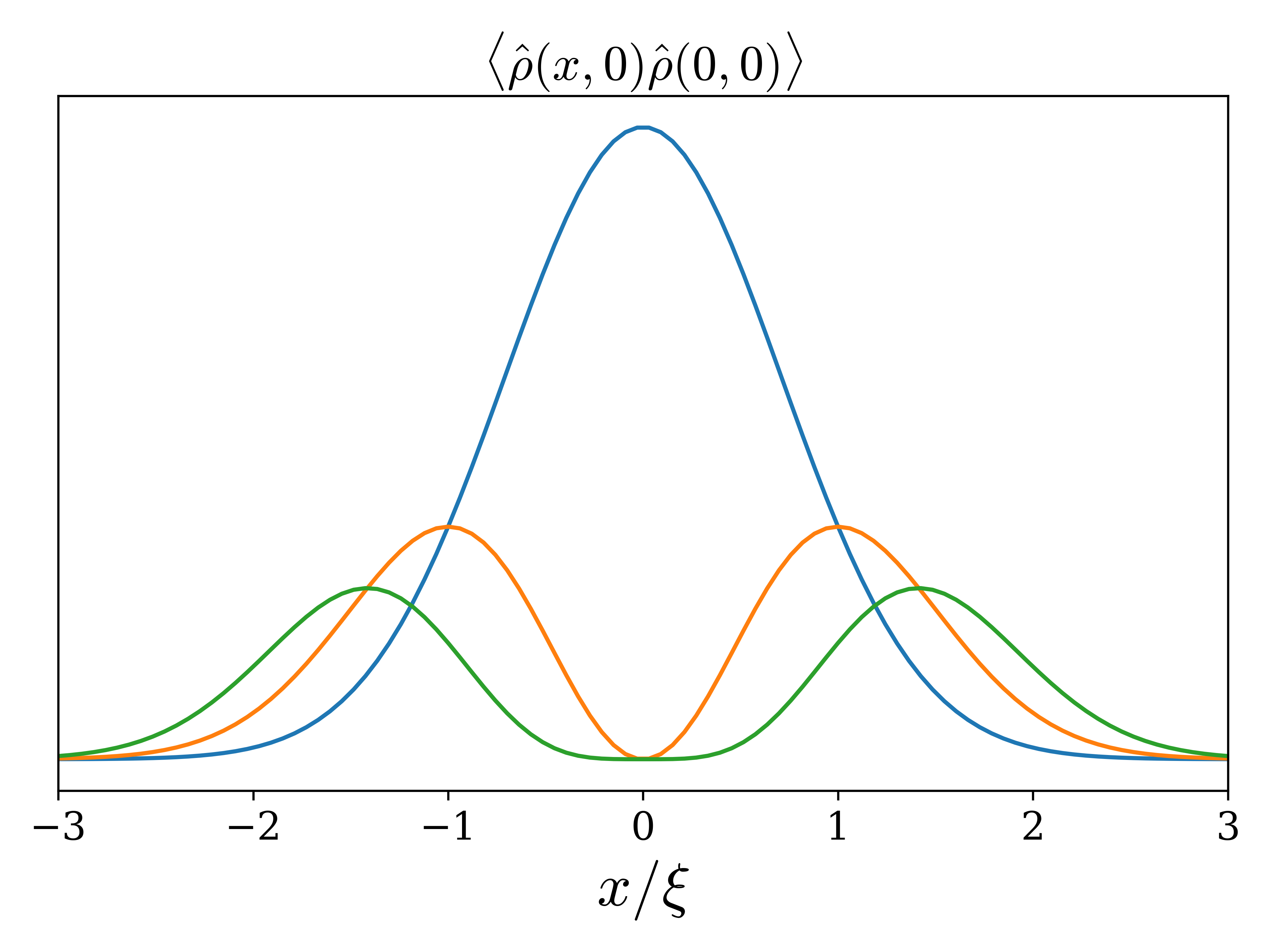}
\\
\includegraphics[width=0.9\linewidth]{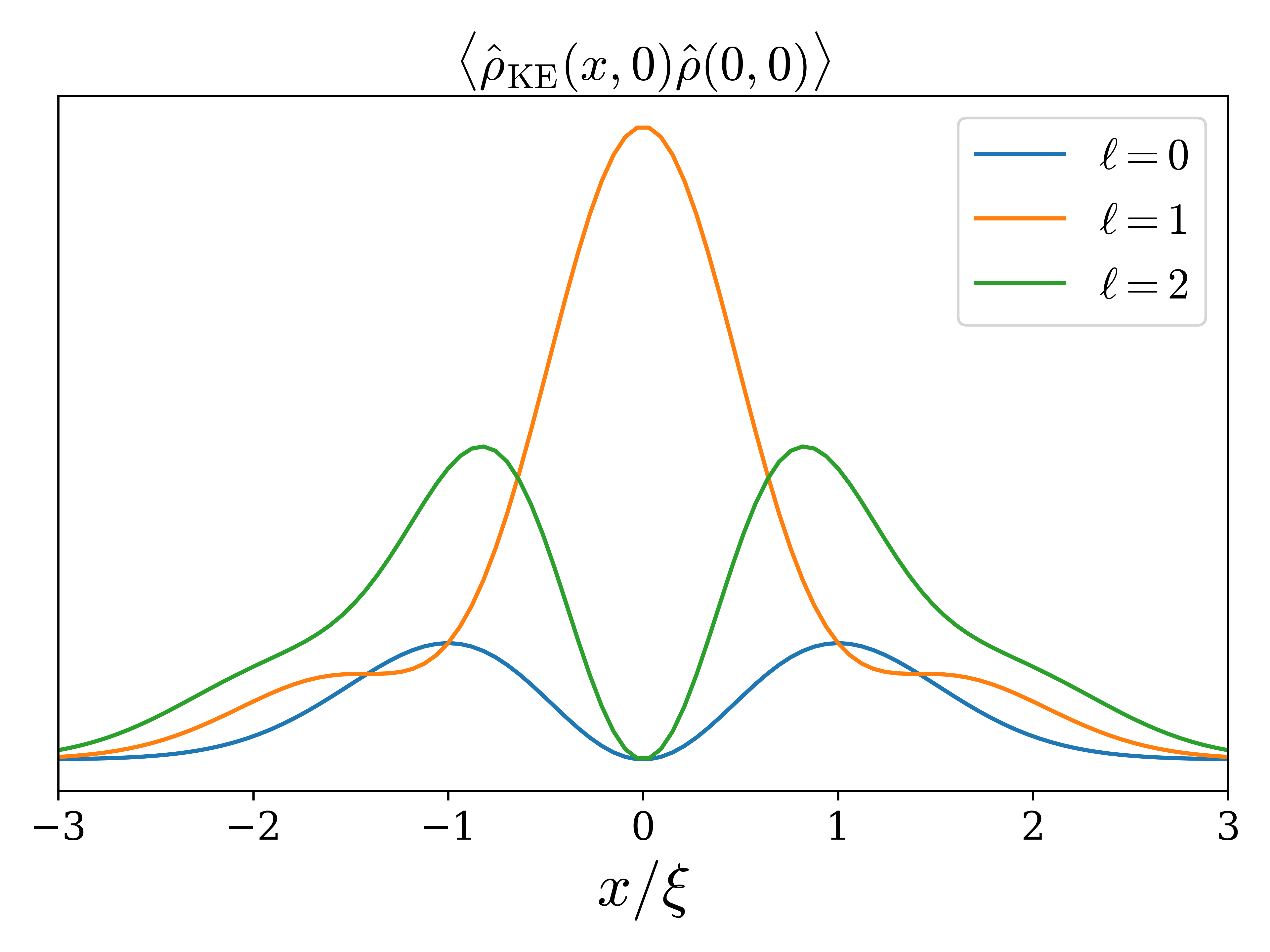}
    \caption{Density-density and kinetic-energy--density correlators measured with the phase-space microscope in 2-body bound states $\psi_{\rm pair}({\bm r}) \sim (x+{\rm i} y)^\ell \exp(-r^2/2\xi^2)$, for $\ell =0 , 1, 2$.
    The shapes of the curves are qualitatively different between $\ell=1$ and $\ell=2$ in the KE-density correlator, but not so in the density-density correlator.
    }

       \label{fig:boundstates}
%\end{figure*}
\end{figure}

On the BEC side of the BEC-BCS crossover, the averaged-mode phase-space microscope can help distinguish different relative angular momentum states.
For instance, as shown in Fig.~\ref{fig:boundstates}, $\langle \hat{\rho}_{\text{KE},\uparrow}(x,0) \hat{\rho}_{\downarrow}(0,0) \rangle$ is qualitatively different in $\ell=1$ states compared to $\ell=2$ states (there is a zero at the origin in the latter case), while the density-density correlations do not look significantly different and precise measurements are needed. 
Indeed, measurements of kinetic energy densities produce $|\nabla \psi_{\rm pair}({\bm r})|^2$, which contains extra phase information not available in simple density-density correlation functions, even when no obvious qualitative differences are present.
If $\psi_\text{pair}({\bm r}) \propto \ee^{\ii \ell \theta_{\bm r}}$ for angular coordinate $\theta_{\bm r}$, then $|\nabla \psi_{\text{pair}}|^2 = (\partial_r |\psi_\text{pair}|)^2 + \ell^2|\psi_\text{pair}|^2/r^2.$
This will corroborate and improve upon measurements of $\ell^2$ obtained through power fitting in density-density correlations.

%\begin{figure*}

On the BCS side of the crossover, $\langle \hat\rho_{\text{KE}, \uparrow}({\bm r}) \hat\rho_\downarrow({\bm 0}) \rangle_\text{c}$ produces $|\nabla g_{\uparrow\downarrow}({\bm r})|^2$.
The gradient gives information on the phase winding in a similar fashion to the BEC case.
Note that this simple relation only holds for a Gaussian state such as the BCS state, so deviations from $\langle \hat\rho_{\text{KE}, \uparrow}({\bm r}) \hat\rho_\downarrow({\bm 0}) \rangle_\text{c} = |\nabla g_{\uparrow\downarrow}({\bm r})|^2$ signal beyond-mean-field features in the state.
This is similar in spirit to how the QGM measurements of three-body correlations in the BEC-BCS crossover can be used to detect beyond-mean-field effects through deviations from predictions of Wick's theorem~\cite{daix_observing_2026, Jongh2025}, but the PSM measurements only require two-body correlations to be measured.

\begin{figure*}
    \includegraphics[width=0.9\linewidth]{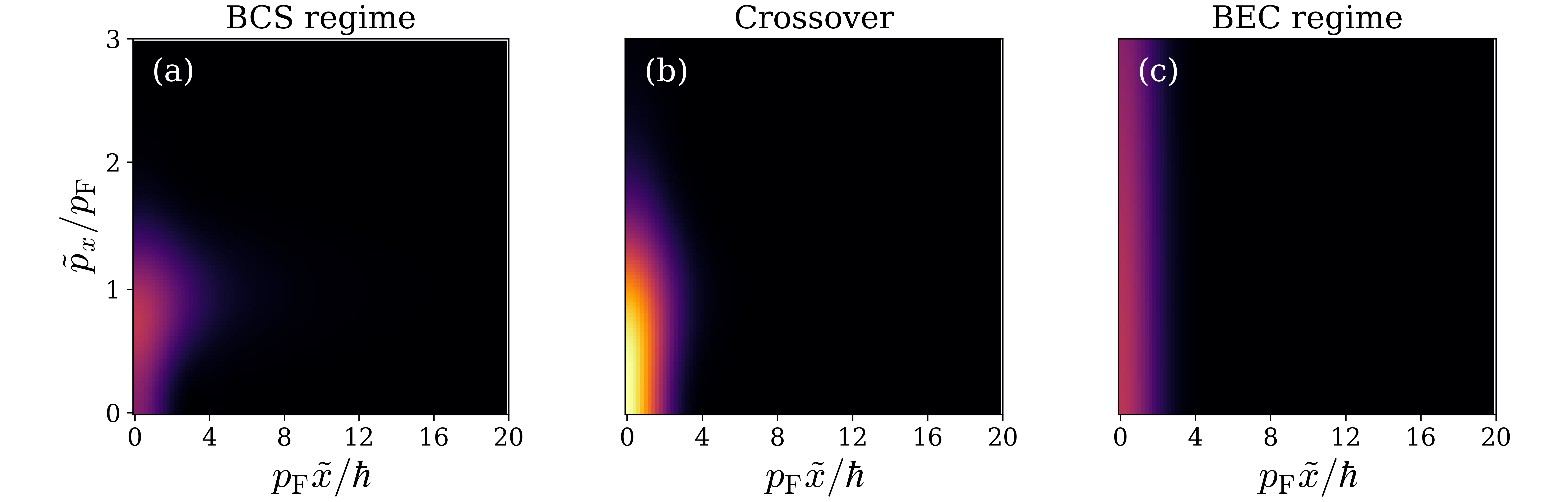}

    \caption{Phase-space correlation functions analogous to Fig.~\ref{fig:cross}(a-c), only with a finer position space resolution $\sigma=0.07\lambda_\text{F}$.
    Again, the BCS and BEC limits [panels (a) and (c)] are enhanced to make them visible.
    One sees in the BEC regime that the correlation functions becomes much more concentrated near $x=0$ compared to Fig.~\ref{fig:cross}(c), as noted in the main text.
    In the BCS regime, the Fermi surface is not clearly imaged, even though the state is identical to that in Fig.~\ref{fig:cross}(a).
    The shape of the correlation function in the crossover regime is also changed substantially.}
    \label{fig:cross_pos}
\end{figure*}

\section{General features of the phase-space image in the crossover} \label{sec:cross}
Firstly, we present a version of Fig.~\ref{fig:cross} with a finer position space resolution in Fig.~\ref{fig:cross_pos} to show how the choice of the resolution is crucial in determining the information one obtains in PSM measurements.
With the finer resolution, one no longer sees the Fermi surface in the BCS regime; instead one observes tight confinement in the BEC regime.

To understand exactly what role is played by the resolution, we discuss more generally features in phase space seen in the crossover.
We first note a few general properties of the POVM measurements.
The correlation functions are normalized (in the next expression we included the disconnected parts), in the sense that their integrals over phase space produce the trace of the TBDM 
\begin{equation}
    \int \langle \hat\rho_\uparrow (\tilde {\bm r}, \tilde {\bm p}) \hat\rho_\downarrow (\tilde {\bm r}', \tilde {\bm p}') \rangle \,\frac{\mathrm d\tilde {\bm r} \,\mathrm d\tilde {\bm p} \,\mathrm d\tilde {\bm r}' \,\mathrm d\tilde {\bm p}'}{(2\pi\hbar)^{4}} = \operatorname{tr} \rho^{(2)}_{\uparrow\downarrow, \uparrow\downarrow},
\end{equation}
where we specialized to the up-down spin combination.
Integrating over connected correlators gives, correspondingly, $\operatorname{tr} \big(\rho^{(2)}_{\uparrow\downarrow, \uparrow\downarrow} - \rho^{(1)}_{\uparrow\uparrow} \otimes \rho^{(1)}_{\downarrow\downarrow}\big)$.
For the BCS state, this trace is $\sum_{\bm{p}} |u_{\bm p} v_{\bm p}^*|^2$.

At any point in phase space, the correlation function is also bounded, such that
\begin{equation}
    \langle \hat\rho_\uparrow (\tilde {\bm r}, \tilde {\bm p}) \hat\rho_\downarrow (\tilde {\bm r}', \tilde {\bm p}') \rangle_\text{c} \leq \operatorname{tr} \big(\rho^{(2)}_{\uparrow\downarrow, \uparrow\downarrow} - \rho^{(1)}_{\uparrow\uparrow} \otimes \rho^{(1)}_{\downarrow\downarrow}\big).
\end{equation}
Note that the trace is usually (as one sees in the BCS case) proportional to the particle number or volume.
A more useful bound can be found when (like the zero-temperature BCS state) there is a condensate of pairs whose centre-of-mass momentum is zero.
In such a case, the momentum space correlation functions contain Kronecker delta functions, such as $\langle \hat{n}_{{\bm p}\uparrow} \hat{n}_{{\bm p}'\downarrow} \rangle \propto \delta_{{\bm p} + {\bm p}'}$.
And since the phase-space coherent states have weights beyond ${\bm p} + {\bm p}' = {\bm 0}$, the maximum amplitude is correspondingly smaller.
Starting from Eq.~\eqref{eq:bcs_povm}, one finds using a Cauchy--Schwarz inequality that the maximum amplitude satisfies
\begin{multline}
    \langle \hat\rho_\uparrow (\tilde {\bm r}, \tilde {\bm p}) \hat\rho_\downarrow (\tilde {\bm r}', \tilde {\bm p}') \rangle_\text{c} \leq  \frac{(2\pi)^{2}}{V} \frac{\sigma^2}{2 \pi} \ee^{-\sigma^2 (\tilde {\bm p}+\tilde {\bm p}')^2/2\hbar^2} \times {} \\
    {}\times \sum_{\bm{q}} |u_{\bm q} v_{\bm q}^*|^2,
    \label{eq:bcs_bound}
\end{multline}
where $V$ is the volume. This bound is saturated if the function $u_{\bm q} v_{\bm q}^*$ is proportional to a coherent state $\ee^{-(\sigma|\tilde {\bm p}-{\bm q}|)^2/2\hbar^2-(\sigma|\tilde {\bm p}'+{\bm q}|)^2/2\hbar^2} \ee^{-\ii\bm q\cdot (\tilde {\bm r}-\tilde {\bm r}')/\hbar}$ labeled by $\tilde {\bm r}, \tilde {\bm r}', \tilde {\bm p}, \tilde {\bm p}'$.

One can measure the maximum amplitude of correlations in phase space and compare them to the bounds.
The maximum amplitude in phase space
\begin{equation}
\max_{\sigma} \max_{\tilde {\bm r}, \tilde {\bm p}} \langle \hat\rho_\uparrow (\tilde {\bm r}, \tilde {\bm p}) \hat\rho_\downarrow (\bm 0, -\tilde {\bm p}) \rangle_\text{c}
\label{eq:crossover_max}
\end{equation}
is constant at $1/4$ on the BCS side and the crossover, and decays in the BEC region.
On the other hand, the saturation
\begin{equation}
\max_{\sigma} \frac{(2\pi)^3\hbar^4V}{\sigma^2}\frac{\max_{\tilde {\bm r}, \tilde {\bm p}} \langle \hat\rho_\uparrow (\tilde {\bm r}, \tilde {\bm p}) \hat\rho_\downarrow (\bm 0, -\tilde {\bm p}) \rangle_\text{c}}{\int \langle \hat\rho_\uparrow (\tilde {\bm r}, \tilde {\bm p}) \hat\rho_\downarrow (\tilde {\bm r}', \tilde {\bm p}') \rangle_\text{c} \, \dd\tilde {\bm r}\, \dd\tilde {\bm r}' \, \dd\tilde {\bm p}\, \dd\tilde {\bm p}'}
\label{eq:crossover_sat}
\end{equation}
approaches 1 for the BEC side and the crossover, and decays in the BCS region.
Thus, the crossover is the region where the amplitude is itself large and also saturates the theoretical bound.

\begin{figure}
    \centering
    \includegraphics[width=0.9\linewidth]{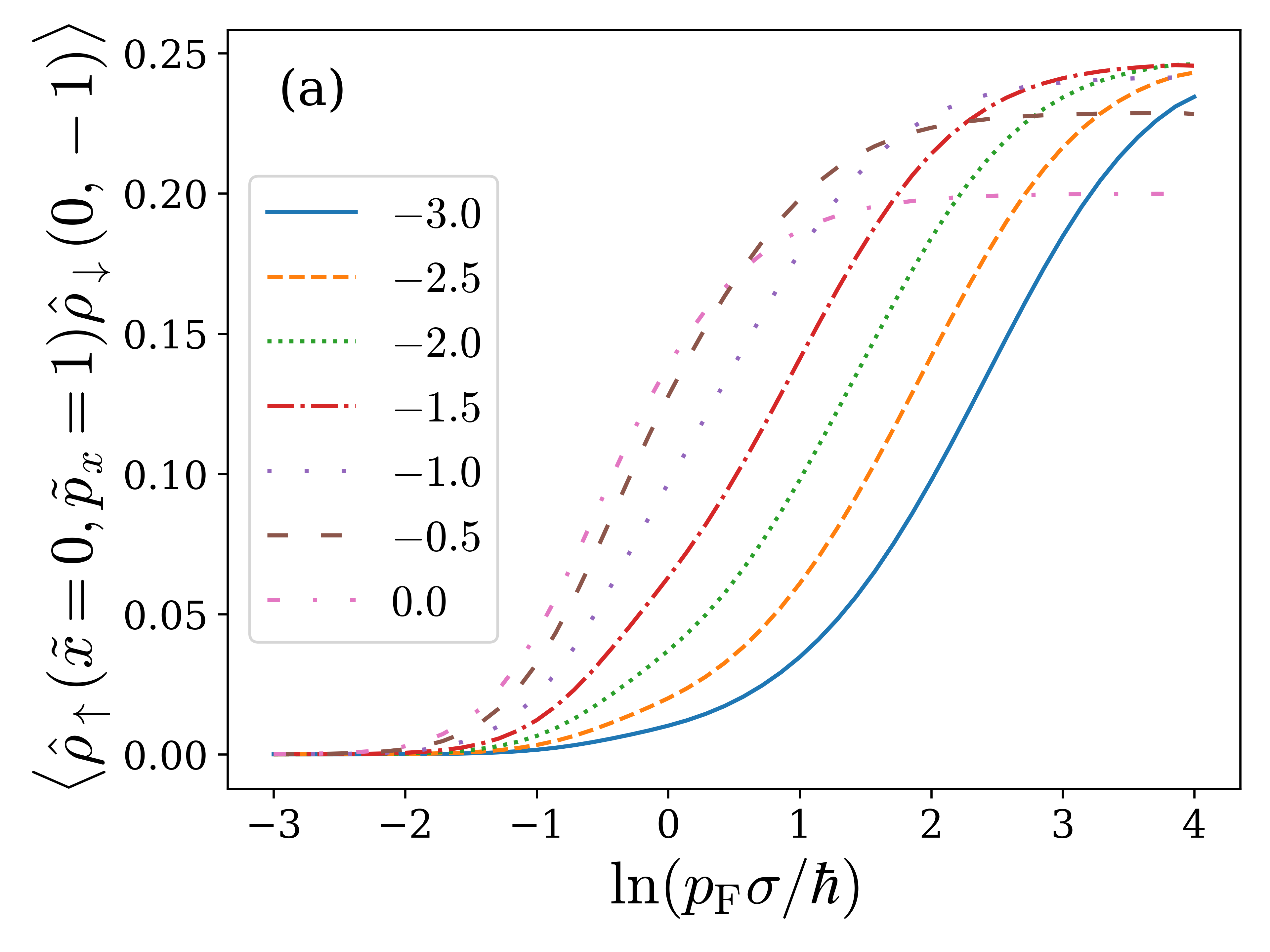} \\
    \includegraphics[width=0.9\linewidth]{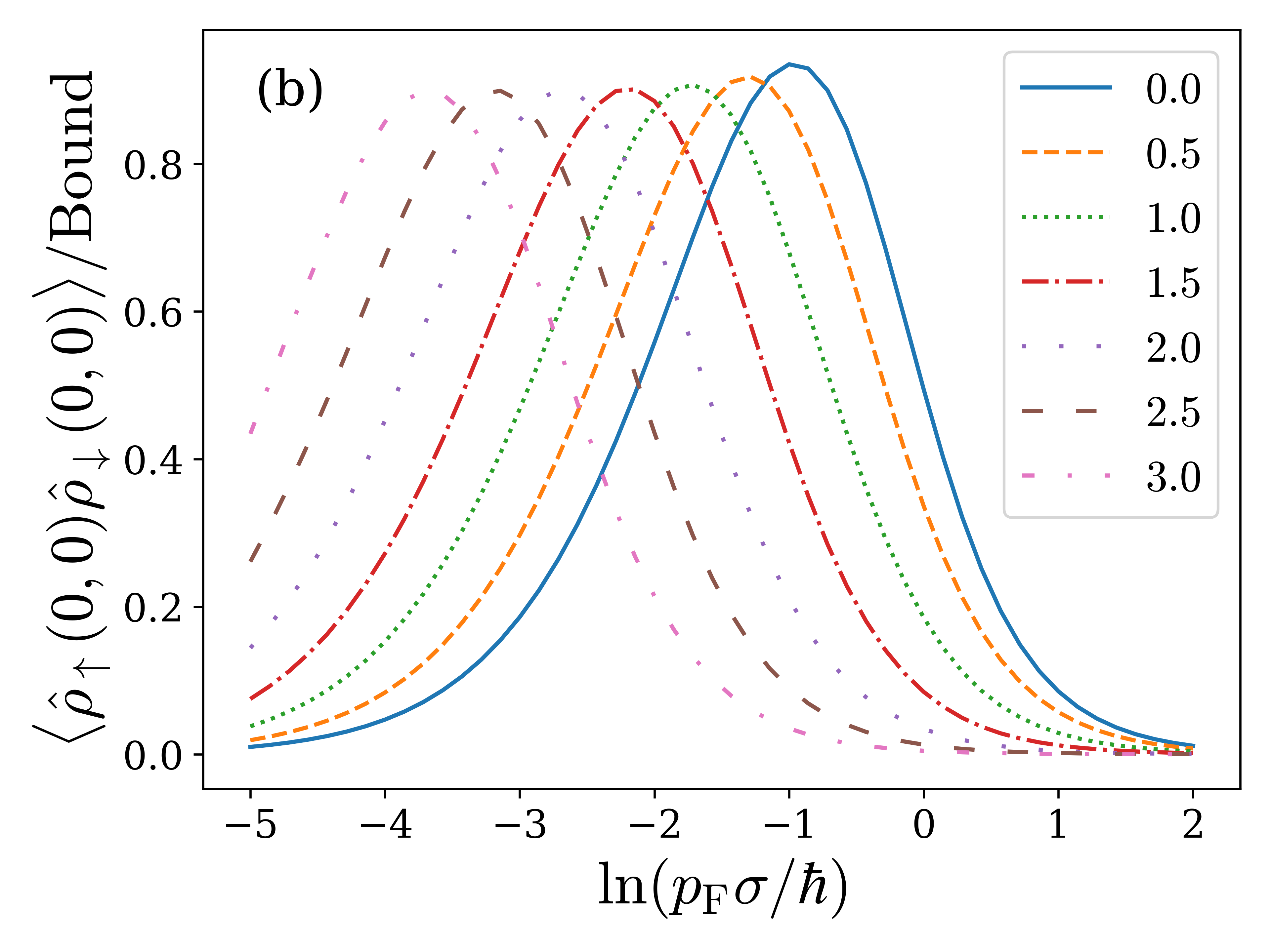}
    
    \caption{The quantities to be maximized in Eqs.~\eqref{eq:crossover_max} and \eqref{eq:crossover_sat}, shown as functions of $\sigma$ on a logarithmic scale. The different lines are at different points in the crossover, with their values of $\ln(\Delta/2E_\text{F})$ indicated in the legends. In panel (a), we plot the (approximate) maxima of $\langle \hat\rho_\uparrow (\tilde {\bm r}, \tilde {\bm p}) \hat\rho_\downarrow (\bm 0, -\tilde {\bm p}) \rangle_\text{c}$, which in the BCS regime shown are around $|\tilde{\bm p}|=p_\text{F}, \tilde{\bm r}={\bm 0}$. These correlation functions attain maximum value when $\sigma\to \infty$. It can also be seen that the curves uniformly spaced in $\ln\Delta$ are likewise evenly spaced in $\ln\sigma$. In panel (b), we plot the approximate maxima divided by the bound (Eq.~\eqref{eq:bcs_bound}) for the BEC regime. Note that the maxima in the BEC regime are around $\tilde{\bm p}={\bm 0}, \tilde{\bm r}={\bm 0}$. Again, the values of $\sigma$ at which the saturation is maximized are equally spaced (logarithmically) for evenly spaced $\ln\Delta$.}
    \label{fig:sigma_choice}
\end{figure}

Additionally, one can also consider at which microscope parameter $\sigma$ the amplitudes are maximized, \textit{i.e.},
\begin{equation}
\argmax_{\sigma} \frac{(2\pi)^3\hbar^4V}{\sigma^2}\frac{\max_{\tilde {\bm r}, \tilde {\bm p}} \langle \hat\rho_\uparrow (\tilde {\bm r}, \tilde {\bm p}) \hat\rho_\downarrow (\bm 0, -\tilde {\bm p}) \rangle_\text{c}}{\int \langle \hat\rho_\uparrow (\tilde {\bm r}, \tilde {\bm p}) \hat\rho_\downarrow (\tilde {\bm r}', \tilde {\bm p}') \rangle_\text{c}\, \dd\tilde {\bm r}\, \dd\tilde {\bm r}' \, \dd\tilde {\bm p}\, \dd\tilde {\bm p}'}.
\end{equation}
In the BEC regime, the optimal $\sigma$ was found to be $\propto 1/\Delta$ or proportional to the pair sizes.
Similarly, in the BCS regime, $\max_{\tilde {\bm r}, \tilde {\bm p}}\langle \hat\rho_\uparrow (\tilde {\bm r}, \tilde {\bm p}) \hat\rho_\downarrow (\bm 0, -\tilde {\bm p}) \rangle$ increases monotonically as $\sigma$ increases.
Tracking the $\sigma$ at which this quantity reaches 50\% of the maximum value, one also finds that this $\sigma \propto 1/\Delta$.
Thus, by scanning the PSM parameters, information about the state can also be extracted.
See Fig.~\ref{fig:sigma_choice}.

Since the crossover region maximizes both the correlation amplitude and the saturation of the bound in Eq.~\eqref{eq:bcs_bound}, one speculates that it has the ``tightest'' phase-space distribution.
One can define the ``tightness'' mathematically as an $L^2$-norm of the POVM measurement
\begin{equation}
\int |\langle \hat\rho_\uparrow (\tilde {\bm r}, \tilde {\bm p}) \hat\rho_\downarrow(\tilde {\bm r}', \tilde {\bm p}') \rangle|^2 \,\frac{\dd\tilde {\bm r}\, \dd\tilde {\bm r}' \, \dd\tilde {\bm p}\, \dd\tilde {\bm p}'}{(2\pi\hbar)^4}.
\end{equation}
The $L^2$-norm is difficult to calculate, but one expects it to be peaked in the crossover region for a combination of reasons discussed above:
the BEC regime has a broad and shallow distribution in phase space, whereas the BCS regime has a smaller total weight and a more complex shape in momentum space. %\note{CW: the discussions of the bounds are interesting! Can one connect this to the PSD required for condensation in the BEC limit?}

Going beyond the ideal case, the PSM offers greater versatility.
We can consider three different ways in which the simple picture fails.
Firstly, there can be spatial inhomogeneity, as we address in Sec.~\ref{sec:inhom}.
Secondly, there can be non-zero temperature, so that some pairs are non-condensed with finite centre-of-mass momenta.
This is something that can be conveniently observed with the PSM using the ring protocol.
Taking the limit that $\sigma\to\infty$ and $\tilde{p}_\text{a}=\tilde{p}_\text{a}'=\gamma |\tilde {\bm p}|^2$, the ring protocol picks up contributions from CoM momenta up to $2\tilde p$, where the thermal pairs contribute fringes (similar to those seen in Fig.~\ref{fig:cross}) that become blurred at higher temperatures.
Finally, there can be beyond-mean-field pair fluctuation effects.
This is most easily observed in the averaged mode, as we noted, $\langle \hat\rho_{\text{KE}, \uparrow}({\bm r}) \hat\rho_\downarrow({\bm 0}) \rangle_\text{c} = |\nabla g_{\uparrow\downarrow}({\bm r})|^2$ in a BCS state, so failure of this relation could indicate such pair fluctuations.
Quantum fluctuations in the pairing field $\Delta_{\bm q}$ will also degrade the fringes in the ring protocol, and by observing the decay of the fringes, one can deduce the relative strengths of these fluctuations.

%\section{Measurements of the BCS Wavefunction}
%\label{app:bcs}
\section{General phase-space correlations in inhomogeneous superconductors}
\label{app:inhom_bcs}
In this section, we consider the phase-space correlation functions of general inhomogeneous superconducting states.
We take states of the form
\begin{equation}
    |\Psi \rangle \propto \exp\biggl(\int Z({\bm r}, {\bm r}') \hat{c}_{{\bm r}\uparrow}^\dagger \hat{c}_{{\bm r}'\downarrow}^\dagger \, \dd^2r \, \dd^2r' \biggr)|{\text{vac}}\rangle .
\end{equation}
{This is still a Gaussian state and Wick's theorem applies.
Thus, calculating correlation functions reduces to calculating the anomalous expectation value $\langle \hat{c}_{{\bm r}\uparrow}^\dagger \hat{c}_{{\bm r}'\downarrow}^\dagger  \rangle$.
By commuting operators through the exponentials, this can be evaluated to be}
\begin{equation}
    \langle \hat{c}_{{\bm r}\uparrow}^\dagger \hat{c}_{{\bm r}'\downarrow}^\dagger  \rangle \propto [Z^\dagger (I+Z Z^\dagger)^{-1} ]({\bm r}', {\bm r}),
\end{equation}
where we are treating $Z$ as a matrix and $I$ is the identity matrix~\cite{ring_nuclear_2004,dobaczewski_mean_field_1996}.
In the usual homogeneous BCS state, $Z_{\bm k} = v_{\bm k} / u_{\bm k}$ in Fourier space, and this expectation value reduces to $u_{\bm k} v_{\bm k}^*$, as we discussed in the main text.
Correspondingly, the phase-space correlation function is the Husimi Q function of this function.

If one assumes that the inhomogeneous state is comprised of different condensates, for instance (in CoM and relative coordinates) $Z({\bm r}_\text{rel}, {\bm R}) = (\ee^{\ii {\bm Q} \cdot {\bm R}} + \ee^{-\ii {\bm Q} \cdot {\bm R}})Z_\text{rel} ({\bm r}_\text{rel})$, then the above formula predicts the different harmonics in the correlation function.
One possible scenario justifying the truncation to the lowest harmonics is if the eigenvalues of $Z Z^\dagger$ are much smaller than $1$.
In the uniform BCS state, this would correspond to $v_{\bm k} / u_{\bm k} \ll 1$ for all $\bm k$, or the BEC limit.
Nevertheless, as a first approximation, this may also be done in the BCS limit.

\section{Measuring weights in different LLs}
\label{app:ll_weight}
The Glauber--Sudarshan P distribution of a density matrix $\hat{\rho}$ is defined as
\begin{equation}
    \hat{\rho} = \int P(\tilde x, \tilde p) |\tilde x, \tilde p\rangle \langle \tilde x, \tilde p| \, \frac{\dd \tilde x\, \dd \tilde p}{2\pi\hbar}.
\end{equation}
The Husimi Q distribution is defined as $Q(\tilde x, \tilde p) = \langle \tilde x, \tilde p |\hat{\rho}| \tilde x, \tilde p \rangle$.
The overlap between two states $\psi$ and $\phi$, described with $P_\psi$ and $Q_\phi$ respectively, is given by
\begin{multline}
    \operatorname{tr} \hat\rho_\psi \hat\rho_\phi = \int \frac{\dd \tilde x\, \dd \tilde p}{2\pi\hbar} \,P_\psi(\tilde x, \tilde p) \operatorname{tr} |\tilde x, \tilde p\rangle \langle \tilde x, \tilde p|\hat\rho_\phi = \\
    \int \frac{\dd \tilde x\, \dd \tilde p}{2\pi\hbar} \,P_\psi(\tilde x, \tilde p) Q_\phi(\tilde x, \tilde p),
\end{multline}
proving Eq.~\eqref{eq:overlap}.
In the many-body case with many coherent state labels $\{\tilde x_{i=1\dots N},\tilde p_i\}$, one defines correspondingly $Q(\{\tilde x_i, \tilde p_i\}) = \langle \{\tilde x_i, \tilde p_i\} | \hat{\rho} | \{\tilde x_i, \tilde p_i\} \rangle$ (as in main text) and $\hat{\rho} = \int P(\{\tilde x_i, \tilde p_i\}) |\{\tilde x_i, \tilde p_i\}\rangle \langle \{\tilde x_i, \tilde p_i\}| \, \prod_{i=1}^N \frac{\dd \tilde x_i \, \dd \tilde p_i}{2\pi\hbar}$, satisfying
\begin{equation}
    \operatorname{tr} \hat\rho_\psi \hat\rho_\phi = \int \prod_{i=1}^N \frac{\dd \tilde x_i \, \dd \tilde p_i}{2\pi\hbar} \,P_\psi(\{\tilde x_i, \tilde p_i\}) Q_\phi(\{\tilde x_i, \tilde p_i\}).
\end{equation}
To calculate the overlap of the relative wavefunction with the zero angular momentum states, we only need to calculate the Glauber--Sudarshan P distribution of said states.
These P distributions will be singular, but the integrals remain well-defined, as we show below.

In the symmetric gauge where ${\bm A} = \frac{B}{2}(-y, x)$, eigenstates in a magnetic field are described by two harmonic oscillators, one for the cyclotronic motion (Landau levels) and one for the guiding centre motion.
The zero angular momentum states are Fock states with equal occupation in both oscillators.
Using the known forms of the P distribution for Fock states, in the $n$th Landau level, in terms of the oscillator coordinates $\alpha$ and $\beta$
\begin{equation}
	P_n(\alpha, \beta) = \frac{\ee^{|\alpha|^2 + |\beta|^2}}{(n!)^2} (\partial_\alpha \partial_{\alpha^*} \partial_\beta \partial_{\beta^*})^n \delta(\alpha) \delta(\beta),
    \label{eq:glauber}
\end{equation}
where $\delta$ are Dirac delta functions.
This can then be rewritten in terms of the coordinates $x, y, p_x, p_y$, where the momentum is the canonical momentum, using the standard definitions $\alpha = \frac{1}{\sqrt2} [\ell_\text{B} (p_x-\ii p_y)/\hbar - (y + \ii x)/2\ell_\text{B}]$ and $\beta = \frac{1}{\sqrt2} [\ell_\text{B} (p_x+\ii p_y)/\hbar + (y - \ii x)/2\ell_\text{B}]$.

Eq.~\eqref{eq:glauber} is particularly simple for the first few Landau levels.
For the LLL, one has $P_0(\alpha, \beta) = \delta(\alpha) \delta(\beta)$, and the weight of the relative wavefunction in the LLL is simply given by $Q(0, 0)$.
For the first LL,
\begin{equation}
    P_1(\alpha, \beta) = \ee^{|\alpha|^2 + |\beta|^2} \partial_\alpha \partial_{\alpha^*} \partial_\beta \partial_{\beta^*} \delta(\alpha) \delta(\beta),
\end{equation}
and the weight in the first LL is given by
\begin{equation}
    \partial_\alpha \partial_{\alpha^*} \partial_\beta \partial_{\beta^*} [\ee^{|\alpha|^2 + |\beta|^2} Q(\alpha, \beta) ] \Big|_{\alpha,\alpha^*,\beta,\beta^*=0},
\end{equation}
which are the derivatives of the Husimi Q function at the origin.
In general, the weight in the $n$th LL is
\begin{equation}
    \frac{1}{(n!)^2}(\partial_\alpha \partial_{\alpha^*} \partial_\beta \partial_{\beta^*})^n [\ee^{|\alpha|^2 + |\beta|^2} Q(\alpha, \beta) ] \Big|_{\alpha,\alpha^*,\beta,\beta^*=0},
\end{equation}
involving higher order derivatives when $n$ is large.
Clearly, the higher order derivatives are more susceptible to noise, making measurements of weights in higher LLs difficult.


\begin{thebibliography}{93}%
\makeatletter
\providecommand \@ifxundefined [1]{%
 \@ifx{#1\undefined}
}%
\providecommand \@ifnum [1]{%
 \ifnum #1\expandafter \@firstoftwo
 \else \expandafter \@secondoftwo
 \fi
}%
\providecommand \@ifx [1]{%
 \ifx #1\expandafter \@firstoftwo
 \else \expandafter \@secondoftwo
 \fi
}%
\providecommand \natexlab [1]{#1}%
\providecommand \enquote  [1]{``#1''}%
\providecommand \bibnamefont  [1]{#1}%
\providecommand \bibfnamefont [1]{#1}%
\providecommand \citenamefont [1]{#1}%
\providecommand \href@noop [0]{\@secondoftwo}%
\providecommand \href [0]{\begingroup \@sanitize@url \@href}%
\providecommand \@href[1]{\@@startlink{#1}\@@href}%
\providecommand \@@href[1]{\endgroup#1\@@endlink}%
\providecommand \@sanitize@url [0]{\catcode `\\12\catcode `\$12\catcode
  `\&12\catcode `\#12\catcode `\^12\catcode `\_12\catcode `\%12\relax}%
\providecommand \@@startlink[1]{}%
\providecommand \@@endlink[0]{}%
\providecommand \url  [0]{\begingroup\@sanitize@url \@url }%
\providecommand \@url [1]{\endgroup\@href {#1}{\urlprefix }}%
\providecommand \urlprefix  [0]{URL }%
\providecommand \Eprint [0]{\href }%
\providecommand \doibase [0]{https://doi.org/}%
\providecommand \selectlanguage [0]{\@gobble}%
\providecommand \bibinfo  [0]{\@secondoftwo}%
\providecommand \bibfield  [0]{\@secondoftwo}%
\providecommand \translation [1]{[#1]}%
\providecommand \BibitemOpen [0]{}%
\providecommand \bibitemStop [0]{}%
\providecommand \bibitemNoStop [0]{.\EOS\space}%
\providecommand \EOS [0]{\spacefactor3000\relax}%
\providecommand \BibitemShut  [1]{\csname bibitem#1\endcsname}%
\let\auto@bib@innerbib\@empty
%</preamble>
\bibitem [{\citenamefont {Bakr}\ \emph {et~al.}(2010)\citenamefont {Bakr},
  \citenamefont {Peng}, \citenamefont {Tai}, \citenamefont {Ma}, \citenamefont
  {Simon}, \citenamefont {Gillen}, \citenamefont {F{\"{o}}lling}, \citenamefont
  {Pollet},\ and\ \citenamefont {Greiner}}]{Bakr2010}%
  \BibitemOpen
  \bibfield  {author} {\bibinfo {author} {\bibfnamefont {W.~S.}\ \bibnamefont
  {Bakr}}, \bibinfo {author} {\bibfnamefont {A.}~\bibnamefont {Peng}}, \bibinfo
  {author} {\bibfnamefont {M.~E.}\ \bibnamefont {Tai}}, \bibinfo {author}
  {\bibfnamefont {R.}~\bibnamefont {Ma}}, \bibinfo {author} {\bibfnamefont
  {J.}~\bibnamefont {Simon}}, \bibinfo {author} {\bibfnamefont {J.~I.}\
  \bibnamefont {Gillen}}, \bibinfo {author} {\bibfnamefont {S.}~\bibnamefont
  {F{\"{o}}lling}}, \bibinfo {author} {\bibfnamefont {L.}~\bibnamefont
  {Pollet}},\ and\ \bibinfo {author} {\bibfnamefont {M.}~\bibnamefont
  {Greiner}},\ }\bibfield  {title} {\bibinfo {title} {{Probing the
  superfluid-to-Mott insulator transition at the single-atom level}},\ }\href
  {https://doi.org/10.1126/science.1192368} {\bibfield  {journal} {\bibinfo
  {journal} {Science}\ }\textbf {\bibinfo {volume} {329}},\ \bibinfo {pages}
  {547} (\bibinfo {year} {2010})}\BibitemShut {NoStop}%
\bibitem [{\citenamefont {Sherson}\ \emph {et~al.}(2010)\citenamefont
  {Sherson}, \citenamefont {Weitenberg}, \citenamefont {Endres}, \citenamefont
  {Cheneau}, \citenamefont {Bloch},\ and\ \citenamefont {Kuhr}}]{Sherson2010}%
  \BibitemOpen
  \bibfield  {author} {\bibinfo {author} {\bibfnamefont {J.~F.}\ \bibnamefont
  {Sherson}}, \bibinfo {author} {\bibfnamefont {C.}~\bibnamefont {Weitenberg}},
  \bibinfo {author} {\bibfnamefont {M.}~\bibnamefont {Endres}}, \bibinfo
  {author} {\bibfnamefont {M.}~\bibnamefont {Cheneau}}, \bibinfo {author}
  {\bibfnamefont {I.}~\bibnamefont {Bloch}},\ and\ \bibinfo {author}
  {\bibfnamefont {S.}~\bibnamefont {Kuhr}},\ }\bibfield  {title} {\bibinfo
  {title} {Single-atom-resolved fluorescence imaging of an atomic {M}ott
  insulator},\ }\href {https://doi.org/10.1038/nature09378} {\bibfield
  {journal} {\bibinfo  {journal} {Nature}\ }\textbf {\bibinfo {volume} {467}},\
  \bibinfo {pages} {68–72} (\bibinfo {year} {2010})}\BibitemShut {NoStop}%
\bibitem [{\citenamefont {Endres}\ \emph {et~al.}(2011)\citenamefont {Endres},
  \citenamefont {Cheneau}, \citenamefont {Fukuhara}, \citenamefont
  {Weitenberg}, \citenamefont {Schauß}, \citenamefont {Gross}, \citenamefont
  {Mazza}, \citenamefont {Bañuls}, \citenamefont {Pollet}, \citenamefont
  {Bloch},\ and\ \citenamefont {Kuhr}}]{Endres2011}%
  \BibitemOpen
  \bibfield  {author} {\bibinfo {author} {\bibfnamefont {M.}~\bibnamefont
  {Endres}}, \bibinfo {author} {\bibfnamefont {M.}~\bibnamefont {Cheneau}},
  \bibinfo {author} {\bibfnamefont {T.}~\bibnamefont {Fukuhara}}, \bibinfo
  {author} {\bibfnamefont {C.}~\bibnamefont {Weitenberg}}, \bibinfo {author}
  {\bibfnamefont {P.}~\bibnamefont {Schauß}}, \bibinfo {author} {\bibfnamefont
  {C.}~\bibnamefont {Gross}}, \bibinfo {author} {\bibfnamefont
  {L.}~\bibnamefont {Mazza}}, \bibinfo {author} {\bibfnamefont {M.~C.}\
  \bibnamefont {Bañuls}}, \bibinfo {author} {\bibfnamefont {L.}~\bibnamefont
  {Pollet}}, \bibinfo {author} {\bibfnamefont {I.}~\bibnamefont {Bloch}},\ and\
  \bibinfo {author} {\bibfnamefont {S.}~\bibnamefont {Kuhr}},\ }\bibfield
  {title} {\bibinfo {title} {Observation of correlated particle-hole pairs and
  string order in low-dimensional {M}ott insulators},\ }\href
  {https://doi.org/10.1126/science.1209284} {\bibfield  {journal} {\bibinfo
  {journal} {Science}\ }\textbf {\bibinfo {volume} {334}},\ \bibinfo {pages}
  {200} (\bibinfo {year} {2011})},\ \Eprint
  {https://arxiv.org/abs/https://www.science.org/doi/pdf/10.1126/science.1209284}
  {https://www.science.org/doi/pdf/10.1126/science.1209284} \BibitemShut
  {NoStop}%
\bibitem [{\citenamefont {Gross}\ and\ \citenamefont {Bakr}(2021)}]{Gross2021}%
  \BibitemOpen
  \bibfield  {author} {\bibinfo {author} {\bibfnamefont {C.}~\bibnamefont
  {Gross}}\ and\ \bibinfo {author} {\bibfnamefont {W.~S.}\ \bibnamefont
  {Bakr}},\ }\bibfield  {title} {\bibinfo {title} {{Quantum gas microscopy for
  single atom and spin detection}},\ }\href
  {https://doi.org/10.1038/s41567-021-01370-5} {\bibfield  {journal} {\bibinfo
  {journal} {Nature Physics}\ }\textbf {\bibinfo {volume} {17}},\ \bibinfo
  {pages} {1316} (\bibinfo {year} {2021})}\BibitemShut {NoStop}%
\bibitem [{\citenamefont {Asteria}\ \emph {et~al.}(2021)\citenamefont
  {Asteria}, \citenamefont {Zahn}, \citenamefont {Kosch}, \citenamefont
  {Sengstock},\ and\ \citenamefont {Weitenberg}}]{Asteria2021}%
  \BibitemOpen
  \bibfield  {author} {\bibinfo {author} {\bibfnamefont {L.}~\bibnamefont
  {Asteria}}, \bibinfo {author} {\bibfnamefont {H.~P.}\ \bibnamefont {Zahn}},
  \bibinfo {author} {\bibfnamefont {M.~N.}\ \bibnamefont {Kosch}}, \bibinfo
  {author} {\bibfnamefont {K.}~\bibnamefont {Sengstock}},\ and\ \bibinfo
  {author} {\bibfnamefont {C.}~\bibnamefont {Weitenberg}},\ }\bibfield  {title}
  {\bibinfo {title} {{Quantum gas magnifier for sub-lattice-resolved imaging of
  3D quantum systems}},\ }\href {https://doi.org/10.1038/s41586-021-04011-2}
  {\bibfield  {journal} {\bibinfo  {journal} {Nature}\ }\textbf {\bibinfo
  {volume} {599}},\ \bibinfo {pages} {571} (\bibinfo {year}
  {2021})}\BibitemShut {NoStop}%
\bibitem [{\citenamefont {Brandstetter}\ \emph {et~al.}(2025)\citenamefont
  {Brandstetter}, \citenamefont {Heintze}, \citenamefont {Hill}, \citenamefont
  {Preiss}, \citenamefont {Ga\l{}ka},\ and\ \citenamefont
  {Jochim}}]{Brandstetter2025}%
  \BibitemOpen
  \bibfield  {author} {\bibinfo {author} {\bibfnamefont {S.}~\bibnamefont
  {Brandstetter}}, \bibinfo {author} {\bibfnamefont {C.}~\bibnamefont
  {Heintze}}, \bibinfo {author} {\bibfnamefont {P.}~\bibnamefont {Hill}},
  \bibinfo {author} {\bibfnamefont {P.~M.}\ \bibnamefont {Preiss}}, \bibinfo
  {author} {\bibfnamefont {M.}~\bibnamefont {Ga\l{}ka}},\ and\ \bibinfo
  {author} {\bibfnamefont {S.}~\bibnamefont {Jochim}},\ }\bibfield  {title}
  {\bibinfo {title} {Magnifying the wave function of interacting fermionic
  atoms},\ }\href {https://doi.org/10.1103/wdjr-m2hg} {\bibfield  {journal}
  {\bibinfo  {journal} {Phys. Rev. Lett.}\ }\textbf {\bibinfo {volume} {135}},\
  \bibinfo {pages} {103401} (\bibinfo {year} {2025})}\BibitemShut {NoStop}%
\bibitem [{\citenamefont {Murthy}\ and\ \citenamefont
  {Jochim}(2019)}]{Murthy2019}%
  \BibitemOpen
  \bibfield  {author} {\bibinfo {author} {\bibfnamefont {P.~A.}\ \bibnamefont
  {Murthy}}\ and\ \bibinfo {author} {\bibfnamefont {S.}~\bibnamefont
  {Jochim}},\ }\bibfield  {title} {\bibinfo {title} {Direct imaging of the
  order parameter of an atomic superfluid using matterwave optics},\ }\href
  {https://arxiv.org/abs/1911.10824} {\  (\bibinfo {year} {2019})},\ \Eprint
  {https://arxiv.org/abs/1911.10824} {arXiv:1911.10824 [cond-mat.quant-gas]}
  \BibitemShut {NoStop}%
\bibitem [{\citenamefont {Brüggenjürgen}\ \emph {et~al.}(2026)\citenamefont
  {Brüggenjürgen}, \citenamefont {Fischer},\ and\ \citenamefont
  {Weitenberg}}]{Bruggenjurgen2026}%
  \BibitemOpen
  \bibfield  {author} {\bibinfo {author} {\bibfnamefont {J.~C.}\ \bibnamefont
  {Brüggenjürgen}}, \bibinfo {author} {\bibfnamefont {M.~S.}\ \bibnamefont
  {Fischer}},\ and\ \bibinfo {author} {\bibfnamefont {C.}~\bibnamefont
  {Weitenberg}},\ }\bibfield  {title} {\bibinfo {title} {A phase microscope for
  quantum gases},\ }\href {https://doi.org/10.1126/science.adt1712} {\bibfield
  {journal} {\bibinfo  {journal} {Science}\ }\textbf {\bibinfo {volume}
  {393}},\ \bibinfo {pages} {167} (\bibinfo {year} {2026})}\BibitemShut
  {NoStop}%
\bibitem [{\citenamefont {Weitenberg}\ \emph {et~al.}(2026)\citenamefont
  {Weitenberg}, \citenamefont {Asteria}, \citenamefont {Carlsson},
  \citenamefont {Bohrdt},\ and\ \citenamefont {Grusdt}}]{Weitenberg2026}%
  \BibitemOpen
  \bibfield  {author} {\bibinfo {author} {\bibfnamefont {C.}~\bibnamefont
  {Weitenberg}}, \bibinfo {author} {\bibfnamefont {L.}~\bibnamefont {Asteria}},
  \bibinfo {author} {\bibfnamefont {O.}~\bibnamefont {Carlsson}}, \bibinfo
  {author} {\bibfnamefont {A.}~\bibnamefont {Bohrdt}},\ and\ \bibinfo {author}
  {\bibfnamefont {F.}~\bibnamefont {Grusdt}},\ }\bibfield  {title} {\bibinfo
  {title} {Protocols for a many-body phase microscope: From coherences and
  $d$-wave superconductivity to green's functions},\ }\href
  {https://doi.org/10.1103/99mf-c8tv} {\bibfield  {journal} {\bibinfo
  {journal} {PRX Quantum}\ }\textbf {\bibinfo {volume} {7}},\ \bibinfo {pages}
  {033068} (\bibinfo {year} {2026})}\BibitemShut {NoStop}%
\bibitem [{\citenamefont {Cooper}\ \emph {et~al.}(2026)\citenamefont {Cooper},
  \citenamefont {Yang},\ and\ \citenamefont
  {Weitenberg}}]{cooper_phase-space_2026}%
  \BibitemOpen
  \bibfield  {author} {\bibinfo {author} {\bibfnamefont {N.~R.}\ \bibnamefont
  {Cooper}}, \bibinfo {author} {\bibfnamefont {Y.}~\bibnamefont {Yang}},\ and\
  \bibinfo {author} {\bibfnamefont {C.}~\bibnamefont {Weitenberg}},\ }\bibfield
   {title} {\bibinfo {title} {Phase-space microscopes for quantum gases:
  Imaging conjugate variables and momentum-weighted densities},\ }\href
  {https://doi.org/10.1103/12rn-rlg8} {\bibfield  {journal} {\bibinfo
  {journal} {Phys. Rev. Lett.}\ }\textbf {\bibinfo {volume} {137}},\ \bibinfo
  {pages} {173402} (\bibinfo {year} {2026})}\BibitemShut {NoStop}%
\bibitem [{\citenamefont {Peres}(1995)}]{Peres1995}%
  \BibitemOpen
  \bibfield  {author} {\bibinfo {author} {\bibfnamefont {A.}~\bibnamefont
  {Peres}},\ }\href@noop {} {\emph {\bibinfo {title} {Quantum Theory: Concepts
  and Methods}}},\ \bibinfo {series} {Fundamental Theories of Physics},
  Vol.~\bibinfo {volume} {57}\ (\bibinfo  {publisher} {Kluwer Academic
  Publishers},\ \bibinfo {address} {Dordrecht},\ \bibinfo {year}
  {1995})\BibitemShut {NoStop}%
\bibitem [{\citenamefont {Arthurs}\ and\ \citenamefont
  {Kelly~Jr}(1965)}]{Arthurs1965}%
  \BibitemOpen
  \bibfield  {author} {\bibinfo {author} {\bibfnamefont {E.}~\bibnamefont
  {Arthurs}}\ and\ \bibinfo {author} {\bibfnamefont {J.~L.}\ \bibnamefont
  {Kelly~Jr}},\ }\bibfield  {title} {\bibinfo {title} {The concept of
  experimental accuracy and simultaneous measurements of position and
  momentum},\ }\href@noop {} {\bibfield  {journal} {\bibinfo  {journal} {Bell
  System Technical Journal}\ }\textbf {\bibinfo {volume} {44}},\ \bibinfo
  {pages} {725} (\bibinfo {year} {1965})}\BibitemShut {NoStop}%
\bibitem [{\citenamefont {Afek}\ \emph {et~al.}(2017)\citenamefont {Afek},
  \citenamefont {Coslovsky}, \citenamefont {Courvoisier}, \citenamefont
  {Livneh},\ and\ \citenamefont {Davidson}}]{Afek2017}%
  \BibitemOpen
  \bibfield  {author} {\bibinfo {author} {\bibfnamefont {G.}~\bibnamefont
  {Afek}}, \bibinfo {author} {\bibfnamefont {J.}~\bibnamefont {Coslovsky}},
  \bibinfo {author} {\bibfnamefont {A.}~\bibnamefont {Courvoisier}}, \bibinfo
  {author} {\bibfnamefont {O.}~\bibnamefont {Livneh}},\ and\ \bibinfo {author}
  {\bibfnamefont {N.}~\bibnamefont {Davidson}},\ }\bibfield  {title} {\bibinfo
  {title} {Observing power-law dynamics of position-velocity correlation in
  anomalous diffusion},\ }\href
  {https://doi.org/10.1103/PhysRevLett.119.060602} {\bibfield  {journal}
  {\bibinfo  {journal} {Phys. Rev. Lett.}\ }\textbf {\bibinfo {volume} {119}},\
  \bibinfo {pages} {060602} (\bibinfo {year} {2017})}\BibitemShut {NoStop}%
\bibitem [{\citenamefont {Pauli}(1933)}]{pauli_1933}%
  \BibitemOpen
  \bibfield  {author} {\bibinfo {author} {\bibfnamefont {W.}~\bibnamefont
  {Pauli}},\ }in\ \href@noop {} {\emph {\bibinfo {booktitle} {Handbuch der
  {Physik}}}},\ Vol.\ \bibinfo {volume} {24, Pt.~1},\ \bibinfo {editor} {edited
  by\ \bibinfo {editor} {\bibfnamefont {H.}~\bibnamefont {Geiger}}\ and\
  \bibinfo {editor} {\bibfnamefont {K.}~\bibnamefont {Scheel}}}\ (\bibinfo
  {publisher} {Springer},\ \bibinfo {address} {Berlin},\ \bibinfo {year}
  {1933})\ p.~\bibinfo {pages} {98}\BibitemShut {NoStop}%
\bibitem [{\citenamefont {Corbett}(2006)}]{corbett_pauli_2006}%
  \BibitemOpen
  \bibfield  {author} {\bibinfo {author} {\bibfnamefont {J.}~\bibnamefont
  {Corbett}},\ }\bibfield  {title} {\bibinfo {title} {The pauli problem, state
  reconstruction and quantum-real numbers},\ }\href
  {https://doi.org/10.1016/S0034-4877(06)80008-X} {\bibfield  {journal}
  {\bibinfo  {journal} {Reports on Mathematical Physics}\ }\textbf {\bibinfo
  {volume} {57}},\ \bibinfo {pages} {53} (\bibinfo {year} {2006})}\BibitemShut
  {NoStop}%
\bibitem [{\citenamefont {Zheng}(2024)}]{zheng_pauli_2024}%
  \BibitemOpen
  \bibfield  {author} {\bibinfo {author} {\bibfnamefont {H.}~\bibnamefont
  {Zheng}},\ }\bibfield  {title} {\bibinfo {title} {{The {Pauli} problem and
  wave function lifting: reconstruction of quantum states from physical
  observables}},\ }\href {https://doi.org/10.3934/mine.2024025} {\bibfield
  {journal} {\bibinfo  {journal} {Mathematics in Engineering}\ }\textbf
  {\bibinfo {volume} {6}},\ \bibinfo {pages} {648} (\bibinfo {year}
  {2024})}\BibitemShut {NoStop}%
\bibitem [{\citenamefont {Husimi}(1940)}]{husimi1940density}%
  \BibitemOpen
  \bibfield  {author} {\bibinfo {author} {\bibfnamefont {K.}~\bibnamefont
  {Husimi}},\ }\bibfield  {title} {\bibinfo {title} {Some formal properties of
  the density matrix},\ }\href@noop {} {\bibfield  {journal} {\bibinfo
  {journal} {Proceedings of the Physico-Mathematical Society of Japan. 3rd
  Series}\ }\textbf {\bibinfo {volume} {22}},\ \bibinfo {pages} {264} (\bibinfo
  {year} {1940})}\BibitemShut {NoStop}%
\bibitem [{\citenamefont {Leonhardt}(2010)}]{leonhardt2010essential}%
  \BibitemOpen
  \bibfield  {author} {\bibinfo {author} {\bibfnamefont {U.}~\bibnamefont
  {Leonhardt}},\ }\href@noop {} {\emph {\bibinfo {title} {Essential Quantum
  Optics: From Quantum Measurements to Black Holes}}}\ (\bibinfo  {publisher}
  {Cambridge University Press},\ \bibinfo {address} {Cambridge},\ \bibinfo
  {year} {2010})\BibitemShut {NoStop}%
\bibitem [{\citenamefont
  {Czerwinski}(2022)}]{czerwinskiSelectedConceptsQuantum2022}%
  \BibitemOpen
  \bibfield  {author} {\bibinfo {author} {\bibfnamefont {A.}~\bibnamefont
  {Czerwinski}},\ }\bibfield  {title} {\bibinfo {title} {Selected {Concepts} of
  {Quantum} {State} {Tomography}},\ }\href {https://doi.org/10.3390/opt3030026}
  {\bibfield  {journal} {\bibinfo  {journal} {Optics}\ }\textbf {\bibinfo
  {volume} {3}},\ \bibinfo {pages} {268} (\bibinfo {year} {2022})}\BibitemShut
  {NoStop}%
\bibitem [{\citenamefont {Cramer}\ \emph {et~al.}(2010)\citenamefont {Cramer},
  \citenamefont {Plenio}, \citenamefont {Flammia}, \citenamefont {Somma},
  \citenamefont {Gross}, \citenamefont {Bartlett}, \citenamefont
  {Landon-Cardinal}, \citenamefont {Poulin},\ and\ \citenamefont
  {Liu}}]{cramerEfficientQuantumState2010}%
  \BibitemOpen
  \bibfield  {author} {\bibinfo {author} {\bibfnamefont {M.}~\bibnamefont
  {Cramer}}, \bibinfo {author} {\bibfnamefont {M.~B.}\ \bibnamefont {Plenio}},
  \bibinfo {author} {\bibfnamefont {S.~T.}\ \bibnamefont {Flammia}}, \bibinfo
  {author} {\bibfnamefont {R.}~\bibnamefont {Somma}}, \bibinfo {author}
  {\bibfnamefont {D.}~\bibnamefont {Gross}}, \bibinfo {author} {\bibfnamefont
  {S.~D.}\ \bibnamefont {Bartlett}}, \bibinfo {author} {\bibfnamefont
  {O.}~\bibnamefont {Landon-Cardinal}}, \bibinfo {author} {\bibfnamefont
  {D.}~\bibnamefont {Poulin}},\ and\ \bibinfo {author} {\bibfnamefont {Y.-K.}\
  \bibnamefont {Liu}},\ }\bibfield  {title} {\bibinfo {title} {Efficient
  quantum state tomography},\ }\href {https://doi.org/10.1038/ncomms1147}
  {\bibfield  {journal} {\bibinfo  {journal} {Nature Communications}\ }\textbf
  {\bibinfo {volume} {1}},\ \bibinfo {pages} {149} (\bibinfo {year}
  {2010})}\BibitemShut {NoStop}%
\bibitem [{\citenamefont {Lundeen}\ \emph {et~al.}(2011)\citenamefont
  {Lundeen}, \citenamefont {Sutherland}, \citenamefont {Patel}, \citenamefont
  {Stewart},\ and\ \citenamefont
  {Bamber}}]{lundeenDirectMeasurementQuantum2011}%
  \BibitemOpen
  \bibfield  {author} {\bibinfo {author} {\bibfnamefont {J.~S.}\ \bibnamefont
  {Lundeen}}, \bibinfo {author} {\bibfnamefont {B.}~\bibnamefont {Sutherland}},
  \bibinfo {author} {\bibfnamefont {A.}~\bibnamefont {Patel}}, \bibinfo
  {author} {\bibfnamefont {C.}~\bibnamefont {Stewart}},\ and\ \bibinfo {author}
  {\bibfnamefont {C.}~\bibnamefont {Bamber}},\ }\bibfield  {title} {\bibinfo
  {title} {Direct measurement of the quantum wavefunction},\ }\href
  {https://doi.org/10.1038/nature10120} {\bibfield  {journal} {\bibinfo
  {journal} {Nature}\ }\textbf {\bibinfo {volume} {474}},\ \bibinfo {pages}
  {188} (\bibinfo {year} {2011})}\BibitemShut {NoStop}%
\bibitem [{\citenamefont {Hofmann}\ and\ \citenamefont
  {Takeuchi}(2004)}]{hofmannQuantumstateTomographySpinl2004}%
  \BibitemOpen
  \bibfield  {author} {\bibinfo {author} {\bibfnamefont {H.~F.}\ \bibnamefont
  {Hofmann}}\ and\ \bibinfo {author} {\bibfnamefont {S.}~\bibnamefont
  {Takeuchi}},\ }\bibfield  {title} {\bibinfo {title} {Quantum-state tomography
  for spin-l systems},\ }\href {https://doi.org/10.1103/PhysRevA.69.042108}
  {\bibfield  {journal} {\bibinfo  {journal} {Physical Review A}\ }\textbf
  {\bibinfo {volume} {69}},\ \bibinfo {pages} {042108} (\bibinfo {year}
  {2004})},\ \bibinfo {note} {arXiv:quant-ph/0310003}\BibitemShut {NoStop}%
\bibitem [{\citenamefont {Kanem}\ \emph {et~al.}(2005)\citenamefont {Kanem},
  \citenamefont {Maneshi}, \citenamefont {Myrskog},\ and\ \citenamefont
  {Steinberg}}]{Kanem2005}%
  \BibitemOpen
  \bibfield  {author} {\bibinfo {author} {\bibfnamefont {J.~F.}\ \bibnamefont
  {Kanem}}, \bibinfo {author} {\bibfnamefont {S.}~\bibnamefont {Maneshi}},
  \bibinfo {author} {\bibfnamefont {S.~H.}\ \bibnamefont {Myrskog}},\ and\
  \bibinfo {author} {\bibfnamefont {A.~M.}\ \bibnamefont {Steinberg}},\
  }\bibfield  {title} {\bibinfo {title} {Phase space tomography of classical
  and nonclassical vibrational states of atoms in an optical lattice},\ }\href
  {https://doi.org/10.1088/1464-4266/7/12/037} {\bibfield  {journal} {\bibinfo
  {journal} {Journal of Optics B: Quantum and Semiclassical Optics}\ }\textbf
  {\bibinfo {volume} {7}},\ \bibinfo {pages} {S705} (\bibinfo {year}
  {2005})}\BibitemShut {NoStop}%
\bibitem [{\citenamefont {Brown}\ \emph {et~al.}(2023)\citenamefont {Brown},
  \citenamefont {Muleady}, \citenamefont {Dworschack}, \citenamefont
  {Lewis-Swan}, \citenamefont {Rey}, \citenamefont {Romero-Isart},\ and\
  \citenamefont {Regal}}]{Brown2023}%
  \BibitemOpen
  \bibfield  {author} {\bibinfo {author} {\bibfnamefont {M.~O.}\ \bibnamefont
  {Brown}}, \bibinfo {author} {\bibfnamefont {S.~R.}\ \bibnamefont {Muleady}},
  \bibinfo {author} {\bibfnamefont {W.~J.}\ \bibnamefont {Dworschack}},
  \bibinfo {author} {\bibfnamefont {R.~J.}\ \bibnamefont {Lewis-Swan}},
  \bibinfo {author} {\bibfnamefont {A.~M.}\ \bibnamefont {Rey}}, \bibinfo
  {author} {\bibfnamefont {O.}~\bibnamefont {Romero-Isart}},\ and\ \bibinfo
  {author} {\bibfnamefont {C.~A.}\ \bibnamefont {Regal}},\ }\bibfield  {title}
  {\bibinfo {title} {{Time-of-flight quantum tomography of an atom in an
  optical tweezer}},\ }\href {https://doi.org/10.1038/s41567-022-01890-8}
  {\bibfield  {journal} {\bibinfo  {journal} {Nature Physics}\ }\textbf
  {\bibinfo {volume} {19}},\ \bibinfo {pages} {569} (\bibinfo {year}
  {2023})}\BibitemShut {NoStop}%
\bibitem [{Note1()}]{Note1}%
  \BibitemOpen
  \bibinfo {note} {In simple cases, such as the ground state or states with
  symmetry constraints, position and momentum space data can be
  sufficient.}\BibitemShut {Stop}%
\bibitem [{\citenamefont {Zachos}\ \emph {et~al.}(2005)\citenamefont {Zachos},
  \citenamefont {Fairlie},\ and\ \citenamefont {Curtright}}]{Zachos2005}%
  \BibitemOpen
  \bibinfo {editor} {\bibfnamefont {C.~K.}\ \bibnamefont {Zachos}}, \bibinfo
  {editor} {\bibfnamefont {D.~B.}\ \bibnamefont {Fairlie}},\ and\ \bibinfo
  {editor} {\bibfnamefont {T.~L.}\ \bibnamefont {Curtright}},\ eds.,\ \href
  {https://doi.org/10.1142/5287} {\emph {\bibinfo {title} {Quantum Mechanics in
  Phase Space: An Overview with Selected Papers}}},\ \bibinfo {series} {World
  Scientific Series in 20th Century Physics}, Vol.~\bibinfo {volume} {34}\
  (\bibinfo  {publisher} {World Scientific Publishing Company},\ \bibinfo
  {year} {2005})\BibitemShut {NoStop}%
\bibitem [{\citenamefont {Sugita}(2003)}]{sugita_moments_2003}%
  \BibitemOpen
  \bibfield  {author} {\bibinfo {author} {\bibfnamefont {A.}~\bibnamefont
  {Sugita}},\ }\bibfield  {title} {\bibinfo {title} {Moments of generalized
  {Husimi} distributions and complexity of many-body quantum states},\ }\href
  {https://doi.org/10.1088/0305-4470/36/34/310} {\bibfield  {journal} {\bibinfo
   {journal} {Journal of Physics A: Mathematical and General}\ }\textbf
  {\bibinfo {volume} {36}},\ \bibinfo {pages} {9081} (\bibinfo {year}
  {2003})}\BibitemShut {NoStop}%
\bibitem [{\citenamefont {Simon}(2000)}]{simon_peres-horodecki_2000}%
  \BibitemOpen
  \bibfield  {author} {\bibinfo {author} {\bibfnamefont {R.}~\bibnamefont
  {Simon}},\ }\bibfield  {title} {\bibinfo {title} {Peres-{Horodecki}
  {Separability} {Criterion} for {Continuous} {Variable} {Systems}},\ }\href
  {https://doi.org/10.1103/PhysRevLett.84.2726} {\bibfield  {journal} {\bibinfo
   {journal} {Physical Review Letters}\ }\textbf {\bibinfo {volume} {84}},\
  \bibinfo {pages} {2726} (\bibinfo {year} {2000})}\BibitemShut {NoStop}%
\bibitem [{\citenamefont {Duan}\ \emph {et~al.}(2000)\citenamefont {Duan},
  \citenamefont {Giedke}, \citenamefont {Cirac},\ and\ \citenamefont
  {Zoller}}]{duan_inseparability_2000}%
  \BibitemOpen
  \bibfield  {author} {\bibinfo {author} {\bibfnamefont {L.-M.}\ \bibnamefont
  {Duan}}, \bibinfo {author} {\bibfnamefont {G.}~\bibnamefont {Giedke}},
  \bibinfo {author} {\bibfnamefont {J.~I.}\ \bibnamefont {Cirac}},\ and\
  \bibinfo {author} {\bibfnamefont {P.}~\bibnamefont {Zoller}},\ }\bibfield
  {title} {\bibinfo {title} {Inseparability {Criterion} for {Continuous}
  {Variable} {Systems}},\ }\href@noop {} {\bibfield  {journal} {\bibinfo
  {journal} {Physical Review Letters}\ }\textbf {\bibinfo {volume} {84}},\
  \bibinfo {pages} {2722} (\bibinfo {year} {2000})}\BibitemShut {NoStop}%
\bibitem [{\citenamefont {Mancini}\ \emph {et~al.}(2002)\citenamefont
  {Mancini}, \citenamefont {Giovannetti}, \citenamefont {Vitali},\ and\
  \citenamefont {Tombesi}}]{mancini_entangling_2002}%
  \BibitemOpen
  \bibfield  {author} {\bibinfo {author} {\bibfnamefont {S.}~\bibnamefont
  {Mancini}}, \bibinfo {author} {\bibfnamefont {V.}~\bibnamefont
  {Giovannetti}}, \bibinfo {author} {\bibfnamefont {D.}~\bibnamefont
  {Vitali}},\ and\ \bibinfo {author} {\bibfnamefont {P.}~\bibnamefont
  {Tombesi}},\ }\bibfield  {title} {\bibinfo {title} {Entangling {Macroscopic}
  {Oscillators} {Exploiting} {Radiation} {Pressure}},\ }\href
  {https://doi.org/10.1103/PhysRevLett.88.120401} {\bibfield  {journal}
  {\bibinfo  {journal} {Physical Review Letters}\ }\textbf {\bibinfo {volume}
  {88}},\ \bibinfo {pages} {120401} (\bibinfo {year} {2002})}\BibitemShut
  {NoStop}%
\bibitem [{\citenamefont {Agarwal}\ and\ \citenamefont
  {Biswas}(2005)}]{agarwal_inseparability_2005}%
  \BibitemOpen
  \bibfield  {author} {\bibinfo {author} {\bibfnamefont {G.~S.}\ \bibnamefont
  {Agarwal}}\ and\ \bibinfo {author} {\bibfnamefont {A.}~\bibnamefont
  {Biswas}},\ }\bibfield  {title} {\bibinfo {title} {Inseparability
  inequalities for higher order moments for bipartite systems},\ }\href
  {https://doi.org/10.1088/1367-2630/7/1/211} {\bibfield  {journal} {\bibinfo
  {journal} {New Journal of Physics}\ }\textbf {\bibinfo {volume} {7}},\
  \bibinfo {pages} {211} (\bibinfo {year} {2005})}\BibitemShut {NoStop}%
\bibitem [{\citenamefont {Shchukin}\ and\ \citenamefont
  {Vogel}(2005)}]{ShchukinVogel2005}%
  \BibitemOpen
  \bibfield  {author} {\bibinfo {author} {\bibfnamefont {E.}~\bibnamefont
  {Shchukin}}\ and\ \bibinfo {author} {\bibfnamefont {W.}~\bibnamefont
  {Vogel}},\ }\bibfield  {title} {\bibinfo {title} {Inseparability criteria for
  continuous bipartite quantum states},\ }\href
  {https://doi.org/10.1103/PhysRevLett.95.230502} {\bibfield  {journal}
  {\bibinfo  {journal} {Phys. Rev. Lett.}\ }\textbf {\bibinfo {volume} {95}},\
  \bibinfo {pages} {230502} (\bibinfo {year} {2005})},\ \Eprint
  {https://arxiv.org/abs/quant-ph/0508132} {quant-ph/0508132} \BibitemShut
  {NoStop}%
\bibitem [{\citenamefont {Miranowicz}\ \emph {et~al.}(2009)\citenamefont
  {Miranowicz}, \citenamefont {Piani}, \citenamefont {Horodecki},\ and\
  \citenamefont {Horodecki}}]{Miranowicz2009}%
  \BibitemOpen
  \bibfield  {author} {\bibinfo {author} {\bibfnamefont {A.}~\bibnamefont
  {Miranowicz}}, \bibinfo {author} {\bibfnamefont {M.}~\bibnamefont {Piani}},
  \bibinfo {author} {\bibfnamefont {P.}~\bibnamefont {Horodecki}},\ and\
  \bibinfo {author} {\bibfnamefont {R.}~\bibnamefont {Horodecki}},\ }\bibfield
  {title} {\bibinfo {title} {Inseparability criteria based on matrices of
  moments},\ }\href {https://doi.org/10.1103/PhysRevA.80.052303} {\bibfield
  {journal} {\bibinfo  {journal} {Phys. Rev. A}\ }\textbf {\bibinfo {volume}
  {80}},\ \bibinfo {pages} {052303} (\bibinfo {year} {2009})},\ \Eprint
  {https://arxiv.org/abs/quant-ph/0605001} {quant-ph/0605001} \BibitemShut
  {NoStop}%
\bibitem [{\citenamefont {Julsgaard}\ \emph {et~al.}(2001)\citenamefont
  {Julsgaard}, \citenamefont {Kozhekin},\ and\ \citenamefont
  {Polzik}}]{Julsgaard2001}%
  \BibitemOpen
  \bibfield  {author} {\bibinfo {author} {\bibfnamefont {B.}~\bibnamefont
  {Julsgaard}}, \bibinfo {author} {\bibfnamefont {A.}~\bibnamefont
  {Kozhekin}},\ and\ \bibinfo {author} {\bibfnamefont {E.~S.}\ \bibnamefont
  {Polzik}},\ }\bibfield  {title} {\bibinfo {title} {Experimental long-lived
  entanglement of two macroscopic objects},\ }\href
  {https://doi.org/10.1038/35096524} {\bibfield  {journal} {\bibinfo  {journal}
  {Nature}\ }\textbf {\bibinfo {volume} {413}},\ \bibinfo {pages} {400}
  (\bibinfo {year} {2001})}\BibitemShut {NoStop}%
\bibitem [{\citenamefont {Peise}\ \emph {et~al.}(2015)\citenamefont {Peise},
  \citenamefont {Kruse}, \citenamefont {Lange}, \citenamefont {L{\"u}cke},
  \citenamefont {Pezz{\`e}}, \citenamefont {Arlt}, \citenamefont {Ertmer},
  \citenamefont {Hammerer}, \citenamefont {Santos}, \citenamefont {Smerzi},\
  and\ \citenamefont {Klempt}}]{Peise2015}%
  \BibitemOpen
  \bibfield  {author} {\bibinfo {author} {\bibfnamefont {J.}~\bibnamefont
  {Peise}}, \bibinfo {author} {\bibfnamefont {I.}~\bibnamefont {Kruse}},
  \bibinfo {author} {\bibfnamefont {K.}~\bibnamefont {Lange}}, \bibinfo
  {author} {\bibfnamefont {B.}~\bibnamefont {L{\"u}cke}}, \bibinfo {author}
  {\bibfnamefont {L.}~\bibnamefont {Pezz{\`e}}}, \bibinfo {author}
  {\bibfnamefont {J.}~\bibnamefont {Arlt}}, \bibinfo {author} {\bibfnamefont
  {W.}~\bibnamefont {Ertmer}}, \bibinfo {author} {\bibfnamefont
  {K.}~\bibnamefont {Hammerer}}, \bibinfo {author} {\bibfnamefont
  {L.}~\bibnamefont {Santos}}, \bibinfo {author} {\bibfnamefont
  {A.}~\bibnamefont {Smerzi}},\ and\ \bibinfo {author} {\bibfnamefont
  {C.}~\bibnamefont {Klempt}},\ }\bibfield  {title} {\bibinfo {title}
  {Satisfying the {Einstein--Podolsky--Rosen} criterion with massive
  particles},\ }\href {https://doi.org/10.1038/ncomms9984} {\bibfield
  {journal} {\bibinfo  {journal} {Nature Communications}\ }\textbf {\bibinfo
  {volume} {6}},\ \bibinfo {pages} {8984} (\bibinfo {year} {2015})}\BibitemShut
  {NoStop}%
\bibitem [{\citenamefont {Fadel}\ \emph {et~al.}(2018)\citenamefont {Fadel},
  \citenamefont {Zibold}, \citenamefont {D{\'e}camps},\ and\ \citenamefont
  {Treutlein}}]{Fadel2018}%
  \BibitemOpen
  \bibfield  {author} {\bibinfo {author} {\bibfnamefont {M.}~\bibnamefont
  {Fadel}}, \bibinfo {author} {\bibfnamefont {T.}~\bibnamefont {Zibold}},
  \bibinfo {author} {\bibfnamefont {B.}~\bibnamefont {D{\'e}camps}},\ and\
  \bibinfo {author} {\bibfnamefont {P.}~\bibnamefont {Treutlein}},\ }\bibfield
  {title} {\bibinfo {title} {Spatial entanglement patterns and
  {Einstein--Podolsky--Rosen} steering in {Bose--Einstein} condensates},\
  }\href {https://doi.org/10.1126/science.aao1850} {\bibfield  {journal}
  {\bibinfo  {journal} {Science}\ }\textbf {\bibinfo {volume} {360}},\ \bibinfo
  {pages} {409} (\bibinfo {year} {2018})}\BibitemShut {NoStop}%
\bibitem [{\citenamefont {Kunkel}\ \emph {et~al.}(2018)\citenamefont {Kunkel},
  \citenamefont {Pr{\"u}fer}, \citenamefont {Strobel}, \citenamefont
  {Linnemann}, \citenamefont {Fr{\"o}lian}, \citenamefont {Gasenzer},
  \citenamefont {G{\"a}rttner},\ and\ \citenamefont {Oberthaler}}]{Kunkel2018}%
  \BibitemOpen
  \bibfield  {author} {\bibinfo {author} {\bibfnamefont {P.}~\bibnamefont
  {Kunkel}}, \bibinfo {author} {\bibfnamefont {M.}~\bibnamefont {Pr{\"u}fer}},
  \bibinfo {author} {\bibfnamefont {H.}~\bibnamefont {Strobel}}, \bibinfo
  {author} {\bibfnamefont {D.}~\bibnamefont {Linnemann}}, \bibinfo {author}
  {\bibfnamefont {A.}~\bibnamefont {Fr{\"o}lian}}, \bibinfo {author}
  {\bibfnamefont {T.}~\bibnamefont {Gasenzer}}, \bibinfo {author}
  {\bibfnamefont {M.}~\bibnamefont {G{\"a}rttner}},\ and\ \bibinfo {author}
  {\bibfnamefont {M.~K.}\ \bibnamefont {Oberthaler}},\ }\bibfield  {title}
  {\bibinfo {title} {Spatially distributed multipartite entanglement enables
  {EPR} steering of atomic clouds},\ }\href
  {https://doi.org/10.1126/science.aao2254} {\bibfield  {journal} {\bibinfo
  {journal} {Science}\ }\textbf {\bibinfo {volume} {360}},\ \bibinfo {pages}
  {413} (\bibinfo {year} {2018})}\BibitemShut {NoStop}%
\bibitem [{\citenamefont {Lange}\ \emph {et~al.}(2018)\citenamefont {Lange},
  \citenamefont {Peise}, \citenamefont {L{\"u}cke}, \citenamefont {Kruse},
  \citenamefont {Vitagliano}, \citenamefont {Apellaniz}, \citenamefont
  {Kleinmann}, \citenamefont {T{\'o}th},\ and\ \citenamefont
  {Klempt}}]{Lange2018}%
  \BibitemOpen
  \bibfield  {author} {\bibinfo {author} {\bibfnamefont {K.}~\bibnamefont
  {Lange}}, \bibinfo {author} {\bibfnamefont {J.}~\bibnamefont {Peise}},
  \bibinfo {author} {\bibfnamefont {B.}~\bibnamefont {L{\"u}cke}}, \bibinfo
  {author} {\bibfnamefont {I.}~\bibnamefont {Kruse}}, \bibinfo {author}
  {\bibfnamefont {G.}~\bibnamefont {Vitagliano}}, \bibinfo {author}
  {\bibfnamefont {I.}~\bibnamefont {Apellaniz}}, \bibinfo {author}
  {\bibfnamefont {M.}~\bibnamefont {Kleinmann}}, \bibinfo {author}
  {\bibfnamefont {G.}~\bibnamefont {T{\'o}th}},\ and\ \bibinfo {author}
  {\bibfnamefont {C.}~\bibnamefont {Klempt}},\ }\bibfield  {title} {\bibinfo
  {title} {Entanglement between two spatially separated atomic modes},\ }\href
  {https://doi.org/10.1126/science.aao2035} {\bibfield  {journal} {\bibinfo
  {journal} {Science}\ }\textbf {\bibinfo {volume} {360}},\ \bibinfo {pages}
  {416} (\bibinfo {year} {2018})}\BibitemShut {NoStop}%
\bibitem [{\citenamefont {Tura}\ \emph {et~al.}(2014)\citenamefont {Tura},
  \citenamefont {Augusiak}, \citenamefont {Sainz}, \citenamefont {Vértesi},
  \citenamefont {Lewenstein},\ and\ \citenamefont {Acín}}]{Tura2014}%
  \BibitemOpen
  \bibfield  {author} {\bibinfo {author} {\bibfnamefont {J.}~\bibnamefont
  {Tura}}, \bibinfo {author} {\bibfnamefont {R.}~\bibnamefont {Augusiak}},
  \bibinfo {author} {\bibfnamefont {A.~B.}\ \bibnamefont {Sainz}}, \bibinfo
  {author} {\bibfnamefont {T.}~\bibnamefont {Vértesi}}, \bibinfo {author}
  {\bibfnamefont {M.}~\bibnamefont {Lewenstein}},\ and\ \bibinfo {author}
  {\bibfnamefont {A.}~\bibnamefont {Acín}},\ }\bibfield  {title} {\bibinfo
  {title} {Detecting nonlocality in many-body quantum states},\ }\href
  {https://doi.org/10.1126/science.1247715} {\bibfield  {journal} {\bibinfo
  {journal} {Science}\ }\textbf {\bibinfo {volume} {344}},\ \bibinfo {pages}
  {1256} (\bibinfo {year} {2014})},\ \Eprint
  {https://arxiv.org/abs/https://www.science.org/doi/pdf/10.1126/science.1247715}
  {https://www.science.org/doi/pdf/10.1126/science.1247715} \BibitemShut
  {NoStop}%
\bibitem [{\citenamefont {Fr\'erot}\ \emph {et~al.}(2022)\citenamefont
  {Fr\'erot}, \citenamefont {Baccari},\ and\ \citenamefont
  {Ac\'{\i}n}}]{Frerot2022}%
  \BibitemOpen
  \bibfield  {author} {\bibinfo {author} {\bibfnamefont {I.}~\bibnamefont
  {Fr\'erot}}, \bibinfo {author} {\bibfnamefont {F.}~\bibnamefont {Baccari}},\
  and\ \bibinfo {author} {\bibfnamefont {A.}~\bibnamefont {Ac\'{\i}n}},\
  }\bibfield  {title} {\bibinfo {title} {Unveiling quantum entanglement in
  many-body systems from partial information},\ }\href
  {https://doi.org/10.1103/PRXQuantum.3.010342} {\bibfield  {journal} {\bibinfo
   {journal} {PRX Quantum}\ }\textbf {\bibinfo {volume} {3}},\ \bibinfo {pages}
  {010342} (\bibinfo {year} {2022})}\BibitemShut {NoStop}%
\bibitem [{\citenamefont {Islam}\ \emph {et~al.}(2015)\citenamefont {Islam},
  \citenamefont {Ma}, \citenamefont {Preiss}, \citenamefont {Lukin},
  \citenamefont {Rispoli},\ and\ \citenamefont {Greiner}}]{Islam2015}%
  \BibitemOpen
  \bibfield  {author} {\bibinfo {author} {\bibfnamefont {R.}~\bibnamefont
  {Islam}}, \bibinfo {author} {\bibfnamefont {R.}~\bibnamefont {Ma}}, \bibinfo
  {author} {\bibfnamefont {M.}~\bibnamefont {Preiss}, \bibfnamefont {Philipp M.
  Eric~Tai}}, \bibinfo {author} {\bibfnamefont {A.}~\bibnamefont {Lukin}},
  \bibinfo {author} {\bibfnamefont {M.}~\bibnamefont {Rispoli}},\ and\ \bibinfo
  {author} {\bibfnamefont {M.}~\bibnamefont {Greiner}},\ }\bibfield  {title}
  {\bibinfo {title} {{Measuring entanglement entropy in a quantum many-body
  system}},\ }\href {https://doi.org/10.1038/nature15750} {\bibfield  {journal}
  {\bibinfo  {journal} {Nature}\ }\textbf {\bibinfo {volume} {528}},\ \bibinfo
  {pages} {77} (\bibinfo {year} {2015})}\BibitemShut {NoStop}%
\bibitem [{\citenamefont {Ketterle}\ and\ \citenamefont
  {Zwierlein}(2008)}]{Ketterle2008}%
  \BibitemOpen
  \bibfield  {author} {\bibinfo {author} {\bibfnamefont {W.}~\bibnamefont
  {Ketterle}}\ and\ \bibinfo {author} {\bibfnamefont {M.~W.}\ \bibnamefont
  {Zwierlein}},\ }\href {https://doi.org/10.1393/ncr/i2008-10033-1} {\bibfield
  {journal} {\bibinfo  {journal} {La Rivista del Nuovo Cimento}\ }\textbf
  {\bibinfo {volume} {31}},\ \bibinfo {pages} {247–422} (\bibinfo {year}
  {2008})}\BibitemShut {NoStop}%
\bibitem [{\citenamefont {Chen}\ \emph {et~al.}(2024)\citenamefont {Chen},
  \citenamefont {Wang}, \citenamefont {Boyack}, \citenamefont {Yang},\ and\
  \citenamefont {Levin}}]{chen_when_2024}%
  \BibitemOpen
  \bibfield  {author} {\bibinfo {author} {\bibfnamefont {Q.}~\bibnamefont
  {Chen}}, \bibinfo {author} {\bibfnamefont {Z.}~\bibnamefont {Wang}}, \bibinfo
  {author} {\bibfnamefont {R.}~\bibnamefont {Boyack}}, \bibinfo {author}
  {\bibfnamefont {S.}~\bibnamefont {Yang}},\ and\ \bibinfo {author}
  {\bibfnamefont {K.}~\bibnamefont {Levin}},\ }\bibfield  {title} {\bibinfo
  {title} {When superconductivity crosses over: {From} {BCS} to {BEC}},\ }\href
  {https://doi.org/10.1103/RevModPhys.96.025002} {\bibfield  {journal}
  {\bibinfo  {journal} {Reviews of Modern Physics}\ }\textbf {\bibinfo {volume}
  {96}},\ \bibinfo {pages} {025002} (\bibinfo {year} {2024})}\BibitemShut
  {NoStop}%
\bibitem [{\citenamefont {Yao}\ \emph {et~al.}(2025)\citenamefont {Yao},
  \citenamefont {Chi}, \citenamefont {Wang}, \citenamefont {Fletcher},\ and\
  \citenamefont {Zwierlein}}]{Yao2025}%
  \BibitemOpen
  \bibfield  {author} {\bibinfo {author} {\bibfnamefont {R.}~\bibnamefont
  {Yao}}, \bibinfo {author} {\bibfnamefont {S.}~\bibnamefont {Chi}}, \bibinfo
  {author} {\bibfnamefont {M.}~\bibnamefont {Wang}}, \bibinfo {author}
  {\bibfnamefont {R.~J.}\ \bibnamefont {Fletcher}},\ and\ \bibinfo {author}
  {\bibfnamefont {M.}~\bibnamefont {Zwierlein}},\ }\bibfield  {title} {\bibinfo
  {title} {{Measuring Pair Correlations in Bose and Fermi Gases via
  Atom-Resolved Microscopy}},\ }\href
  {https://doi.org/10.1103/PhysRevLett.134.183402} {\bibfield  {journal}
  {\bibinfo  {journal} {Phys. Rev. Lett.}\ }\textbf {\bibinfo {volume} {134}},\
  \bibinfo {pages} {183402} (\bibinfo {year} {2025})}\BibitemShut {NoStop}%
\bibitem [{\citenamefont {Bergschneider}\ \emph {et~al.}(2019)\citenamefont
  {Bergschneider}, \citenamefont {Klinkhamer}, \citenamefont {Becher},
  \citenamefont {Klemt}, \citenamefont {Palm}, \citenamefont {Zürn},
  \citenamefont {Jochim},\ and\ \citenamefont {Preiss}}]{Bergschneider2019}%
  \BibitemOpen
  \bibfield  {author} {\bibinfo {author} {\bibfnamefont {A.}~\bibnamefont
  {Bergschneider}}, \bibinfo {author} {\bibfnamefont {V.~M.}\ \bibnamefont
  {Klinkhamer}}, \bibinfo {author} {\bibfnamefont {J.~H.}\ \bibnamefont
  {Becher}}, \bibinfo {author} {\bibfnamefont {R.}~\bibnamefont {Klemt}},
  \bibinfo {author} {\bibfnamefont {L.}~\bibnamefont {Palm}}, \bibinfo {author}
  {\bibfnamefont {G.}~\bibnamefont {Zürn}}, \bibinfo {author} {\bibfnamefont
  {S.}~\bibnamefont {Jochim}},\ and\ \bibinfo {author} {\bibfnamefont {P.~M.}\
  \bibnamefont {Preiss}},\ }\bibfield  {title} {\bibinfo {title} {{Experimental
  characterization of two-particle entanglement through position and momentum
  correlations}},\ }\href {https://doi.org/10.1038/s41567-019-0508-6}
  {\bibfield  {journal} {\bibinfo  {journal} {Nature Physics}\ }\textbf
  {\bibinfo {volume} {15}},\ \bibinfo {pages} {640} (\bibinfo {year}
  {2019})}\BibitemShut {NoStop}%
\bibitem [{\citenamefont {Lunt}\ \emph {et~al.}(2024)\citenamefont {Lunt},
  \citenamefont {Hill}, \citenamefont {Reiter}, \citenamefont {Preiss},
  \citenamefont {Ga\l{}ka},\ and\ \citenamefont {Jochim}}]{Lunt2024}%
  \BibitemOpen
  \bibfield  {author} {\bibinfo {author} {\bibfnamefont {P.}~\bibnamefont
  {Lunt}}, \bibinfo {author} {\bibfnamefont {P.}~\bibnamefont {Hill}}, \bibinfo
  {author} {\bibfnamefont {J.}~\bibnamefont {Reiter}}, \bibinfo {author}
  {\bibfnamefont {P.~M.}\ \bibnamefont {Preiss}}, \bibinfo {author}
  {\bibfnamefont {M.}~\bibnamefont {Ga\l{}ka}},\ and\ \bibinfo {author}
  {\bibfnamefont {S.}~\bibnamefont {Jochim}},\ }\bibfield  {title} {\bibinfo
  {title} {Realization of a laughlin state of two rapidly rotating fermions},\
  }\href {https://doi.org/10.1103/PhysRevLett.133.253401} {\bibfield  {journal}
  {\bibinfo  {journal} {Phys. Rev. Lett.}\ }\textbf {\bibinfo {volume} {133}},\
  \bibinfo {pages} {253401} (\bibinfo {year} {2024})}\BibitemShut {NoStop}%
\bibitem [{\citenamefont {Becher}\ \emph {et~al.}(2020)\citenamefont {Becher},
  \citenamefont {Sindici}, \citenamefont {Klemt}, \citenamefont {Jochim},
  \citenamefont {Daley},\ and\ \citenamefont {Preiss}}]{Becher2020}%
  \BibitemOpen
  \bibfield  {author} {\bibinfo {author} {\bibfnamefont {J.~H.}\ \bibnamefont
  {Becher}}, \bibinfo {author} {\bibfnamefont {E.}~\bibnamefont {Sindici}},
  \bibinfo {author} {\bibfnamefont {R.}~\bibnamefont {Klemt}}, \bibinfo
  {author} {\bibfnamefont {S.}~\bibnamefont {Jochim}}, \bibinfo {author}
  {\bibfnamefont {A.~J.}\ \bibnamefont {Daley}},\ and\ \bibinfo {author}
  {\bibfnamefont {P.~M.}\ \bibnamefont {Preiss}},\ }\bibfield  {title}
  {\bibinfo {title} {Measurement of identical particle entanglement and the
  influence of antisymmetrization},\ }\href
  {https://doi.org/10.1103/PhysRevLett.125.180402} {\bibfield  {journal}
  {\bibinfo  {journal} {Phys. Rev. Lett.}\ }\textbf {\bibinfo {volume} {125}},\
  \bibinfo {pages} {180402} (\bibinfo {year} {2020})}\BibitemShut {NoStop}%
\bibitem [{\citenamefont {Holten}\ \emph {et~al.}(2022)\citenamefont {Holten},
  \citenamefont {Bayha}, \citenamefont {Subramanian}, \citenamefont
  {Brandstetter}, \citenamefont {Heintze}, \citenamefont {Lunt}, \citenamefont
  {Preiss},\ and\ \citenamefont {Jochim}}]{Holten2022}%
  \BibitemOpen
  \bibfield  {author} {\bibinfo {author} {\bibfnamefont {M.}~\bibnamefont
  {Holten}}, \bibinfo {author} {\bibfnamefont {L.}~\bibnamefont {Bayha}},
  \bibinfo {author} {\bibfnamefont {K.}~\bibnamefont {Subramanian}}, \bibinfo
  {author} {\bibfnamefont {S.}~\bibnamefont {Brandstetter}}, \bibinfo {author}
  {\bibfnamefont {C.}~\bibnamefont {Heintze}}, \bibinfo {author} {\bibfnamefont
  {P.}~\bibnamefont {Lunt}}, \bibinfo {author} {\bibfnamefont {P.~M.}\
  \bibnamefont {Preiss}},\ and\ \bibinfo {author} {\bibfnamefont
  {S.}~\bibnamefont {Jochim}},\ }\bibfield  {title} {\bibinfo {title}
  {{Observation of Cooper pairs in a mesoscopic two-dimensional Fermi gas}},\
  }\href {https://doi.org/10.1038/s41586-022-04678-1} {\bibfield  {journal}
  {\bibinfo  {journal} {Nature}\ }\textbf {\bibinfo {volume} {606}},\ \bibinfo
  {pages} {287} (\bibinfo {year} {2022})}\BibitemShut {NoStop}%
\bibitem [{\citenamefont {Obeso-Jureidini}\ and\ \citenamefont
  {Romero-Rochín}(2022)}]{obeso-jureidini_density_2022}%
  \BibitemOpen
  \bibfield  {author} {\bibinfo {author} {\bibfnamefont {J.~C.}\ \bibnamefont
  {Obeso-Jureidini}}\ and\ \bibinfo {author} {\bibfnamefont {V.}~\bibnamefont
  {Romero-Rochín}},\ }\bibfield  {title} {\bibinfo {title} {Density
  correlation functions and the spatial structure of the two-dimensional
  {BEC}-{BCS} crossover},\ }\href {https://doi.org/10.1103/PhysRevA.105.043307}
  {\bibfield  {journal} {\bibinfo  {journal} {Physical Review A}\ }\textbf
  {\bibinfo {volume} {105}},\ \bibinfo {pages} {043307} (\bibinfo {year}
  {2022})}\BibitemShut {NoStop}%
\bibitem [{Note2()}]{Note2}%
  \BibitemOpen
  \bibinfo {note} {This can be achieved by applying a de-focused Gaussian beam
  in the Fourier plane, where a potential of the form $Az + Br^2 + Czr^2$ is
  created (where $A,B,C$ are constants). This contains the required term
  $V(x,y,z) =C (x^2+y^2)z$, while the gradient $Az$ simply leads to a global
  shift in $z$ direction and the quadratic term $Br^2$ can be compensated by
  the length of the matter-wave lens.}\BibitemShut {Stop}%
\bibitem [{\citenamefont {Fuchs}\ \emph {et~al.}(2004)\citenamefont {Fuchs},
  \citenamefont {Recati},\ and\ \citenamefont {Zwerger}}]{Fuchs2004}%
  \BibitemOpen
  \bibfield  {author} {\bibinfo {author} {\bibfnamefont {J.~N.}\ \bibnamefont
  {Fuchs}}, \bibinfo {author} {\bibfnamefont {A.}~\bibnamefont {Recati}},\ and\
  \bibinfo {author} {\bibfnamefont {W.}~\bibnamefont {Zwerger}},\ }\bibfield
  {title} {\bibinfo {title} {Exactly solvable model of the {BCS-BEC}
  crossover},\ }\href {https://doi.org/10.1103/PhysRevLett.93.090408}
  {\bibfield  {journal} {\bibinfo  {journal} {Phys. Rev. Lett.}\ }\textbf
  {\bibinfo {volume} {93}},\ \bibinfo {pages} {090408} (\bibinfo {year}
  {2004})}\BibitemShut {NoStop}%
\bibitem [{\citenamefont {Struck}\ \emph {et~al.}(2011)\citenamefont {Struck},
  \citenamefont {{\"{O}}lschl{\"{a}}ger}, \citenamefont {{Le Targat}},
  \citenamefont {Soltan-Panahi}, \citenamefont {Eckardt}, \citenamefont
  {Lewenstein}, \citenamefont {Windpassinger},\ and\ \citenamefont
  {Sengstock}}]{Struck2011}%
  \BibitemOpen
  \bibfield  {author} {\bibinfo {author} {\bibfnamefont {J.}~\bibnamefont
  {Struck}}, \bibinfo {author} {\bibfnamefont {C.}~\bibnamefont
  {{\"{O}}lschl{\"{a}}ger}}, \bibinfo {author} {\bibfnamefont {R.}~\bibnamefont
  {{Le Targat}}}, \bibinfo {author} {\bibfnamefont {P.}~\bibnamefont
  {Soltan-Panahi}}, \bibinfo {author} {\bibfnamefont {A.}~\bibnamefont
  {Eckardt}}, \bibinfo {author} {\bibfnamefont {M.}~\bibnamefont {Lewenstein}},
  \bibinfo {author} {\bibfnamefont {P.}~\bibnamefont {Windpassinger}},\ and\
  \bibinfo {author} {\bibfnamefont {K.}~\bibnamefont {Sengstock}},\ }\bibfield
  {title} {\bibinfo {title} {{Quantum simulation of frustrated classical
  magnetism in triangular optical lattices}},\ }\href
  {https://doi.org/10.1126/science.1207239} {\bibfield  {journal} {\bibinfo
  {journal} {Science}\ }\textbf {\bibinfo {volume} {333}},\ \bibinfo {pages}
  {996} (\bibinfo {year} {2011})}\BibitemShut {NoStop}%
\bibitem [{\citenamefont {Hasan}\ \emph {et~al.}(2025)\citenamefont {Hasan},
  \citenamefont {Donini}, \citenamefont {Shanokprasith}, \citenamefont
  {Braund}, \citenamefont {Marozsak}, \citenamefont {Epping}, \citenamefont
  {Reed}, \citenamefont {Melchner}, \citenamefont {Harte},\ and\ \citenamefont
  {Schneider}}]{hasan2025}%
  \BibitemOpen
  \bibfield  {author} {\bibinfo {author} {\bibfnamefont {M.}~\bibnamefont
  {Hasan}}, \bibinfo {author} {\bibfnamefont {L.}~\bibnamefont {Donini}},
  \bibinfo {author} {\bibfnamefont {S.}~\bibnamefont {Shanokprasith}}, \bibinfo
  {author} {\bibfnamefont {D.}~\bibnamefont {Braund}}, \bibinfo {author}
  {\bibfnamefont {T.}~\bibnamefont {Marozsak}}, \bibinfo {author}
  {\bibfnamefont {M.}~\bibnamefont {Epping}}, \bibinfo {author} {\bibfnamefont
  {D.}~\bibnamefont {Reed}}, \bibinfo {author} {\bibfnamefont {M.}~\bibnamefont
  {Melchner}}, \bibinfo {author} {\bibfnamefont {T.}~\bibnamefont {Harte}},\
  and\ \bibinfo {author} {\bibfnamefont {U.}~\bibnamefont {Schneider}},\ }\href
  {https://arxiv.org/abs/2509.20352} {\bibinfo {title} {{Superfluid-Mott}
  transition in a frustrated triangular optical lattice}} (\bibinfo {year}
  {2025}),\ \Eprint {https://arxiv.org/abs/2509.20352} {arXiv:2509.20352
  [cond-mat.quant-gas]} \BibitemShut {NoStop}%
\bibitem [{\citenamefont {Ozawa}\ \emph {et~al.}(2023)\citenamefont {Ozawa},
  \citenamefont {Yamamoto},\ and\ \citenamefont {Fukuhara}}]{Ozawa2023}%
  \BibitemOpen
  \bibfield  {author} {\bibinfo {author} {\bibfnamefont {H.}~\bibnamefont
  {Ozawa}}, \bibinfo {author} {\bibfnamefont {R.}~\bibnamefont {Yamamoto}},\
  and\ \bibinfo {author} {\bibfnamefont {T.}~\bibnamefont {Fukuhara}},\
  }\bibfield  {title} {\bibinfo {title} {Observation of chiral-mode domains in
  a frustrated xy model on optical triangular lattices},\ }\href
  {https://doi.org/10.1103/PhysRevResearch.5.L042026} {\bibfield  {journal}
  {\bibinfo  {journal} {Phys. Rev. Res.}\ }\textbf {\bibinfo {volume} {5}},\
  \bibinfo {pages} {L042026} (\bibinfo {year} {2023})}\BibitemShut {NoStop}%
\bibitem [{\citenamefont {Yu}\ \emph {et~al.}(2024)\citenamefont {Yu},
  \citenamefont {Bhave}, \citenamefont {Reeve}, \citenamefont {Song},\ and\
  \citenamefont {Schneider}}]{Yu2024}%
  \BibitemOpen
  \bibfield  {author} {\bibinfo {author} {\bibfnamefont {J.-C.}\ \bibnamefont
  {Yu}}, \bibinfo {author} {\bibfnamefont {S.}~\bibnamefont {Bhave}}, \bibinfo
  {author} {\bibfnamefont {L.}~\bibnamefont {Reeve}}, \bibinfo {author}
  {\bibfnamefont {B.}~\bibnamefont {Song}},\ and\ \bibinfo {author}
  {\bibfnamefont {U.}~\bibnamefont {Schneider}},\ }\bibfield  {title} {\bibinfo
  {title} {{Observing the two-dimensional Bose glass in an optical
  quasicrystal}},\ }\href {https://doi.org/10.1038/s41586-024-07875-2}
  {\bibfield  {journal} {\bibinfo  {journal} {Nature}\ }\textbf {\bibinfo
  {volume} {633}},\ \bibinfo {pages} {338} (\bibinfo {year}
  {2024})}\BibitemShut {NoStop}%
\bibitem [{\citenamefont {Koehn}\ \emph {et~al.}(2026)\citenamefont {Koehn},
  \citenamefont {Parsonage}, \citenamefont {Duncan}, \citenamefont {Kirton},
  \citenamefont {Daley}, \citenamefont {Hilker}, \citenamefont {Haller},
  \citenamefont {La~Rooij},\ and\ \citenamefont {Kuhr}}]{Koehn2026}%
  \BibitemOpen
  \bibfield  {author} {\bibinfo {author} {\bibfnamefont {L.}~\bibnamefont
  {Koehn}}, \bibinfo {author} {\bibfnamefont {C.}~\bibnamefont {Parsonage}},
  \bibinfo {author} {\bibfnamefont {C.~W.}\ \bibnamefont {Duncan}}, \bibinfo
  {author} {\bibfnamefont {P.}~\bibnamefont {Kirton}}, \bibinfo {author}
  {\bibfnamefont {A.~J.}\ \bibnamefont {Daley}}, \bibinfo {author}
  {\bibfnamefont {T.}~\bibnamefont {Hilker}}, \bibinfo {author} {\bibfnamefont
  {E.}~\bibnamefont {Haller}}, \bibinfo {author} {\bibfnamefont
  {A.}~\bibnamefont {La~Rooij}},\ and\ \bibinfo {author} {\bibfnamefont
  {S.}~\bibnamefont {Kuhr}},\ }\bibfield  {title} {\bibinfo {title}
  {{Quantum-gas microscopy and Talbot interferometry of the Bose-glass
  phase}},\ }\href {https://doi.org/10.1103/xb42-j6px} {\bibfield  {journal}
  {\bibinfo  {journal} {Phys. Rev. A}\ }\textbf {\bibinfo {volume} {113}},\
  \bibinfo {pages} {043303} (\bibinfo {year} {2026})}\BibitemShut {NoStop}%
\bibitem [{\citenamefont {Fulde}\ and\ \citenamefont
  {Ferrell}(1964)}]{fulde_superconductivity_1964}%
  \BibitemOpen
  \bibfield  {author} {\bibinfo {author} {\bibfnamefont {P.}~\bibnamefont
  {Fulde}}\ and\ \bibinfo {author} {\bibfnamefont {R.~A.}\ \bibnamefont
  {Ferrell}},\ }\bibfield  {title} {\bibinfo {title} {Superconductivity in a
  strong spin-exchange field},\ }\href
  {https://doi.org/10.1103/PhysRev.135.A550} {\bibfield  {journal} {\bibinfo
  {journal} {Phys. Rev.}\ }\textbf {\bibinfo {volume} {135}},\ \bibinfo {pages}
  {A550} (\bibinfo {year} {1964})}\BibitemShut {NoStop}%
\bibitem [{\citenamefont {Larkin}\ and\ \citenamefont
  {Ovchinnikov}(1964)}]{larkin_nonuniform_1964}%
  \BibitemOpen
  \bibfield  {author} {\bibinfo {author} {\bibfnamefont {A.~I.}\ \bibnamefont
  {Larkin}}\ and\ \bibinfo {author} {\bibfnamefont {Y.~N.}\ \bibnamefont
  {Ovchinnikov}},\ }\bibfield  {title} {\bibinfo {title} {Nonuniform state of
  superconductors},\ }\href@noop {} {\bibfield  {journal} {\bibinfo  {journal}
  {Zh. Eksp. Teor. Fiz.}\ }\textbf {\bibinfo {volume} {47}},\ \bibinfo {pages}
  {1136} (\bibinfo {year} {1964})}\BibitemShut {NoStop}%
\bibitem [{\citenamefont {Agterberg}\ \emph {et~al.}(2020)\citenamefont
  {Agterberg}, \citenamefont {Davis}, \citenamefont {Edkins}, \citenamefont
  {Fradkin}, \citenamefont {Van~Harlingen}, \citenamefont {Kivelson},
  \citenamefont {Lee}, \citenamefont {Radzihovsky}, \citenamefont {Tranquada},\
  and\ \citenamefont {Wang}}]{agterberg_pdw_2020}%
  \BibitemOpen
  \bibfield  {author} {\bibinfo {author} {\bibfnamefont {D.~F.}\ \bibnamefont
  {Agterberg}}, \bibinfo {author} {\bibfnamefont {J.~S.}\ \bibnamefont
  {Davis}}, \bibinfo {author} {\bibfnamefont {S.~D.}\ \bibnamefont {Edkins}},
  \bibinfo {author} {\bibfnamefont {E.}~\bibnamefont {Fradkin}}, \bibinfo
  {author} {\bibfnamefont {D.~J.}\ \bibnamefont {Van~Harlingen}}, \bibinfo
  {author} {\bibfnamefont {S.~A.}\ \bibnamefont {Kivelson}}, \bibinfo {author}
  {\bibfnamefont {P.~A.}\ \bibnamefont {Lee}}, \bibinfo {author} {\bibfnamefont
  {L.}~\bibnamefont {Radzihovsky}}, \bibinfo {author} {\bibfnamefont {J.~M.}\
  \bibnamefont {Tranquada}},\ and\ \bibinfo {author} {\bibfnamefont
  {Y.}~\bibnamefont {Wang}},\ }\bibfield  {title} {\bibinfo {title} {The
  physics of pair-density waves: Cuprate superconductors and beyond},\ }\href
  {https://doi.org/https://doi.org/10.1146/annurev-conmatphys-031119-050711}
  {\bibfield  {journal} {\bibinfo  {journal} {Annual Review of Condensed Matter
  Physics}\ }\textbf {\bibinfo {volume} {11}},\ \bibinfo {pages} {231}
  (\bibinfo {year} {2020})}\BibitemShut {NoStop}%
\bibitem [{\citenamefont {Scalettar}\ \emph {et~al.}(1989)\citenamefont
  {Scalettar}, \citenamefont {Loh}, \citenamefont {Gubernatis}, \citenamefont
  {Moreo}, \citenamefont {White}, \citenamefont {Scalapino}, \citenamefont
  {Sugar},\ and\ \citenamefont {Dagotto}}]{scalettar_phase_1989}%
  \BibitemOpen
  \bibfield  {author} {\bibinfo {author} {\bibfnamefont {R.~T.}\ \bibnamefont
  {Scalettar}}, \bibinfo {author} {\bibfnamefont {E.~Y.}\ \bibnamefont {Loh}},
  \bibinfo {author} {\bibfnamefont {J.~E.}\ \bibnamefont {Gubernatis}},
  \bibinfo {author} {\bibfnamefont {A.}~\bibnamefont {Moreo}}, \bibinfo
  {author} {\bibfnamefont {S.~R.}\ \bibnamefont {White}}, \bibinfo {author}
  {\bibfnamefont {D.~J.}\ \bibnamefont {Scalapino}}, \bibinfo {author}
  {\bibfnamefont {R.~L.}\ \bibnamefont {Sugar}},\ and\ \bibinfo {author}
  {\bibfnamefont {E.}~\bibnamefont {Dagotto}},\ }\bibfield  {title} {\bibinfo
  {title} {{Phase diagram of the two-dimensional negative-U Hubbard model}},\
  }\href {https://doi.org/10.1103/PhysRevLett.62.1407} {\bibfield  {journal}
  {\bibinfo  {journal} {Phys. Rev. Lett.}\ }\textbf {\bibinfo {volume} {62}},\
  \bibinfo {pages} {1407} (\bibinfo {year} {1989})}\BibitemShut {NoStop}%
\bibitem [{\citenamefont {Rosenberg}\ \emph {et~al.}(2017)\citenamefont
  {Rosenberg}, \citenamefont {Shi},\ and\ \citenamefont
  {Zhang}}]{rosenbergUltracoldAtomsSquare2017}%
  \BibitemOpen
  \bibfield  {author} {\bibinfo {author} {\bibfnamefont {P.}~\bibnamefont
  {Rosenberg}}, \bibinfo {author} {\bibfnamefont {H.}~\bibnamefont {Shi}},\
  and\ \bibinfo {author} {\bibfnamefont {S.}~\bibnamefont {Zhang}},\ }\bibfield
   {title} {\bibinfo {title} {Ultracold {Atoms} in a {Square} {Lattice} with
  {Spin}-{Orbit} {Coupling}: {Charge} {Order}, {Superfluidity}, and
  {Topological} {Signatures}},\ }\href
  {https://doi.org/10.1103/PhysRevLett.119.265301} {\bibfield  {journal}
  {\bibinfo  {journal} {Physical Review Letters}\ }\textbf {\bibinfo {volume}
  {119}},\ \bibinfo {pages} {265301} (\bibinfo {year} {2017})}\BibitemShut
  {NoStop}%
\bibitem [{\citenamefont {Bourgund}\ \emph {et~al.}(2025)\citenamefont
  {Bourgund}, \citenamefont {Chalopin}, \citenamefont {Bojović}, \citenamefont
  {Schlömer}, \citenamefont {Wang}, \citenamefont {Franz}, \citenamefont
  {Hirthe}, \citenamefont {Bohrdt}, \citenamefont {Grusdt}, \citenamefont
  {Bloch},\ and\ \citenamefont {Hilker}}]{bourgund_formation_2025}%
  \BibitemOpen
  \bibfield  {author} {\bibinfo {author} {\bibfnamefont {D.}~\bibnamefont
  {Bourgund}}, \bibinfo {author} {\bibfnamefont {T.}~\bibnamefont {Chalopin}},
  \bibinfo {author} {\bibfnamefont {P.}~\bibnamefont {Bojović}}, \bibinfo
  {author} {\bibfnamefont {H.}~\bibnamefont {Schlömer}}, \bibinfo {author}
  {\bibfnamefont {S.}~\bibnamefont {Wang}}, \bibinfo {author} {\bibfnamefont
  {T.}~\bibnamefont {Franz}}, \bibinfo {author} {\bibfnamefont
  {S.}~\bibnamefont {Hirthe}}, \bibinfo {author} {\bibfnamefont
  {A.}~\bibnamefont {Bohrdt}}, \bibinfo {author} {\bibfnamefont
  {F.}~\bibnamefont {Grusdt}}, \bibinfo {author} {\bibfnamefont
  {I.}~\bibnamefont {Bloch}},\ and\ \bibinfo {author} {\bibfnamefont {T.~A.}\
  \bibnamefont {Hilker}},\ }\bibfield  {title} {\bibinfo {title} {Formation of
  individual stripes in a mixed-dimensional cold-atom {Fermi}–{Hubbard}
  system},\ }\href {https://doi.org/10.1038/s41586-024-08270-7} {\bibfield
  {journal} {\bibinfo  {journal} {Nature}\ }\textbf {\bibinfo {volume} {637}},\
  \bibinfo {pages} {57} (\bibinfo {year} {2025})}\BibitemShut {NoStop}%
\bibitem [{\citenamefont {Kendrick}\ \emph {et~al.}(2025)\citenamefont
  {Kendrick}, \citenamefont {Kale}, \citenamefont {Gang}, \citenamefont
  {Deters}, \citenamefont {Lebrat}, \citenamefont {Young},\ and\ \citenamefont
  {Greiner}}]{Kendrick2025}%
  \BibitemOpen
  \bibfield  {author} {\bibinfo {author} {\bibfnamefont {L.~H.}\ \bibnamefont
  {Kendrick}}, \bibinfo {author} {\bibfnamefont {A.}~\bibnamefont {Kale}},
  \bibinfo {author} {\bibfnamefont {Y.}~\bibnamefont {Gang}}, \bibinfo {author}
  {\bibfnamefont {A.~D.}\ \bibnamefont {Deters}}, \bibinfo {author}
  {\bibfnamefont {M.}~\bibnamefont {Lebrat}}, \bibinfo {author} {\bibfnamefont
  {A.~W.}\ \bibnamefont {Young}},\ and\ \bibinfo {author} {\bibfnamefont
  {M.}~\bibnamefont {Greiner}},\ }\bibfield  {title} {\bibinfo {title}
  {Pseudogap in a {Fermi-Hubbard} quantum simulator},\ }\href
  {https://arxiv.org/abs/2509.18075} {\  (\bibinfo {year} {2025})},\ \Eprint
  {https://arxiv.org/abs/2509.18075} {arXiv:2509.18075 [cond-mat.quant-gas]}
  \BibitemShut {NoStop}%
\bibitem [{\citenamefont {Chen}\ \emph {et~al.}(2005)\citenamefont {Chen},
  \citenamefont {Stajic}, \citenamefont {Tan},\ and\ \citenamefont
  {Levin}}]{chen_bcsbec_2005}%
  \BibitemOpen
  \bibfield  {author} {\bibinfo {author} {\bibfnamefont {Q.}~\bibnamefont
  {Chen}}, \bibinfo {author} {\bibfnamefont {J.}~\bibnamefont {Stajic}},
  \bibinfo {author} {\bibfnamefont {S.}~\bibnamefont {Tan}},\ and\ \bibinfo
  {author} {\bibfnamefont {K.}~\bibnamefont {Levin}},\ }\bibfield  {title}
  {\bibinfo {title} {{BCS}–{BEC} crossover: {From} high temperature
  superconductors to ultracold superfluids},\ }\href
  {https://doi.org/10.1016/j.physrep.2005.02.005} {\bibfield  {journal}
  {\bibinfo  {journal} {Physics Reports}\ }\textbf {\bibinfo {volume} {412}},\
  \bibinfo {pages} {1} (\bibinfo {year} {2005})}\BibitemShut {NoStop}%
\bibitem [{\citenamefont {Ohashi}\ \emph {et~al.}(2020)\citenamefont {Ohashi},
  \citenamefont {Tajima},\ and\ \citenamefont {Van~Wyk}}]{ohashi_bcsbec_2020}%
  \BibitemOpen
  \bibfield  {author} {\bibinfo {author} {\bibfnamefont {Y.}~\bibnamefont
  {Ohashi}}, \bibinfo {author} {\bibfnamefont {H.}~\bibnamefont {Tajima}},\
  and\ \bibinfo {author} {\bibfnamefont {P.}~\bibnamefont {Van~Wyk}},\
  }\bibfield  {title} {\bibinfo {title} {{BCS}–{BEC} crossover in cold atomic
  and in nuclear systems},\ }\href {https://doi.org/10.1016/j.ppnp.2019.103739}
  {\bibfield  {journal} {\bibinfo  {journal} {Progress in Particle and Nuclear
  Physics}\ }\textbf {\bibinfo {volume} {111}},\ \bibinfo {pages} {103739}
  (\bibinfo {year} {2020})}\BibitemShut {NoStop}%
\bibitem [{\citenamefont {Cooper}(2008)}]{Cooper2008}%
  \BibitemOpen
  \bibfield  {author} {\bibinfo {author} {\bibfnamefont {N.}~\bibnamefont
  {Cooper}},\ }\bibfield  {title} {\bibinfo {title} {Rapidly rotating atomic
  gases},\ }\href {https://doi.org/10.1080/00018730802564122} {\bibfield
  {journal} {\bibinfo  {journal} {Advances in Physics}\ }\textbf {\bibinfo
  {volume} {57}},\ \bibinfo {pages} {539} (\bibinfo {year} {2008})},\ \Eprint
  {https://arxiv.org/abs/https://doi.org/10.1080/00018730802564122}
  {https://doi.org/10.1080/00018730802564122} \BibitemShut {NoStop}%
\bibitem [{\citenamefont {Tan}(2008)}]{tan_large_2008}%
  \BibitemOpen
  \bibfield  {author} {\bibinfo {author} {\bibfnamefont {S.}~\bibnamefont
  {Tan}},\ }\bibfield  {title} {\bibinfo {title} {Large momentum part of a
  strongly correlated {Fermi} gas},\ }\href
  {https://doi.org/10.1016/j.aop.2008.03.005} {\bibfield  {journal} {\bibinfo
  {journal} {Annals of Physics}\ }\textbf {\bibinfo {volume} {323}},\ \bibinfo
  {pages} {2971} (\bibinfo {year} {2008})}\BibitemShut {NoStop}%
\bibitem [{\citenamefont {Hofmann}\ and\ \citenamefont
  {Zwerger}(2024)}]{hofmann_scale_2024}%
  \BibitemOpen
  \bibfield  {author} {\bibinfo {author} {\bibfnamefont {J.}~\bibnamefont
  {Hofmann}}\ and\ \bibinfo {author} {\bibfnamefont {W.}~\bibnamefont
  {Zwerger}},\ }\bibfield  {title} {\bibinfo {title} {Scale {Invariance} in the
  {Lowest} {Landau} {Level}},\ }\href {https://doi.org/10.5802/crphys.137}
  {\bibfield  {journal} {\bibinfo  {journal} {Comptes Rendus. Physique}\
  }\textbf {\bibinfo {volume} {24}},\ \bibinfo {pages} {69} (\bibinfo {year}
  {2024})}\BibitemShut {NoStop}%
\bibitem [{\citenamefont {Whitehead}\ \emph {et~al.}(2016)\citenamefont
  {Whitehead}, \citenamefont {Schonenberg}, \citenamefont {Kongsuwan},
  \citenamefont {Needs},\ and\ \citenamefont
  {Conduit}}]{whitehead_pseudopotential_2016}%
  \BibitemOpen
  \bibfield  {author} {\bibinfo {author} {\bibfnamefont {T.~M.}\ \bibnamefont
  {Whitehead}}, \bibinfo {author} {\bibfnamefont {L.~M.}\ \bibnamefont
  {Schonenberg}}, \bibinfo {author} {\bibfnamefont {N.}~\bibnamefont
  {Kongsuwan}}, \bibinfo {author} {\bibfnamefont {R.~J.}\ \bibnamefont
  {Needs}},\ and\ \bibinfo {author} {\bibfnamefont {G.~J.}\ \bibnamefont
  {Conduit}},\ }\bibfield  {title} {\bibinfo {title} {Pseudopotential for the
  two-dimensional contact interaction},\ }\href
  {https://doi.org/10.1103/PhysRevA.93.042702} {\bibfield  {journal} {\bibinfo
  {journal} {Physical Review A}\ }\textbf {\bibinfo {volume} {93}},\ \bibinfo
  {pages} {042702} (\bibinfo {year} {2016})}\BibitemShut {NoStop}%
\bibitem [{\citenamefont {He}\ and\ \citenamefont
  {Zhou}(2019)}]{he_contact_2019}%
  \BibitemOpen
  \bibfield  {author} {\bibinfo {author} {\bibfnamefont {M.}~\bibnamefont
  {He}}\ and\ \bibinfo {author} {\bibfnamefont {Q.}~\bibnamefont {Zhou}},\
  }\bibfield  {title} {\bibinfo {title} {$s$-wave contacts of quantum gases in
  quasi-one-dimensional and quasi-two-dimensional traps},\ }\href
  {https://doi.org/10.1103/PhysRevA.100.012701} {\bibfield  {journal} {\bibinfo
   {journal} {Phys. Rev. A}\ }\textbf {\bibinfo {volume} {100}},\ \bibinfo
  {pages} {012701} (\bibinfo {year} {2019})}\BibitemShut {NoStop}%
\bibitem [{\citenamefont {Spasic-Mlacak}\ and\ \citenamefont
  {Cooper}(2025)}]{Spasic-Mlacak2025}%
  \BibitemOpen
  \bibfield  {author} {\bibinfo {author} {\bibfnamefont {D.}~\bibnamefont
  {Spasic-Mlacak}}\ and\ \bibinfo {author} {\bibfnamefont {N.~R.}\ \bibnamefont
  {Cooper}},\ }\bibfield  {title} {\bibinfo {title} {Correlations of the
  current density in many-body landau level states},\ }\href
  {https://doi.org/10.1103/cn8q-zwdk} {\bibfield  {journal} {\bibinfo
  {journal} {Phys. Rev. Res.}\ }\textbf {\bibinfo {volume} {7}},\ \bibinfo
  {pages} {L032032} (\bibinfo {year} {2025})}\BibitemShut {NoStop}%
\bibitem [{\citenamefont {MacDonald}\ and\ \citenamefont
  {Girvin}(1988)}]{macdonald_density_1988}%
  \BibitemOpen
  \bibfield  {author} {\bibinfo {author} {\bibfnamefont {A.~H.}\ \bibnamefont
  {MacDonald}}\ and\ \bibinfo {author} {\bibfnamefont {S.~M.}\ \bibnamefont
  {Girvin}},\ }\bibfield  {title} {\bibinfo {title} {Density matrices for
  states in the lowest {Landau} level of a two-dimensional electron gas},\
  }\href {https://doi.org/10.1103/PhysRevB.38.6295} {\bibfield  {journal}
  {\bibinfo  {journal} {Physical Review B}\ }\textbf {\bibinfo {volume} {38}},\
  \bibinfo {pages} {6295} (\bibinfo {year} {1988})}\BibitemShut {NoStop}%
\bibitem [{\citenamefont {Read}\ and\ \citenamefont {Cooper}(2003)}]{Read2003}%
  \BibitemOpen
  \bibfield  {author} {\bibinfo {author} {\bibfnamefont {N.}~\bibnamefont
  {Read}}\ and\ \bibinfo {author} {\bibfnamefont {N.~R.}\ \bibnamefont
  {Cooper}},\ }\bibfield  {title} {\bibinfo {title} {Free expansion of
  lowest-landau-level states of trapped atoms: A wave-function microscope},\
  }\href {https://doi.org/10.1103/PhysRevA.68.035601} {\bibfield  {journal}
  {\bibinfo  {journal} {Phys. Rev. A}\ }\textbf {\bibinfo {volume} {68}},\
  \bibinfo {pages} {035601} (\bibinfo {year} {2003})}\BibitemShut {NoStop}%
\bibitem [{\citenamefont {Aidelsburger}\ \emph {et~al.}(2015)\citenamefont
  {Aidelsburger}, \citenamefont {Lohse}, \citenamefont {Schweizer},
  \citenamefont {Atala}, \citenamefont {Barreiro}, \citenamefont {Nascimbène},
  \citenamefont {Cooper}, \citenamefont {Bloch},\ and\ \citenamefont
  {Goldman}}]{aidelsburger_measuring_2015}%
  \BibitemOpen
  \bibfield  {author} {\bibinfo {author} {\bibfnamefont {M.}~\bibnamefont
  {Aidelsburger}}, \bibinfo {author} {\bibfnamefont {M.}~\bibnamefont {Lohse}},
  \bibinfo {author} {\bibfnamefont {C.}~\bibnamefont {Schweizer}}, \bibinfo
  {author} {\bibfnamefont {M.}~\bibnamefont {Atala}}, \bibinfo {author}
  {\bibfnamefont {J.~T.}\ \bibnamefont {Barreiro}}, \bibinfo {author}
  {\bibfnamefont {S.}~\bibnamefont {Nascimbène}}, \bibinfo {author}
  {\bibfnamefont {N.~R.}\ \bibnamefont {Cooper}}, \bibinfo {author}
  {\bibfnamefont {I.}~\bibnamefont {Bloch}},\ and\ \bibinfo {author}
  {\bibfnamefont {N.}~\bibnamefont {Goldman}},\ }\bibfield  {title} {\bibinfo
  {title} {Measuring the {Chern} number of {Hofstadter} bands with ultracold
  bosonic atoms},\ }\href {https://doi.org/10.1038/nphys3171} {\bibfield
  {journal} {\bibinfo  {journal} {Nature Physics}\ }\textbf {\bibinfo {volume}
  {11}},\ \bibinfo {pages} {162} (\bibinfo {year} {2015})}\BibitemShut
  {NoStop}%
\bibitem [{\citenamefont {Asteria}\ \emph {et~al.}(2019)\citenamefont
  {Asteria}, \citenamefont {Tran}, \citenamefont {Ozawa}, \citenamefont
  {Tarnowski}, \citenamefont {Rem}, \citenamefont {Fläschner}, \citenamefont
  {Sengstock}, \citenamefont {Goldman},\ and\ \citenamefont
  {Weitenberg}}]{asteria_measuring_2019}%
  \BibitemOpen
  \bibfield  {author} {\bibinfo {author} {\bibfnamefont {L.}~\bibnamefont
  {Asteria}}, \bibinfo {author} {\bibfnamefont {D.~T.}\ \bibnamefont {Tran}},
  \bibinfo {author} {\bibfnamefont {T.}~\bibnamefont {Ozawa}}, \bibinfo
  {author} {\bibfnamefont {M.}~\bibnamefont {Tarnowski}}, \bibinfo {author}
  {\bibfnamefont {B.~S.}\ \bibnamefont {Rem}}, \bibinfo {author} {\bibfnamefont
  {N.}~\bibnamefont {Fläschner}}, \bibinfo {author} {\bibfnamefont
  {K.}~\bibnamefont {Sengstock}}, \bibinfo {author} {\bibfnamefont
  {N.}~\bibnamefont {Goldman}},\ and\ \bibinfo {author} {\bibfnamefont
  {C.}~\bibnamefont {Weitenberg}},\ }\bibfield  {title} {\bibinfo {title}
  {Measuring quantized circular dichroism in ultracold topological matter},\
  }\href {https://doi.org/10.1038/s41567-019-0417-8} {\bibfield  {journal}
  {\bibinfo  {journal} {Nature Physics}\ }\textbf {\bibinfo {volume} {15}},\
  \bibinfo {pages} {449} (\bibinfo {year} {2019})}\BibitemShut {NoStop}%
\bibitem [{\citenamefont {Price}\ and\ \citenamefont
  {Cooper}(2012)}]{Price_2012}%
  \BibitemOpen
  \bibfield  {author} {\bibinfo {author} {\bibfnamefont {H.~M.}\ \bibnamefont
  {Price}}\ and\ \bibinfo {author} {\bibfnamefont {N.~R.}\ \bibnamefont
  {Cooper}},\ }\bibfield  {title} {\bibinfo {title} {Mapping the berry
  curvature from semiclassical dynamics in optical lattices},\ }\href
  {https://doi.org/10.1103/PhysRevA.85.033620} {\bibfield  {journal} {\bibinfo
  {journal} {Phys. Rev. A}\ }\textbf {\bibinfo {volume} {85}},\ \bibinfo
  {pages} {033620} (\bibinfo {year} {2012})}\BibitemShut {NoStop}%
\bibitem [{\citenamefont {Fläschner}\ \emph {et~al.}(2016)\citenamefont
  {Fläschner}, \citenamefont {Rem}, \citenamefont {Tarnowski}, \citenamefont
  {Vogel}, \citenamefont {Lühmann}, \citenamefont {Sengstock},\ and\
  \citenamefont {Weitenberg}}]{flaschner_experimental_2016}%
  \BibitemOpen
  \bibfield  {author} {\bibinfo {author} {\bibfnamefont {N.}~\bibnamefont
  {Fläschner}}, \bibinfo {author} {\bibfnamefont {B.~S.}\ \bibnamefont {Rem}},
  \bibinfo {author} {\bibfnamefont {M.}~\bibnamefont {Tarnowski}}, \bibinfo
  {author} {\bibfnamefont {D.}~\bibnamefont {Vogel}}, \bibinfo {author}
  {\bibfnamefont {D.-S.}\ \bibnamefont {Lühmann}}, \bibinfo {author}
  {\bibfnamefont {K.}~\bibnamefont {Sengstock}},\ and\ \bibinfo {author}
  {\bibfnamefont {C.}~\bibnamefont {Weitenberg}},\ }\bibfield  {title}
  {\bibinfo {title} {Experimental reconstruction of the {Berry} curvature in a
  {Floquet} {Bloch} band},\ }\href {https://doi.org/10.1126/science.aad4568}
  {\bibfield  {journal} {\bibinfo  {journal} {Science}\ }\textbf {\bibinfo
  {volume} {352}},\ \bibinfo {pages} {1091} (\bibinfo {year}
  {2016})}\BibitemShut {NoStop}%
\bibitem [{\citenamefont {Bianco}\ and\ \citenamefont
  {Resta}(2011)}]{bianco_local_2011}%
  \BibitemOpen
  \bibfield  {author} {\bibinfo {author} {\bibfnamefont {R.}~\bibnamefont
  {Bianco}}\ and\ \bibinfo {author} {\bibfnamefont {R.}~\bibnamefont {Resta}},\
  }\bibfield  {title} {\bibinfo {title} {Mapping topological order in
  coordinate space},\ }\href {https://doi.org/10.1103/PhysRevB.84.241106}
  {\bibfield  {journal} {\bibinfo  {journal} {Phys. Rev. B}\ }\textbf {\bibinfo
  {volume} {84}},\ \bibinfo {pages} {241106(R)} (\bibinfo {year}
  {2011})}\BibitemShut {NoStop}%
\bibitem [{\citenamefont {Caio}\ \emph {et~al.}(2019)\citenamefont {Caio},
  \citenamefont {Möller}, \citenamefont {Cooper},\ and\ \citenamefont
  {Bhaseen}}]{caio_topological_2019}%
  \BibitemOpen
  \bibfield  {author} {\bibinfo {author} {\bibfnamefont {M.~D.}\ \bibnamefont
  {Caio}}, \bibinfo {author} {\bibfnamefont {G.}~\bibnamefont {Möller}},
  \bibinfo {author} {\bibfnamefont {N.~R.}\ \bibnamefont {Cooper}},\ and\
  \bibinfo {author} {\bibfnamefont {M.~J.}\ \bibnamefont {Bhaseen}},\
  }\bibfield  {title} {\bibinfo {title} {Topological marker currents in {Chern}
  insulators},\ }\href {https://doi.org/10.1038/s41567-018-0390-7} {\bibfield
  {journal} {\bibinfo  {journal} {Nature Physics}\ }\textbf {\bibinfo {volume}
  {15}},\ \bibinfo {pages} {257} (\bibinfo {year} {2019})}\BibitemShut
  {NoStop}%
\bibitem [{\citenamefont {Irsigler}\ \emph {et~al.}(2019)\citenamefont
  {Irsigler}, \citenamefont {Zheng},\ and\ \citenamefont
  {Hofstetter}}]{irsigler_chern_2019}%
  \BibitemOpen
  \bibfield  {author} {\bibinfo {author} {\bibfnamefont {B.}~\bibnamefont
  {Irsigler}}, \bibinfo {author} {\bibfnamefont {J.-H.}\ \bibnamefont
  {Zheng}},\ and\ \bibinfo {author} {\bibfnamefont {W.}~\bibnamefont
  {Hofstetter}},\ }\bibfield  {title} {\bibinfo {title} {Microscopic
  characteristics and tomography scheme of the local chern marker},\ }\href
  {https://doi.org/10.1103/PhysRevA.100.023610} {\bibfield  {journal} {\bibinfo
   {journal} {Phys. Rev. A}\ }\textbf {\bibinfo {volume} {100}},\ \bibinfo
  {pages} {023610} (\bibinfo {year} {2019})}\BibitemShut {NoStop}%
\bibitem [{\citenamefont {Chalopin}\ \emph {et~al.}(2020)\citenamefont
  {Chalopin}, \citenamefont {Satoor}, \citenamefont {Evrard}, \citenamefont
  {Makhalov}, \citenamefont {Dalibard}, \citenamefont {Lopes},\ and\
  \citenamefont {Nascimbene}}]{Chalopin2020}%
  \BibitemOpen
  \bibfield  {author} {\bibinfo {author} {\bibfnamefont {T.}~\bibnamefont
  {Chalopin}}, \bibinfo {author} {\bibfnamefont {T.}~\bibnamefont {Satoor}},
  \bibinfo {author} {\bibfnamefont {A.}~\bibnamefont {Evrard}}, \bibinfo
  {author} {\bibfnamefont {V.}~\bibnamefont {Makhalov}}, \bibinfo {author}
  {\bibfnamefont {J.}~\bibnamefont {Dalibard}}, \bibinfo {author}
  {\bibfnamefont {R.}~\bibnamefont {Lopes}},\ and\ \bibinfo {author}
  {\bibfnamefont {S.}~\bibnamefont {Nascimbene}},\ }\bibfield  {title}
  {\bibinfo {title} {{Probing chiral edge dynamics and bulk topology of a
  synthetic Hall system}},\ }\href {https://doi.org/10.1038/s41567-020-0942-5}
  {\bibfield  {journal} {\bibinfo  {journal} {Nature Physics}\ }\textbf
  {\bibinfo {volume} {16}},\ \bibinfo {pages} {1017} (\bibinfo {year}
  {2020})}\BibitemShut {NoStop}%
\bibitem [{\citenamefont {Denschlag}\ \emph {et~al.}(2000)\citenamefont
  {Denschlag}, \citenamefont {Simsarian}, \citenamefont {Feder}, \citenamefont
  {Clark}, \citenamefont {Collins}, \citenamefont {Cubizolles}, \citenamefont
  {Deng}, \citenamefont {Hagley}, \citenamefont {Helmerson}, \citenamefont
  {Reinhardt}, \citenamefont {Rolston}, \citenamefont {Schneider},\ and\
  \citenamefont {Phillips}}]{Denschlag2000}%
  \BibitemOpen
  \bibfield  {author} {\bibinfo {author} {\bibfnamefont {J.}~\bibnamefont
  {Denschlag}}, \bibinfo {author} {\bibfnamefont {J.~E.}\ \bibnamefont
  {Simsarian}}, \bibinfo {author} {\bibfnamefont {D.~L.}\ \bibnamefont
  {Feder}}, \bibinfo {author} {\bibfnamefont {C.~W.}\ \bibnamefont {Clark}},
  \bibinfo {author} {\bibfnamefont {L.~A.}\ \bibnamefont {Collins}}, \bibinfo
  {author} {\bibfnamefont {J.}~\bibnamefont {Cubizolles}}, \bibinfo {author}
  {\bibfnamefont {L.}~\bibnamefont {Deng}}, \bibinfo {author} {\bibfnamefont
  {E.~W.}\ \bibnamefont {Hagley}}, \bibinfo {author} {\bibfnamefont
  {K.}~\bibnamefont {Helmerson}}, \bibinfo {author} {\bibfnamefont {W.~P.}\
  \bibnamefont {Reinhardt}}, \bibinfo {author} {\bibfnamefont {S.~L.}\
  \bibnamefont {Rolston}}, \bibinfo {author} {\bibfnamefont {B.~I.}\
  \bibnamefont {Schneider}},\ and\ \bibinfo {author} {\bibfnamefont {W.~D.}\
  \bibnamefont {Phillips}},\ }\bibfield  {title} {\bibinfo {title} {{Generating
  Solitons by Phase Engineering of a Bose-Einstein Condensate}},\ }\href
  {https://doi.org/10.1126/science.287.5450.97} {\bibfield  {journal} {\bibinfo
   {journal} {Science}\ }\textbf {\bibinfo {volume} {287}},\ \bibinfo {pages}
  {97} (\bibinfo {year} {2000})},\ \Eprint
  {https://arxiv.org/abs/https://www.science.org/doi/pdf/10.1126/science.287.5450.97}
  {https://www.science.org/doi/pdf/10.1126/science.287.5450.97} \BibitemShut
  {NoStop}%
\bibitem [{\citenamefont {Andersen}\ \emph {et~al.}(2006)\citenamefont
  {Andersen}, \citenamefont {Ryu}, \citenamefont {Clad\'e}, \citenamefont
  {Natarajan}, \citenamefont {Vaziri}, \citenamefont {Helmerson},\ and\
  \citenamefont {Phillips}}]{Andersen2006}%
  \BibitemOpen
  \bibfield  {author} {\bibinfo {author} {\bibfnamefont {M.~F.}\ \bibnamefont
  {Andersen}}, \bibinfo {author} {\bibfnamefont {C.}~\bibnamefont {Ryu}},
  \bibinfo {author} {\bibfnamefont {P.}~\bibnamefont {Clad\'e}}, \bibinfo
  {author} {\bibfnamefont {V.}~\bibnamefont {Natarajan}}, \bibinfo {author}
  {\bibfnamefont {A.}~\bibnamefont {Vaziri}}, \bibinfo {author} {\bibfnamefont
  {K.}~\bibnamefont {Helmerson}},\ and\ \bibinfo {author} {\bibfnamefont
  {W.~D.}\ \bibnamefont {Phillips}},\ }\bibfield  {title} {\bibinfo {title}
  {Quantized rotation of atoms from photons with orbital angular momentum},\
  }\href {https://doi.org/10.1103/PhysRevLett.97.170406} {\bibfield  {journal}
  {\bibinfo  {journal} {Phys. Rev. Lett.}\ }\textbf {\bibinfo {volume} {97}},\
  \bibinfo {pages} {170406} (\bibinfo {year} {2006})}\BibitemShut {NoStop}%
\bibitem [{\citenamefont {Moulder}\ \emph {et~al.}(2012)\citenamefont
  {Moulder}, \citenamefont {Beattie}, \citenamefont {Smith}, \citenamefont
  {Tammuz},\ and\ \citenamefont {Hadzibabic}}]{Moulder2012}%
  \BibitemOpen
  \bibfield  {author} {\bibinfo {author} {\bibfnamefont {S.}~\bibnamefont
  {Moulder}}, \bibinfo {author} {\bibfnamefont {S.}~\bibnamefont {Beattie}},
  \bibinfo {author} {\bibfnamefont {R.~P.}\ \bibnamefont {Smith}}, \bibinfo
  {author} {\bibfnamefont {N.}~\bibnamefont {Tammuz}},\ and\ \bibinfo {author}
  {\bibfnamefont {Z.}~\bibnamefont {Hadzibabic}},\ }\bibfield  {title}
  {\bibinfo {title} {{Quantized supercurrent decay in an annular Bose-Einstein
  condensate}},\ }\href {https://doi.org/10.1103/PhysRevA.86.013629} {\bibfield
   {journal} {\bibinfo  {journal} {Physical Review A—Atomic, Molecular, and
  Optical Physics}\ }\textbf {\bibinfo {volume} {86}},\ \bibinfo {pages}
  {013629} (\bibinfo {year} {2012})}\BibitemShut {NoStop}%
\bibitem [{\citenamefont {Shi}\ \emph {et~al.}(2016)\citenamefont {Shi},
  \citenamefont {Rosenberg}, \citenamefont {Chiesa},\ and\ \citenamefont
  {Zhang}}]{shiRashbaSpinOrbitCoupling2016}%
  \BibitemOpen
  \bibfield  {author} {\bibinfo {author} {\bibfnamefont {H.}~\bibnamefont
  {Shi}}, \bibinfo {author} {\bibfnamefont {P.}~\bibnamefont {Rosenberg}},
  \bibinfo {author} {\bibfnamefont {S.}~\bibnamefont {Chiesa}},\ and\ \bibinfo
  {author} {\bibfnamefont {S.}~\bibnamefont {Zhang}},\ }\bibfield  {title}
  {\bibinfo {title} {Rashba {Spin}-{Orbit} {Coupling}, {Strong} {Interactions},
  and the {BCS}-{BEC} {Crossover} in the {Ground} {State} of the
  {Two}-{Dimensional} {Fermi} {Gas}},\ }\href
  {https://doi.org/10.1103/PhysRevLett.117.040401} {\bibfield  {journal}
  {\bibinfo  {journal} {Physical Review Letters}\ }\textbf {\bibinfo {volume}
  {117}},\ \bibinfo {pages} {040401} (\bibinfo {year} {2016})}\BibitemShut
  {NoStop}%
\bibitem [{\citenamefont {Girardeau}(1960)}]{girardeau_relationship_1960}%
  \BibitemOpen
  \bibfield  {author} {\bibinfo {author} {\bibfnamefont {M.}~\bibnamefont
  {Girardeau}},\ }\bibfield  {title} {\bibinfo {title} {Relationship between
  {Systems} of {Impenetrable} {Bosons} and {Fermions} in {One} {Dimension}},\
  }\href {https://doi.org/10.1063/1.1703687} {\bibfield  {journal} {\bibinfo
  {journal} {Journal of Mathematical Physics}\ }\textbf {\bibinfo {volume}
  {1}},\ \bibinfo {pages} {516} (\bibinfo {year} {1960})}\BibitemShut {NoStop}%
\bibitem [{\citenamefont {Lieb}\ and\ \citenamefont
  {Liniger}(1963)}]{lieb_exact_1963}%
  \BibitemOpen
  \bibfield  {author} {\bibinfo {author} {\bibfnamefont {E.~H.}\ \bibnamefont
  {Lieb}}\ and\ \bibinfo {author} {\bibfnamefont {W.}~\bibnamefont {Liniger}},\
  }\bibfield  {title} {\bibinfo {title} {Exact {Analysis} of an {Interacting}
  {Bose} {Gas}. {I}. {The} {General} {Solution} and the {Ground} {State}},\
  }\href {https://doi.org/10.1103/PhysRev.130.1605} {\bibfield  {journal}
  {\bibinfo  {journal} {Physical Review}\ }\textbf {\bibinfo {volume} {130}},\
  \bibinfo {pages} {1605} (\bibinfo {year} {1963})}\BibitemShut {NoStop}%
\bibitem [{\citenamefont {Bandeira}\ \emph {et~al.}(2014)\citenamefont
  {Bandeira}, \citenamefont {Cahill}, \citenamefont {Mixon},\ and\
  \citenamefont {Nelson}}]{bandeira_phase_2014}%
  \BibitemOpen
  \bibfield  {author} {\bibinfo {author} {\bibfnamefont {A.~S.}\ \bibnamefont
  {Bandeira}}, \bibinfo {author} {\bibfnamefont {J.}~\bibnamefont {Cahill}},
  \bibinfo {author} {\bibfnamefont {D.~G.}\ \bibnamefont {Mixon}},\ and\
  \bibinfo {author} {\bibfnamefont {A.~A.}\ \bibnamefont {Nelson}},\ }\bibfield
   {title} {\bibinfo {title} {Saving phase: Injectivity and stability for phase
  retrieval},\ }\href
  {https://doi.org/https://doi.org/10.1016/j.acha.2013.10.002} {\bibfield
  {journal} {\bibinfo  {journal} {Applied and Computational Harmonic Analysis}\
  }\textbf {\bibinfo {volume} {37}},\ \bibinfo {pages} {106} (\bibinfo {year}
  {2014})}\BibitemShut {NoStop}%
\bibitem [{\citenamefont {Greiner}\ \emph {et~al.}(2001)\citenamefont
  {Greiner}, \citenamefont {Bloch}, \citenamefont {Mandel}, \citenamefont
  {Hänsch},\ and\ \citenamefont {Esslinger}}]{greiner_exploring_2001}%
  \BibitemOpen
  \bibfield  {author} {\bibinfo {author} {\bibfnamefont {M.}~\bibnamefont
  {Greiner}}, \bibinfo {author} {\bibfnamefont {I.}~\bibnamefont {Bloch}},
  \bibinfo {author} {\bibfnamefont {O.}~\bibnamefont {Mandel}}, \bibinfo
  {author} {\bibfnamefont {T.~W.}\ \bibnamefont {Hänsch}},\ and\ \bibinfo
  {author} {\bibfnamefont {T.}~\bibnamefont {Esslinger}},\ }\bibfield  {title}
  {\bibinfo {title} {Exploring {Phase} {Coherence} in a {2D} {Lattice} of
  {Bose}-{Einstein} {Condensates}},\ }\href
  {https://doi.org/10.1103/PhysRevLett.87.160405} {\bibfield  {journal}
  {\bibinfo  {journal} {Physical Review Letters}\ }\textbf {\bibinfo {volume}
  {87}},\ \bibinfo {pages} {160405} (\bibinfo {year} {2001})}\BibitemShut
  {NoStop}%
\bibitem [{\citenamefont {Daix}\ \emph {et~al.}(2026)\citenamefont {Daix},
  \citenamefont {Dixmerias}, \citenamefont {He}, \citenamefont {Verstraten},
  \citenamefont {de~Jongh}, \citenamefont {Peaudecerf}, \citenamefont {Zhang},\
  and\ \citenamefont {Yefsah}}]{daix_observing_2026}%
  \BibitemOpen
  \bibfield  {author} {\bibinfo {author} {\bibfnamefont {C.}~\bibnamefont
  {Daix}}, \bibinfo {author} {\bibfnamefont {M.}~\bibnamefont {Dixmerias}},
  \bibinfo {author} {\bibfnamefont {Y.-Y.}\ \bibnamefont {He}}, \bibinfo
  {author} {\bibfnamefont {J.}~\bibnamefont {Verstraten}}, \bibinfo {author}
  {\bibfnamefont {T.}~\bibnamefont {de~Jongh}}, \bibinfo {author}
  {\bibfnamefont {B.}~\bibnamefont {Peaudecerf}}, \bibinfo {author}
  {\bibfnamefont {S.}~\bibnamefont {Zhang}},\ and\ \bibinfo {author}
  {\bibfnamefont {T.}~\bibnamefont {Yefsah}},\ }\bibfield  {title} {\bibinfo
  {title} {Observing spatial charge and spin correlations in a strongly
  interacting fermi gas},\ }\href {https://doi.org/10.1103/2t2k-3ftx}
  {\bibfield  {journal} {\bibinfo  {journal} {Phys. Rev. Lett.}\ }\textbf
  {\bibinfo {volume} {136}},\ \bibinfo {pages} {153402} (\bibinfo {year}
  {2026})}\BibitemShut {NoStop}%
\bibitem [{\citenamefont {de~Jongh}\ \emph {et~al.}(2025)\citenamefont
  {de~Jongh}, \citenamefont {Verstraten}, \citenamefont {Dixmerias},
  \citenamefont {Daix}, \citenamefont {Peaudecerf},\ and\ \citenamefont
  {Yefsah}}]{Jongh2025}%
  \BibitemOpen
  \bibfield  {author} {\bibinfo {author} {\bibfnamefont {T.}~\bibnamefont
  {de~Jongh}}, \bibinfo {author} {\bibfnamefont {J.}~\bibnamefont
  {Verstraten}}, \bibinfo {author} {\bibfnamefont {M.}~\bibnamefont
  {Dixmerias}}, \bibinfo {author} {\bibfnamefont {C.}~\bibnamefont {Daix}},
  \bibinfo {author} {\bibfnamefont {B.}~\bibnamefont {Peaudecerf}},\ and\
  \bibinfo {author} {\bibfnamefont {T.}~\bibnamefont {Yefsah}},\ }\bibfield
  {title} {\bibinfo {title} {Quantum gas microscopy of fermions in the
  continuum},\ }\href {https://doi.org/10.1103/PhysRevLett.134.183403}
  {\bibfield  {journal} {\bibinfo  {journal} {Phys. Rev. Lett.}\ }\textbf
  {\bibinfo {volume} {134}},\ \bibinfo {pages} {183403} (\bibinfo {year}
  {2025})}\BibitemShut {NoStop}%
\bibitem [{\citenamefont {Ring}\ and\ \citenamefont
  {Schuck}(2004)}]{ring_nuclear_2004}%
  \BibitemOpen
  \bibfield  {author} {\bibinfo {author} {\bibfnamefont {P.}~\bibnamefont
  {Ring}}\ and\ \bibinfo {author} {\bibfnamefont {P.}~\bibnamefont {Schuck}},\
  }\href@noop {} {\emph {\bibinfo {title} {The {Nuclear} {Many}-{Body}
  {Problem}}}},\ \bibinfo {edition} {1st}\ ed.,\ Theoretical and {Mathematical}
  {Physics}\ (\bibinfo  {publisher} {Springer Berlin, Heidelberg},\ \bibinfo
  {year} {2004})\BibitemShut {NoStop}%
\bibitem [{\citenamefont {Dobaczewski}\ \emph {et~al.}(1996)\citenamefont
  {Dobaczewski}, \citenamefont {Nazarewicz}, \citenamefont {Werner},
  \citenamefont {Berger}, \citenamefont {Chinn},\ and\ \citenamefont
  {Decharg\'e}}]{dobaczewski_mean_field_1996}%
  \BibitemOpen
  \bibfield  {author} {\bibinfo {author} {\bibfnamefont {J.}~\bibnamefont
  {Dobaczewski}}, \bibinfo {author} {\bibfnamefont {W.}~\bibnamefont
  {Nazarewicz}}, \bibinfo {author} {\bibfnamefont {T.~R.}\ \bibnamefont
  {Werner}}, \bibinfo {author} {\bibfnamefont {J.~F.}\ \bibnamefont {Berger}},
  \bibinfo {author} {\bibfnamefont {C.~R.}\ \bibnamefont {Chinn}},\ and\
  \bibinfo {author} {\bibfnamefont {J.}~\bibnamefont {Decharg\'e}},\ }\bibfield
   {title} {\bibinfo {title} {Mean-field description of ground-state properties
  of drip-line nuclei: Pairing and continuum effects},\ }\href
  {https://doi.org/10.1103/PhysRevC.53.2809} {\bibfield  {journal} {\bibinfo
  {journal} {Phys. Rev. C}\ }\textbf {\bibinfo {volume} {53}},\ \bibinfo
  {pages} {2809} (\bibinfo {year} {1996})}\BibitemShut {NoStop}%
\end{thebibliography}
\end{document}